\documentclass{elsarticle}
\usepackage[margin=1in]{geometry}

\usepackage{soul}

\usepackage{fancyhdr} 
\usepackage{color}
\usepackage{hyperref} 
\usepackage{natbib}

\usepackage{pdfpages} 
\makeatletter
\AtBeginDocument{\let\LS@rot\@undefined}
\makeatother
\usepackage{verbatim} 
\usepackage{amsmath} 
\usepackage{amsthm} 
\usepackage{amssymb}	
\usepackage{graphicx} 

\let\baraccent=\= 
\renewcommand{\=}[1]{\stackrel{#1}{=}} 

\newcommand{\gae}{\lower 2pt \hbox{$\,
    \buildrel{\scriptstyle >}\over {\scriptstyle \sim}\,$}}
\newcommand{\lae}{\lower 2pt \hbox{$\,
    \buildrel{\scriptstyle <}\over {\scriptstyle \sim}\,$}}

\newcommand{\ket}[1]{\big| #1 \big\rangle} 
\newcommand{\bra}[1]{\big\langle #1 \big|} 

\let\adot=\dot 
\renewcommand{\dot}[2]{\v{#1}\cdot \v{#2}} 

\usepackage{xcolor}
\definecolor{violet}{rgb}{0.58, 0.0, 0.83}

\def\beq{\begin{equation}}
\def\eeq{\end{equation}}
\def\bea{\begin{eqnarray}}
\def\eea{\end{eqnarray}}

\begin{document}

\title{Dynamics and Transport at the Threshold of Many-Body Localization} 

\author{Sarang Gopalakrishnan}
\address{Department of Physics and Astronomy, CUNY College of Staten Island,
Staten Island, NY 10314, USA}
\address{Physics Program and Initiative for the Theoretical Sciences,
The Graduate Center, CUNY, New York, NY 10016, USA}

\author{S.~A.~Parameswaran}
\address{Rudolf Peierls Centre for Theoretical Physics,  Clarendon Laboratory, University of Oxford, Oxford OX1 3PU, UK}

\date{\today}

\begin{abstract}
Many-body localization (MBL) describes a class of systems that do not approach thermal
equilibrium under their intrinsic dynamics; MBL and conventional thermalizing systems form
distinct dynamical phases of matter, separated by a phase transition at which equilibrium
statistical mechanics breaks down. True many-body localization is known to occur only under
certain stringent conditions for perfectly isolated one-dimensional systems, with Hamiltonians
that have strictly short-range interactions and lack any continuous non-Abelian symmetries.
However, in practice, even systems that are not strictly MBL can be nearly MBL, with
equilibration rates that are far slower than their other intrinsic timescales; thus, anomalously slow
relaxation occurs in a much broader class of systems than strict localization. In this review we
address transport and dynamics in such nearly-MBL systems from a unified perspective. Our
discussion covers various classes of such systems: (i) disordered and quasiperiodic systems on
the thermal side of the MBL-thermal transition; (ii) systems that are strongly disordered, but
obstructed from localizing because of symmetry, interaction range, or dimensionality; (iii)
multiple-component systems, in which some components would in isolation be MBL but others
are not; and finally (iv) driven systems whose dynamics lead to exponentially slow rates of
heating to infinite temperature. A theme common to many of these problems is that they can be
understood in terms of approximately localized degrees of freedom coupled to a heat bath (or
baths) consisting of thermal degrees of freedom; however, this putative bath is itself nontrivial,
being either small or very slowly relaxing. We discuss anomalous transport, diverging relaxation
times, and other signatures of the proximity to MBL in these systems. We also survey recent theoretical and numerical methods that have been applied to study dynamics on either side of the MBL transition.
\end{abstract}
\maketitle

\tableofcontents

\section{ Introduction: many-body localization\label{sec:intro}}

How an isolated system approaches thermal equilibrium, starting from an arbitrary initial state, is one of the basic questions in many-body physics. This question is puzzling because the initial state contains an extensive amount of information, while the final equilibrium state is well characterized by a small amount of thermodynamic data; nevertheless, the dynamics leading from one to the other is reversible and conserves information content. This puzzle has spawned many research directions, in quantum chaos, random matrix theory, and quantum statistical mechanics~\cite{bohigas, Berry_1977, shankar1985, deutsch_quantum_1991, srednicki_chaos_1994, jarzynski_berry, rigol_thermalization_2008, polkovnikov_colloquium_2011}; a version of it also appears to be relevant to questions of information loss in black holes~\cite{Harlow2013, magan2011, Lashkari2013, Shenker2014, Maldacena2016}. 

In the quantum setting, with which we are concerned here, the eigenstate thermalization hypothesis (ETH) offers one resolution to this puzzle~\cite{deutsch_quantum_1991, srednicki_chaos_1994, jarzynski_berry, rigol_thermalization_2008, polkovnikov_colloquium_2011}. The most basic form of ETH claims that the expectation values of \emph{local} operators are the \emph{same} in all eigenstates with similar values of the energy density and other conserved densities; also, this shared value coincides with the equilibrium prediction. The ETH explains information loss as follows: for simplicity say the initial state of the system was a pure state $|\psi\rangle = \sum_n c_n |n\rangle$, where $|n\rangle$ are the eigenstates of the system's Hamiltonian with eigenvalues $E_n$. Under Hamiltonian evolution, this state becomes $|\psi(t) \rangle = \sum_n c_n e^{i E_n t} |n \rangle$; if one constructs the density matrix $|\psi(t)\rangle \langle \psi(t)|$ and averages out temporal oscillations (on the rationale, e.g., that the measurement is averaged over some time window), one arrives at the ``diagonal ensemble''~\cite{polkovnikov_colloquium_2011} density matrix $\rho_d = \sum_n |c_n|^2 |n \rangle \langle n|$. For simple initial states (e.g., product states), the standard deviation of the energy scales as the square root of system size, so the energy density is sharply defined in the thermodynamic limit. Thus the $|c_n|$ are appreciable only in a narrow energy window. In this narrow window, the ETH predicts that all eigenstates are locally indistinguishable, so the coefficients $|c_n|$ are irrelevant. The information about the initial state is in the phases that we averaged over; for large systems these phases are effectively unmeasurable. 

The ETH offers an elegant resolution to the puzzle of quantum thermalization in isolated quantum systems. Traditional condensed matter systems are far from being isolated, so this resolution was not initially seen as especially practically relevant. However, ultracold atomic gases are much closer to the limit of an isolated system, and their advent has led to a revival of interest in the many-body physics of isolated systems. The question of thermalization gained some urgency when experiments directly looking for quantum thermalization---in the so-called quantum Newton's cradle setup~\cite{kinoshita_quantum_2006}---found no sign of it. The failure of thermalization was related to the approximate integrability of the one-dimensional gases studied in that experiment; much subsequent effort has gone into understanding violations of ETH in integrable systems as well as other classes of quantum systems~\cite{polkovnikov_colloquium_2011}. There are now two well-understood classes of ETH-violating systems: (i)\emph{integrable} quantum systems (which are fine-tuned but approximately realized in many experiments on quasi-one-dimensional materials (see, e.g., Ref.~\cite{coldea}) as well as quantum gases~\cite{simon2011, gring_relaxation_2012, erne2018, dipolar_cradle}); and (ii) many-body localized (MBL) systems~\cite{basko_metalinsulator_2006}, which are a dynamical ``phase'' of strongly random quantum systems.
In the MBL phase, states that are initially away from equilibrium \emph{never} relax; conserved quantities have strictly zero transport coefficients even at nonzero temperatures, and generic autocorrelation functions of local operators saturate at late times to nonequilibrium values, thus retaining some memory of their initial conditions at arbitrarily late times~\cite{nandkishore_mbl_2015}. 
The term ``phase'' is used to emphasize that arbitrary local perturbations of an MBL Hamiltonian preserve its MBL character.

The absence of local relaxation in the MBL phase is related to the fact that the many-body energy eigenstates of an MBL system violate the ETH.
For generic quantum spin chains with quenched randomness, there seem to be at least two phases, a ``thermal'' phase at weak randomness in which all eigenstates satisfy the ETH, and an MBL phase at strong randomness in which all eigenstates violate it (this is sometimes called the fully MBL phase). Whether intermediate phases between these exist is an open question. The thermal and MBL phases can be distinguished by the decay of autocorrelation functions, as noted above, and also by the scaling of the bipartite entanglement entropy (for real-space cuts) of eigenstates at nonzero energy density: in the thermal phase such eigenstates have volume-law entanglement entropy (with a prefactor set by thermodynamics); in the MBL phase, they have area law entanglement. The properties deep in these two phases are relatively well understood and are reviewed in Refs.~\cite{nandkishore_mbl_2015, voskreview, abaninreview}. 

Although the initial arguments for MBL were very general, the bulk of subsequent work has considered one-dimensional lattice systems with spin-$\frac{1}{2}$ degrees of freedom on each lattice site, subject to a (possibly time-periodic) Hamiltonian that contains on-site random potentials and generic nearest-neighbor interactions. For this class of one-dimensional, short-range models the existence of the MBL phase can be proved under certain minimal assumptions~\cite{jzi}. It is unclear whether the MBL phase exists, in the strict sense, under more general conditions: in the continuum, in higher dimensions, in systems with aperiodic but nonrandom on-site potentials, or systems with interactions that decay algebraically with distance. It is also unclear whether systems exist in which some eigenstates are MBL but others are thermal (this is called the ``many-body mobility edge'' scenario~\cite{basko_metalinsulator_2006}). Perturbation theory and small-system numerical studies support MBL in all of these cases; however, these approaches might miss the rare-region effects that were argued to destabilize the MBL phase in many of these situations~\cite{drhms, drh, ldrh, papic2015many}. In any case, the strict MBL phase does not exist in imperfectly isolated systems~\cite{basko_metalinsulator_2006, basko_expt, Nandkishore14}, and no experimental system is perfectly isolated. 

To be testable and widely applicable, therefore, a theory of the MBL phase must also address the properties of systems that are not strictly MBL on the longest timescales. 
This review article is about such ``nearly many-body localized'' systems. 
In nearly MBL systems, the timescales for interaction effects and for transport (or the decay of local autocorrelation functions) are widely separated;
the system looks MBL-like if it is probed between these two timescales. 
We will be more precise about the interaction timescale below, but intuitively what we mean by this is the time taken for quantum information to spread among two neighboring \emph{excitations}: this timescale matters because essentially all the characteristic phenomena in MBL have to do with being at nonzero energy density above the ground state. 
Between the interaction and transport timescales, we expect a nearly MBL system to exhibit the phenomenology of the true MBL phase, which we will briefly summarize below. What sets the transport timescale---and, indeed, whether there is a single transport timescale or a broad spectrum of them---is a question about the mechanisms by which the MBL phase is destabilized; explaining these mechanisms is our central task here.

There are many different ways for a system to be ``nearly MBL''---for instance, by violating any one of the conditions that make strict MBL possible---and so far each of these has been analyzed on its own terms in the literature. And to some extent the instabilities we catalogue really are different in their mechanisms. However, these nearly MBL systems are also broadly similar in some key aspects, which derive from the observation that almost all nearly MBL systems involve localized ``typical'' regions coupled to a sparse, slowly relaxing, and spatially heterogeneous network of thermal degrees of freedom. Delineating these resemblances is the task of this review.

\subsection{Scope and organization of this review}

There are many review articles at this point on the MBL phase and its properties, as well as some of the phenomena discussed here (such as rare-region effects, slow dynamics, and the role of symmetries in MBL)~\cite{nandkishore_mbl_2015, voskreview, abaninreview, ppvreview, Luitz16Review, vasseur2016nonequilibrium, kartiek_review, prelovsek_review, imbrie2017local, de2017many}. The present review covers some of these topics, but differs in perspective in two important ways. First, we are concerned with nearly MBL systems, for which the thermalization timescale is a crucial part of the physics; estimates of thermalization rates are therefore central to our discussion. Since thermalization cannot be theoretically established in essentially any system, this means that in practice we will be estimating the rates of \emph{instabilities} of the localized phase. This leads to the second difference in perspective, which is that we shall assume all apparent instabilities of the localized phase lead to thermalization, unless there are specific theoretical reasons to believe otherwise. This is an explicit \emph{assumption} that runs through the present work: despite some crucial numerical support in toy models~\cite{ldrh, plhc, papic2015many}, by and large the numerical evidence for these instabilities is weak or nonexistent. However, because most numerical evidence is based on exact diagonalization of small ($L \leq 32$ site) systems, or other methods that are either biased toward finding low-entanglement localized states, one would not expect clear numerical evidence even if the instability existed. In the absence of such evidence, these instabilities are admittedly speculative. We choose, nevertheless, to assume that these instabilities exist because this assumption leads to concrete predictions. By contrast, assuming that the putative instabilities are somehow averted yields no specific dynamical predictions. 

The rest of this review is structured as follows. In this section, we will review what is currently believed about the properties of the MBL phase, its dynamics, and its instabilities, particularly the MBL-to-thermal phase transition in one dimension. In Sec.~\ref{sec:thermoMBL} we will briefly review the ``standard'' scenario for one-dimensional systems that are slightly on the thermal side of the MBL transition, in both random and quasiperiodic systems. In Sec.~\ref{sec:unstableMBL} we will turn to systems in which local perturbative considerations suggest MBL but global properties of the system, such as its symmetry or dimensionality, forbid MBL. These systems therefore eventually thermalize, but on anomalously long timescales. In Sec.~\ref{sec:multicomponent} we will extend these considerations to multiple-component systems in which some components are thermal and others are localized. The simplest of these is a small MBL system coupled to a large thermal bath, for which the analysis is relatively straightforward; however, there are many other cases with richer physics. In Sec.~\ref{sec:prethermal} we will briefly consider slow thermalization, emergent conservation laws, and MBL in periodically driven (Floquet) systems; as we will discuss there, Floquet systems offer a particularly simple illustration of some of the mechanisms that block thermalization more generally. In Sec.~\ref{sec:dynamicalsig}  we will tie together these case studies and point out---or, more accurately, re-emphasize--some features they share with one another, as well as with classical glasses. In Sec.~\ref{sec:tools}, we will review various theoretical and algorithmic ideas that have lately been developed to come up with reliable approximations of MBL, nearly MBL, and thermal dynamics. Finally we will conclude in Sec.~\ref{sec:conclusions} with a partial list of open questions.

\subsection{The many-body localized phase}

This section is organized as follows. First, we will introduce the restricted class of models for which the existence of the MBL phase can be proved under modest assumptions. Next, we will provide a perturbative argument for MBL that applies in a far wider variety of models. Finally, we will introduce the nonperturbative ``avalanche'' instability that prevents MBL even in cases where it appears to be perturbatively stable. 

\subsubsection{Models with an MBL phase}\label{sec:intro:mblconservative}

We first introduce a class of models that are generally agreed to have an MBL phase (see, however, Ref.~\cite{vidmar2019}). These are one-dimensional lattice models in which each lattice site hosts a $q$-state system (usually $q = 2$), and the Hamiltonian takes the form
\beq\label{basicham}
\hat H = \hat H_0 + \lambda \hat H_1,
\eeq
where $\hat H_0 = \sum_i  \xi_i \hat h_i$ is a sum of local operators $\hat h_i$ that commute with one another, and $\xi_i$ are spatially uncorrelated random numbers from a reasonably continuous (e.g., box or Gaussian) distribution. The distribution of $\xi_i$ and the spectrum of $\hat h_i$ are assumed to be sufficiently well-behaved that $\hat H_0$ has a nondegenerate spectrum and Poisson energy-level statistics. (We will discuss the case where $\hat H_0$ has symmetry-related degeneracies later.) The nontrivial dynamics is due to the term $\hat H_1$, which contains arbitrary $n$-site operators acting on contiguous lattice sites (usually $n = 2$). The terms $\hat H_0$ and $\hat H_1$ can be either time-independent or periodically modulated in time (leading to ``Floquet'' dynamics~\cite{bukov_universal_2015, kuwahara2016floquet, moessner2017equilibration}). For some purposes the Floquet case---with no conservation laws---is conceptually simpler than the Hamiltonian one, since there are precisely no conserved densities---not even energy---and the thermal steady state is always at ``infinite temperature''~\cite{luca2013, iyer2013many, ponte2015periodically, abanin2016theory, zhang2016floquet, bordia2017periodically}. The form~\eqref{basicham} is thus a little too restrictive; however, it does describe all the models that have been studied to date. 

When $\lambda = 0$ these models trivially have an extensive set of conserved quantities, since $\hat H$ commutes with each of the $\hat h_i$. For $0 \ll \lambda \ll 1$ one can still construct a full set of conserved quantities, which are ``dressed'' versions of the operators $\hat h_i$. Ref.~\cite{jzi, jzi2} shows that when $|\lambda| \ll 1$ these exact, dressed conserved quantities remain local, up to exponential tails. The argument~\cite{jzi} proceeds by applying a sequence of local unitary transformations to the original Hamiltonian to decouple all non-diagonal terms\footnote{The argument relies crucially on an assumption about the many-body level statistics of the full Hamiltonian $H$, i.e., that levels do not attract strongly, in the sense that the probability of finding two levels much closer together than the typical level spacing is not much higher than if the levels were Poisson distributed. This assumption is needed because of the lack of mathematical methods for proving statements about level statistics. From a physical perspective, however, strong generic level attraction would be a far more exotic phenomenon than MBL.}.Thus, one can write down the following form for the Hamiltonian~\eqref{basicham} (specializing now to the case of two-level systems):
\beq\label{eq:lbit}
\hat H_{\mathrm{LIOM}} = \sum_i \epsilon_i \hat \tau^z_i + \sum_{ij} J_{ij} \hat \tau^z_i \hat \tau^z_j + \sum_{ijk} K_{ij} \hat \tau^z_i \hat \tau^z_j \hat \tau^z_k + \ldots
\eeq 
where these operators $\hat \tau^z_i$ are the conserved quantities, also known as local integrals of motion (LIOMs) or l-bits~\cite{huse_phenomenology_2014, serbyn_local_2013}. Any Hamiltonian can be written in the form~\eqref{eq:lbit}; however, generically the $\hat \tau^z_i$ will be highly nonlocal operators. In the MBL phase, the $\hat \tau^z_i$ are quasilocal, in the following sense: if we fix a size $\ell$, (i)~a typical $\hat \tau^z_i$ can be approximated by an operator that lives strictly on $\ell$ sites with an error that scales exponentially in $\ell$, and (ii)~the fraction of operators for which this approximation fails vanishes faster than any power law in $\ell$. The ``diagonal'' interaction terms are quasilocal in the same double sense: typical terms fall off exponentially with distance and the probability of large terms is suppressed, though not necessarily exponentially. It is not clear at present whether the typical decay length for the ``size'' of the LIOM and its effective interactions are the same (or even simply related)~\cite{vrgop}. 

For small enough $|\lambda|$, the typical situation is that a $\hat \tau^z_j$ can unambiguously be identified with a single physical site. Regions where this identification fails are called resonant spots, and play an important role in the phenomenology of the MBL phase.

Systems for which a description of the form~\eqref{eq:lbit} exists are evidently in the MBL phase. Generic local operators will have non-vanishing overlap with nearby LIOMs, so their autocorrelation functions will not decay to zero. The ETH is violated, since to specify an eigenstate one must specify not just a few global conserved quantities but extensively many LIOMs. The entanglement area law for individual eigenstates across typical cuts follows from the observation that the unitary transformation that reduces the Hamiltonian to the form~\eqref{eq:lbit} can be approximated by a finite-depth quantum circuit~\cite{mozgunov, bauer_area_2013} with an error that decreases faster than any power law in the depth.

This introduction to the MBL phase has deliberately been dry; the aim has been to lay out the conditions under which the existence of an MBL phase is relatively uncontroversial. We turn next to a perturbative argument for MBL, which dates back to Fleishman and Anderson~\cite{fleishman}, and was extended to all orders by Basko, Aleiner, and Altshuler~\cite{basko_metalinsulator_2006}; as we shall see, this perturbative argument applies under far more general conditions than those we have stipulated above. 

\subsubsection{Perturbative argument for MBL\label{sec:intro:mblpert}}

Consider a simple model such as the nearest-neighbor Ising model with transverse and longitudinal fields on an arbitrary $d$-dimensional lattice:
\beq
H = \sum_i h_i \sigma^z_i + \Gamma \sigma^x_i + \sum_{\langle ij \rangle} J_{ij} \sigma^z_i \sigma^z_j.
\eeq
We take $h, J$ to typically be of order unity and $\Gamma \ll 1$. When $\Gamma = 0$ the Hamiltonian consists of commuting operators, so any bit-string in the $\sigma^z$-basis is a many-body eigenstate (we shall also refer to these as ``configurations''). We now consider the stability of these states to the addition of a small transverse field; we begin with a typical state in the middle of the spectrum, i.e., at infinite temperature. Suppose we attempt to connect two configurations that differ in the states of $n$ contiguous spins. The typical energy denominator for this process is $n/2^{n}$, while the typical matrix element scales as $\Gamma^n$. Meanwhile, the number of such contiguous shapes also scales exponentially with $n$. Thus, when $\Gamma$ is sufficiently small, none of these rearrangements is typically resonant, and perturbation theory converges in a typical region of the sample. Further, the radius of convergence is temperature-dependent: for states closer to the bottom of the spectrum we have a level spacing $e^{-s(T) n}$, where $s \rightarrow 0$ as $T \rightarrow 0$. 
In the regime where all the states are typically stable against avalanches, the convergence of perturbation theory allows one to define l-bits locally via Schrieffer-Wolff transformations. 

A clarification is in order here about the relation between this perturbative argument and the often-mentioned concept of ``Fock-space localization''~\cite{altshuler_quasiparticle_1997, berkovits1998, monthus_many-body_2010, deluca2014, beaud2017low, logan2019}. In the MBL phase, eigenstates are not localized in configuration space or Fock space, in the sense of having a finite inverse participation ratio~\cite{kramer_RMP}. This is a direct consequence of the perturbative argument above: if one divides the system into $N$ finite-size blocks, the wavefunction in each block will have a large overlap with one configuration $| c \rangle$, but will also have some nondegenerate perturbative corrections. Thus it will take the schematic form $\langle c |\psi_{\mathrm{block}} \sim (1 - \epsilon)$. The global overlap is multiplicative across blocks, and will thus scale as $(1 - \epsilon)^N$, decaying exponentially with $N$, albeit with a much smaller prefactor than in the thermal phase. This argument can also be applied to the IPR in the Fock space of single-particle localized orbitals (see Sec.~\ref{sec:intro:fleish} below). 
The sense in which these states are localized is that presented in Sec.~\ref{sec:intro:mblconservative}: the state can be disentangled with a finite-depth quantum circuit~\cite{bauer_area_2013}, and the LIOMs have a finite inverse participation ratio in the space of single-site operators. This point is generally understood but left implicit in many discussions of the MBL phase. 

The argument sketched out above has three basic failure modes. First, it assumes the energies in a block of size $n$ to be approximately independent numbers, and might fail if there are many exact or near-degeneracies in the spectrum. Second, it relies on the assumption that the dominant rearrangements are of contiguous regions; this is plausible for short-range interactions but not for, say, power-law interactions, which might favor sparse clusters that have a much larger shape entropy. Third, and most generally, even if perturbation theory converges in \emph{typical} regions it will always fail in some fraction of a thermodynamically large system. For MBL to exist, it is crucial that this failure should stay spatially localized and not propagate out into the rest of the system. We will now turn to this question. 

\subsubsection{Rare-region instabilities}\label{mblinstability}

To motivate the rare-region instability in the quantum case, it is helpful to consider a simpler version of this phenomenon that occurs in random classical spin chains~\cite{oph, basko_diffusion}. One could construct such spin chains, for example, by reinterpreting the spin variables in Eq.~\eqref{basicham} as classical spins. We consider initializing the spin chain in a random product state. When the fields are strong, typical spins precess around their effective local field with weak perturbations from their neighbors that can be incorporated via KAM theory~\cite{basko_diffusion}. However, there will some finite density of few-spin clusters where the local precession frequencies are matched to allow for resonant coupling. These clusters evolve chaotically. 
This chaotic evolution acts as noise on the other spins, and renders their motion unpredictable on sufficiently long timescales. In a classical system, any finite density of such local chaotic spots is sufficient to cause global chaos (and is related to the absence of a KAM theorem in the thermodynamic limit). 

In quantum spin chains, the situation is fundamentally different because finite subsystems have a finite level spacing. To ``infect'' their neighbors, this level spacing must be small enough. Indeed, in one dimension one can show (in terms of the phenomenology invoked in Sec.~\ref{sec:intro:mblconservative}) that typical regions are immune to chaotic spots. We consider the simplest such model, consisting of a chaotic spot coupled to a typical region~\cite{drh, ldrh, de2017many}. The full Hamiltonian is given by 
\beq
\mathcal{H} = H_{\mathrm{LIOM}} + H_{\mathrm{RMT}} + g H_{\mathrm{int}},
\eeq
where $H_{\mathrm{LIOM}}$ describes the typical region, in which a rotation into LIOMs (Sec.~\ref{sec:intro:mblconservative}) is possible; $H_{\mathrm{RMT}}$ describes the chaotic region, which we have modeled as a random matrix; and $H_{\mathrm{int}}$ is the boundary coupling between the two regions. A very general form for $H_{\mathrm{int}}$ is the tensor product of an arbitrary operator acting on the RMT Hilbert space and a generic local physical operator on the boundary of the MBL system. We can expand the local operator in terms of LIOMs, as follows:

\beq\label{boundaryterm}
H_{\mathrm{int}} = O \otimes (h^\alpha_i \tau^\alpha_i + J^{\alpha\beta}_{ij} \tau^\alpha_i \tau^\beta_j + \ldots)
\eeq
where summation over repeated indices is assumed, and we can take the coefficients to fall off as \\ $\exp[- \max(i, j, \ldots) / \xi]$ (this follows from the assumption that $H_{\mathrm{LIOM}}$ is localized. For simplicity we keep only the single-LIOM terms in both $H_{\mathrm{LIOM}}$ and $H_{\mathrm{int}}$ in what follows. Thus we are describing the localized region as a noninteracting Anderson insulator. This potentially biases our discussion toward finding MBL stable, but as we will shortly see, even the present model is unstable outside of the one-dimensional short-range limit. 

We are now set up to perturb in $H_{\mathrm{int}}$. In the simple limit we have specialized to, LIOMs are decoupled and can be included sequentially. We incorporate them beginning with those nearest to the bath.  The bath initially has a featureless spectrum with a discrete (but fine) level spacing. The LIOMs located near the interface hybridize with the bath provided that the Golden Rule rate $\sim g^2$ exceeds the level spacing of the bath, $2^{-\ell_{\mathrm{RMT}}}$. We now re-diagonalize the system consisting of the absorbed l-bit and the bath; this causes the effective level spacing of the spectral lines of the operator $O$ to halve, but does not seem to induce any further structure in the spectral function of the operator $O$. We then iterate this procedure, eliminating the couplings between increasingly distant l-bits and the bath. At the $n$th stage, the matrix element is $g^2 \exp(-2n/\xi)$, while the level spacing of the bath is $2^{- (n + \ell_{\mathrm{RMT}})}$. Depending on $\xi$, two outcomes are possible as $n \rightarrow \infty$: for $\xi < \xi_c \equiv \log 2/2$, the Golden Rule ultimately ceases to apply on a scale set by $\ell / (1 - \xi/\xi_c)$, and the system is asymptotically stable against the introduction of a bath. When $\xi > \xi_c$, on the other hand, the Golden Rule applies out to indefinite distances, and a single bath destabilizes an entire MBL chain. 

In one dimension, therefore, MBL can be immune against the introduction of locally chaotic regions, consistent with Imbrie's proof~\cite{jzi}. In higher dimensions, this does not appear to be possible. Instead, when the bath has incorporated spins out to some distance $\Lambda$, the typical level spacing falls off as $2^{- (\ell_{RMT} + \Lambda)^d}$ whereas the matrix element only falls off as $\exp(-\Lambda/\xi)$. Asymptotically, therefore, the bath grows large and effectively classical, and infects the bulk of the MBL system. In Sec.~\ref{sec:unstableMBL} we will revisit this instability, and provide estimates for the critical size of the initial chaotic region, and of the associated length and time scales. 

An obvious objection to the argument above is that our estimates of matrix elements and level spacings depend on treating the bath as a featureless random matrix even after it has absorbed a large number of spins. This clearly cannot be strictly true, as the composite system consisting of the bath and the spins it has added clearly has intrinsic slow timescales. However, we note that the main effect of incorporating a new spin is to \emph{split} the spectral lines of the bath operator $O$ into pairs of lines with roughly equal weight, rather than to redistribute the spectral weight of this operator; and the Thouless time of the larger bath is still (by construction) much shorter than its inverse level spacing, so that its levels are still locally random-matrix like in their correlations. The issue of spectral structure is explored in more detail in Ref.~\cite{potirniche2019}. 

The strongest evidence for the correctness of the avalanche picture, however, comes from the numerical results of Ref.~\cite{ldrh}. This work studies a toy model of the class described above, with an Anderson insulator coupled to a random-matrix bath. Importantly, they take $H_{int.}$ to have the simple form $g \sum_i O \exp(-i/\xi) \sigma^x_i$; this neglects fluctuations in the matrix elements and thus decreases finite size effects. As $\xi$ is increased past the expected value, one sees a clear transition in this toy model: even when the bath is initially too small to couple to distant spins, it abruptly becomes strongly coupled to them when $\xi$ crosses its critical value. This demonstrates that, \emph{in principle}, the avalanche mechanism can destabilize an MBL system, so that certain general objections to the avalanche instability---such as the idea that a small bath cannot destabilize a much larger system---cannot be valid. The numerical situation is less clear-cut in models with realistic fluctuations in the couplings $g$, and/or interactions among the LIOMs. While we expect these modifications to increase finite-size effects, however, no clear scenario exists in which such modifications would avert the instability described above, so we presume that it occurs generally. 

\subsection{MBL in context}

The high level of activity on many-body localization in the past decade has been driven heavily by conceptual puzzles about eigenstate thermalization, entanglement dynamics, and other primarily theoretical considerations. However, the MBL phase and closely related phenomena have a rich prehistory in condensed matter physics. We now review some of the contexts in which these ideas have arisen over the years. {Our selection of topics here is biased toward those that seem relevant to MBL as it is currently understood, i.e., as a finite-temperature dynamical phenomenon. Thus, for example, we have entirely omitted the vast literature on the zero-temperature properties of disordered interacting electrons, since the many subtle issues that arise there seem orthogonal to those we will discuss here.}

\subsubsection{Phonon-less hopping conductivity of electrons in semiconductors}\label{sec:intro:fleish}

The first systems in which Anderson localization was carefully studied were lightly doped disordered semiconductors~\cite{shklovskii_book}. In these systems, at low temperatures, the single-electron states near the Fermi energy are well localized. Transport takes place due to incoherent hopping mechanisms: electrons jump from one localized orbital to another by absorbing or emitting phonons (or some other excitation). The low-temperature physics of tightly localized electrons interacting via the Coulomb interaction exhibits many features in common with glasses, and is known as the ``electron glass''~\cite{pollak2013electron}. Most electron glasses that have been studied are well described by a classical model in which electrons hop incoherently between sites, with activated hopping rates. However, the ``ultrafast'' temporal regime in which quantum many-body effects play an important part in relaxation has also been accessed using pump-probe spectroscopy~\cite{thorsmolle}.

The temperature-dependence of the hopping conductivity is set by phase-space considerations; the nature of the energy bath that facilitates incoherent hopping only determines the prefactor. But Fleishman and Anderson~\cite{fleishman} showed, remarkably, that this prefactor is strictly zero for short-range electron-electron interactions (including the Coulomb interaction in a two-dimensional electron gas). Absent electron-phonon interactions, therefore, there would be \emph{no} hopping transport in this model at low temperatures\footnote{In addition to phoonons, they also neglect spin fluctuations, which as we shall see are an important relaxation channel, and thus technically applies only in the presence of a finite magnetic field.}. Of course, in practice the Coulomb interaction is sufficiently long-range to provide a bath for incoherent hopping.

The perturbative argument of Ref.~\cite{fleishman} already contains the basic intuition behind the MBL phase, and has a slightly different structure than what we presented in Sec.~\ref{sec:intro:mblpert}, so we present it briefly here. Keeping only the low-energy electron states near the Fermi energy, one can write down a reduced Hamiltonian of the form
\beq\label{fermionham}
H = \sum_\alpha c^\dagger_\alpha c_\alpha + \sum_{\alpha\beta\gamma\delta} V_{\alpha\beta\gamma\delta} c^\dagger_\alpha c^\dagger_\beta c_\gamma c_\delta,
\eeq
where $c_\alpha$ annihilates an electron in the localized single-particle orbital $\alpha$, and the interaction term $V_{\alpha\beta\gamma\delta}$ involves a correlated two electron-hop. Thus, $V_{\alpha\beta\gamma\delta}$ decays exponentially in the distance between the localization centers of $\alpha$ and $\gamma$ (as well as $\beta$ and $\delta$), but might decay as a power law in the distance between $\alpha$ and $\beta$. For simplicity we take all of these to be short-range with a characteristic scale $\xi \approx 1$, the localization length. The perturbative argument now proceeds as follows: the matrix element for rearranging a single pair of orbitals is $\sim V$; however, there are only $O(1)$ pairs of orbitals with an appreciable matrix element. For small $V$, none of the energy differences between nearby orbitals are within $V$ of one another, so the interaction does not rearrange the electronic state. Going to longer distances does not help, since the matrix element falls off exponentially with distance and the phase space for resonances only grows polynomially. Going to higher orders in perturbation theory does not help either, as shown by Basko \emph{et al.}~\cite{basko_metalinsulator_2006} and discussed in Sec.~\ref{sec:intro:mblpert}. 

The fact that short-range interactions do not relax a localized system is an instance of the general observation that Fermi's Golden Rule is violated in localized states: even when \emph{on average} the system has a finite density of states at the requisite energy, these states are not accessible from the initial state: the energy denominators and matrix elements are anticorrelated in a way that makes the Golden Rule decay rate vanish~\cite{mott1968, sivan}. We will return to this point in Sec.~\ref{sec:intro:mbldynamics} while discussing response functions in the MBL phase.

\subsubsection{Quasiparticle lifetime in quantum dots}

The Fleishman-Anderson argument combines real-space and Fock-space structure: transitions occur between pairs of many-body states that are connected by few-body \emph{and} spatially local moves. A simpler analysis is possible if one turns to a zero-dimensional problem, which is a quantum dot with $N \gg 1$ energy levels; we are interested in its behavior on energy scales smaller than the Thouless energy $E_{\mathrm{Th}}$ (i.e., the rate at which particles diffuse across the system), so that it is effectively zero-dimensional~\cite{agkl}. (This theoretical setup was inspired by the experiments of Ref.~\cite{sivan1994spectroscopy}.) Recall that the dimensionless conductance of a system is $g \equiv E_{\mathrm{Th}}/\Delta$, where $\Delta$ is a level spacing. Thus to have many levels within an energy window $E_{\mathrm{Th}}$ we require $g \gg 1$, i.e., a metallic grain. To start, we treat Coulomb interactions at the mean-field level. The system consists of $N$ single-particle orbitals, with energies and wavefunctions given by random-matrix theory. We now reinstate interaction effects, to get a Hamiltonian formally identical to Eq.~\eqref{fermionham}, except that now the interaction terms have no spatial structure, but can be regarded as random uncorrelated Gaussian variables. In the presence of interactions, a single quasiparticle state might no longer be stable: instead, a high-energy quasiparticle can decay into a lower-energy state, emitting a quasiparticle-quasihole pair in the process. At the Golden Rule level, there is always a density of states for this process so all quasiparticles have a ``lifetime''~\cite{sivan2}.%

Altshuler et al.~\cite{agkl} (see also Ref.~\cite{fyodorov1997}) observed that this model is in fact a hopping model on a hypercube, in which each lattice site is labeled by a length-$N$ binary string (with 1 denoting occupied and 0 denoting empty orbitals). The energy of a many-body state is found by adding the energies of the occupied quasiparticles, and the matrix elements $V_{\alpha\beta\gamma\delta}$ are hopping amplitudes on the Fock-space hypercube. Since the hypercube is very high-dimensional, one can approximate it as being locally treelike; this approximation, together with neglecting the correlations between the energies of many-body states, allows one to reduce the many-body problem to one of (single-particle) Anderson localization on a tree-like graph; this latter problem was addressed in previous work, and is known to have a localization transition~\cite{abouchacra, chalker_siak, cayleytree3, cayleytree4}. 
At finite $N$, of course, this transition cannot be sharp; moreover, the mapping to the Cayley tree is only approximate. 
There are no truly delocalized states; rather, quasiparticles with a finite lifetime $\tau$ manifest themselves as clusters of $(\Delta \tau)^{-1}$ closely spaced spectral lines of approximately equal intensity, while localized quasiparticles instead correspond to single spectral peaks with exponentially suppressed satellites. 

The nature of this localization problem has been explored in detail, both in the quantum-dot context (see Ref.~\cite{gornyi2016} for a review of this theoretical literature) and as an interesting localization transition in its own right, possibly featuring an intermediate critical phase~\cite{pino_nonergodic, altshuler_nonergodic, biroli2017}. Other problems that are conceptually closely related are the localization transition in the quantum random energy model~\cite{laumann_many-body_2014} and the Fermi-liquid/non-Fermi-liquid transition in coupled Sachdev-Ye-Kitaev models~\cite{symodel, PhysRevB.95.134302}. An important feature of the Cayley-tree localization transition is that, because the phase space grows exponentially with distance (so there are $K^n$ $n$th nearest neighbors), the matrix element must fall off \emph{at least} as fast as $1/K$ at each step if the localized phase is to be stable. There is thus a critical value of the Fock-space localization length beyond which resonances proliferate and delocalize the system. This feature is shared by the MBL transition, as we noted above (Sec.~\ref{mblinstability}).  
\subsubsection{Spin dynamics in many-body systems}

An important piece of physics left out in the Fleishman-Anderson argument is the existence of electron spin. Ironically, Anderson's original work on localization~\cite{pwa} was motivated by experiments on spin diffusion~\cite{feher}; however, understanding localization in spin excitations has proved practically daunting (even the extraction of definite spin diffusion constants is a relatively recent achievement~\cite{zhang_first_1998}). The important conceptual role of spin excitations in relaxing MBL states of the ``charge'' degrees of freedom is a topic we will return to in Sec.~\ref{sec:multicomponent:spins}. In practice, spin dynamics has been considered primarily in the context of electron-spin resonance (ESR) in generic many-body systems, and in a class of magnetic materials with large on-site magnetic moments. In most cases, the relevant spin-spin interactions are dipolar. It seems that dipolar systems cannot exhibit true MBL (Sec.~\ref{sec:unstable:powerlaw}); however, the relaxation times can quite naturally be very slow compared with microscopic timescales, leading to \emph{effective} MBL at strong disorder. 

\emph{Electron-spin resonance and spectral diffusion}.---An early appearance of something like the LIOM model~\eqref{eq:lbit} is in the work by Klauder and Anderson~\cite{klauder_anderson} on spectral diffusion~\cite{portis}. Spectral diffusion is a phenomenon where the precession frequency of a spin fluctuates slowly in time because of the flipping of neighboring spins. In a disordered system, these neighboring spins are generically far off resonance with the spin of interest; thus the only physically relevant part of the spin-spin interaction is the diagonal term, $\sigma^z_i \sigma^z_j$. This diagonal interaction is supplemented by a stochastic Markov spin-flip process to capture spectral diffusion. The resulting phenomenology is closely related to that of MBL in the presence of a bath, which we will revisit in Secs.~\ref{sec:multicomponent:largebath}-\ref{sec:multicomponent:backbath}.

\emph{Long coherence times in disordered dipolar magnets}.---A couple of years before the seminal work~\cite{basko_metalinsulator_2006} on MBL, experiments on the magnetic material LiHo$_x$Y$_{1-x}$F$_4$~\cite{ghosh2002} provided evidence for anomalously long coherence times in solid-state systems. Holmium has a large magnetic moment, which leads to strong magnetic dipole-dipole interactions, and when $x = 1$ the system is a ferromagnet. Diluting the ferromagnet leads first to a spin glass, and then, when $x \leq 0.05$, to an unusual phase of matter. A remarkable feature of this regime is the presence of long-lived coherent oscillations in free induction decay. Free induction decay involves applying a strong rf magnetic field that oscillates at a frequency $\omega$; after this field is turned off, spins that were excited by the field continue to precess for an amount of time that is limited by intrinsic decoherence mechanisms. Ref.~\cite{ghosh2002} found, remarkably, that coherent oscillations with large $Q$-factors (exceeding 100) persist at frequencies well below the scale set by the temperature: these coherent oscillations take place at $5$~Hz when the sample temperature is $0.11$~K (which corresponds to a frequency of many GHz). This suggests the existence of almost isolated degrees of freedom that are far from thermal equilibrium, since in equilibrium the system would have a vanishingly weak magnetic response at frequencies so far below the temperature. The microscopic origin of these degrees of freedom remains controversial, but similar coherent oscillations have now also been seen in other disordered magnets, such as gadolinium gallium garnet~\cite{ghosh2008}. These isolated two-level systems were also recently reconsidered from the point of view of MBL~\cite{silevitch}. 

\emph{Color centers in diamond}.---Recently, dense ensembles of nitrogen-vacancy centers in diamond~\cite{doherty_nitrogen-vacancy_2013} have emerged as a new platform for studying strongly correlated disordered spin systems~\cite{kucsko1, kucsko2}. Nitrogen-vacancy centers are defects in diamond that support electronic spins with long coherence times, and are optically addressable. When the density of nitrogen-vacancy centers is sufficiently high, the dipolar interactions between these spins become important. Moreover, the spins are randomly placed, and experience both random effective fields (i.e., inhomogeneous broadening) and random interactions. In three dimensions, the dipole-dipole interaction asymptotically leads to delocalization (as noted above); however, the nature of polarization decay is strongly non-exponential, as explained in Ref.~\cite{kucsko2}. We will return to this theory Sec.~\ref{sec:unstable:powerlaw}, in our discussion of MBL in the presence of long-range interactions. 

\subsubsection{Two-level systems in glasses}

It has been believed for a long time that the low-temperature thermodynamics and response of structural glasses---or more generally amorphous solids---are dominated by localized two-level systems (TLSs). The microscopic origin of TLSs is still not fully settled, but slowly fluctuating atoms are a major candidate: an atom in a glass is ``caged'' in a potential due to its neighbors, but sometimes this potential has two minima, which the atom tunnels between, thus forming a TLS. Other candidate TLSs are discussed in Ref.~\cite{cole_tls}; they include electrons that tunnel between two accessible states, slowly fluctuating electronic or nuclear spins, and emergent objects such as polarons or spin clusters. The nature of TLSs in amorphous media has lately seen a revival of interest because TLSs turn out to limit the coherence times of superconducting qubits~\cite{cole_tls}. TLSs dominate the low-temperature specific heat of glasses, so their density can be estimated from thermodynamics; to fit experimental results, one also assumes that they have a broad distribution of tunneling rates. This broad distribution is in any case natural given the exponential dependence of tunneling amplitudes on the parameters of the tunneling barrier. A consequence of broadly distributed tunneling rates is that TLSs cause $1/f$ noise in superconducting qubits. 

TLSs in glasses are most often treated as noninteracting; this assumption gives the ``standard tunneling model''~\cite{cole_tls, esquinazi}, which has successfully been used to describe many of their properties. There are many instances in which interactions among TLSs seem to be physically crucial, however. The most important of these is spectral diffusion, discussed in the previous section. Possibly related to this is the fact that if one drives a resonator hard, the absorption saturates much more slowly than the standard tunneling model would predict. This would make sense if the TLSs underwent spectral diffusion, so that the saturated TLSs drift outside the resonance window while fresh TLSs drift into this window~\cite{Faoro2012a, Faoro2015}. Spectral diffusion can be put in either as a stochastic process~\cite{black_halperin} or derived quasi-microscopically from a model of interacting dipoles~\cite{burin1998}. At present, however, a detailed microscopic understanding of TLS-TLS interactions is lacking. 

\subsubsection{Quantum solids}

A very different line of thought that also converged on the idea of MBL was due to Kagan and Maksimov~\cite{kagan1, kagan2}. These works begin by considering the motion of heavy quantum particles---which, for concreteness, we take to be bosons---in a perfect crystalline lattice. An example would be vacancies or interstitials in solid helium. The particles are taken to be heavy enough that the nearest-neighbor density-density interaction greatly exceeds the hopping amplitude; moreover, the interaction is taken to fall off as a rapid power law, $1/R^\alpha$ (which is relevant for the Van der Waals interactions in quantum solids). An $n$-site cluster can only move resonantly, with a matrix element that is suppressed exponentially in $n$. Given the power law falloff of the interaction, however, the energy denominator will only be the same if the initial and final locations of the cluster have the same effective on-site energy. Because of the power-law interactions, the on-site energy depends on the states of spins out to some exponentially large distance. In a given background configuration, these are vanishingly unlikely to be the same in the initial and final states. Thus, in this limit, clusters of particles are effectively immobile, since they can only move via large-scale rearrangements, which become increasingly nonresonant. Once a percolating cluster of sites forms, the system will become essentially immobile, except for a small fraction of atoms rattling in voids in the cluster. This transition is analyzed in considerable detail in Ref.~\cite{kagan2}; it is very similar in spirit to the arguments for disorder-free MBL that we will discuss in Sec.~\ref{sec:unstable:disorderfree}.

\subsubsection{Ultracold atoms and ions}

Soon after the advent of ultracold atomic gases, it was appreciated that such gases offer an especially promising platform for studying nonequilibrium dynamics. An early demonstration of this was in Ref.~\cite{greiner_collapse_2002}, in which a Bose-Einstein condensate was trapped in an optical lattice and the lattice potential was suddenly quenched to a large value. This led to repeated oscillations of the order parameter amplitude, and thus to slow relaxation (compared with the characteristic dynamical timescales). Another major breakthrough was the quantum Newton's cradle experiment~\cite{kinoshita_quantum_2006}, which demonstrated the absence of thermalization in a one-dimensional Bose gas on extremely long timescales. Finally, noninteracting Anderson localization was demonstrated, both in quasiperiodic lattices and in random speckle potentials~\cite{aspect1, aspect2, aspect3, inguscio1, inguscio2, inguscio3}. 

A generic mechanism for slow thermalization in the large-$U$ limit of the Hubbard model (either fermionic or bosonic) was extensively studied in the years following these breakthroughs~\cite{kollath2007, eckstein_thermalization_2009, bernier2011}: ultimately this has to do with the fact that a multiply-occupied site has an excess energy $U$, which cannot easily relax, because the only available energy-conserving relaxation process involves creating a shower of low-energy excitations with characteristic energy scale $t$. This very naturally leads to sluggish relaxation. This slow timescale was experimentally observed in Ref.~\cite{strohmaier_observation_2010}. A related, roughly contemporaneous finding was that in the large-$U$ limit of the Bose-Hubbard model, there exist a tower of edge-localized many-boson states~\cite{haque_flach}. The simplest of these states consists of three particles at the leftmost site of an otherwise empty system. They can only move at third order, with matrix element $t^3/U^2$. At first sight, one might expect this process to happen, just very slowly. However, note that the state with three particles at any site away from the edge gets a second-order, nondegenerate perturbative shift of $-2 t^2/U$ from the virtual hopping of one boson, because bulk sites have two neighbors. Meanwhile, the state at the edge gets a shift of only $-t^2/U$. The difference between these energies exceeds the matrix element, so the configuration is stuck at the edge. One can easily extend this argument to construct localized configurations with $n$ particles at any site at a distance $< n/2$ from the edge.  Ref.~\cite{adhh} provides a unified perspective on slow relaxations in systems of this type.

All of these works considered slow collective thermalization, but were not directly about MBL. The first papers directly bearing on MBL came out in 2015~\cite{kondov_Disorder_2015, Schreiber15}. The former looked at transport, while the latter directly probed the persistence of a nonequilibrium initial state. Subsequent experiments have explored MBL in higher dimensions~\cite{Hild16, Bordia16, bordia2d17}, in driven systems~\cite{bordia_periodically_2016}, in the presence of a bath~\cite{lueschenopen, rubio2018}, and in quasiperiodic potentials with a single-particle mobility edge~\cite{aidelsburger2019}. Three qualitatively different setups have been used so far. In the approach of Ref.~\cite{kondov_Disorder_2015}, a three-dimensional Bose gas is created in a speckle potential (which has short-range correlations in two directions, but a correlation length of multiple lattice sites along the third direction). The experiment involves applying a sudden momentum kick to the atomic cloud and tracking its displacement in response. 

The other experiments use one of two methods. The simpler of these uses the optical gas microscope~\cite{bakr_quantum_2009, sherson_single-atom-resolved_2010}: in these experiments~\cite{Hild16, lukin2019mbl}, the setup is placed in the field of view of the ``microscope'' and one takes site-resolved snapshots of the atomic distribution after waiting for a certain evolution time. This setup offers great flexibility in the accessible observables, allowing one to measure entanglement directly~\cite{islam_measuring_2015, lukin2019mbl}; however, it is limited to modest system sizes and cannot be extended past two dimensions. The other class of experiments is more restrictive but also more scalable. In these experiments, one initializes the atoms in a strong superlattice potential that has twice the wavelength of the ``primary'' optical lattice in which time evolution happens. This way, only every other site of the primary optical lattice is occupied. Then one turns off the superlattice, while leaving on both the primary optical lattice and a second lattice that creates a quasiperiodic potential. (There are also, in general, strong lattice beams in two directions that split up the system into an array of largely independent tubes.) The system evolves from this modulated initial state for some time $t$; then one measures the remaining imbalance between even and odd sites by turning on the superlattice once again, and separately measuring the atom number in even and odd sites. 

A major advantage of these experiments is that they can access relatively large systems at much longer times than current numerical techniques allow: for instance, they can reliably explore systems with many thousands of sites, on timescales that are at least a few hundred times longer than the hopping timescale. By contrast, simulation methods are limited either to shorter times or much smaller systems. In this sense these experiments are a very valuable complement to numerical studies. On the other hand, many of the asymptotic predictions we will discuss here are restricted to a late-time regime that might lie well beyond what is accessible even in these experiments. The basic challenge is that many of the processes in the MBL phase and at the MBL transition occur exponentially slowly; thus, even going out to (e.g.) $1000$ hopping times only allows one to access $\sim 7$-spin rearrangements. Thus we do not expect these experiments to see distinctions between one and more dimensions, for example. 

Finally, we note that MBL has also recently been explored in ion-trap experiments~\cite{monroe2016}. These platforms are in some ways analogous to the gas-microscope experiments, in being highly programmable and also limited in their scalability. However, they also suffer from long-range interactions in general, making it challenging to realize a regime with true MBL in the sense of Sec.~\ref{sec:intro:mblconservative}. Nevertheless, signatures of many-body localization have been observed in this setup.

\subsection{Dynamics of the MBL phase\label{sec:intro:mbldynamics}}

In this section we return to one-dimensional MBL chains for which an effective description of the form Eq.~\eqref{eq:lbit} exists, and study the dynamics of such systems. Unlike properties such as eigenstate entanglement---which are extremely sensitive to any thermalization, however slow---the dynamical properties of MBL systems persist at short and intermediate times even when for some reason the system is thermal in the long-time limit. 
As we have noted above, even systems where MBL is unstable will often have regions in which approximate LIOMs are defined. In the fully-MBL phase, the LIOMs provide a full characterization of the eigenspectrum of the problem: we assume that each of the $2^L$ eigenstates of a system with $L$ sites is uniquely specified by fixing the values of $L$ l-bits~\footnote{Here and throughout this review, unless otherwise specified we assume a two-dimensional on-site Hilbert space.}. In this setting, one can determine many aspects of the time-evolution of local observables by expanding them in the basis of l-bits. 

The specifically \emph{many-body} nature of MBL appears in three distinct properties of the l-bit description that are all absent in the Anderson insulator; all of these properties have important implications for dynamics. First, the l-bits interact, unlike the free-fermion orbitals in an Anderson insulator. These interactions cause generic superpositions of l-bit eigenstates to dephase, leading to slow entanglement growth and related slow dynamics after quenches. Second, generic local operators have matrix elements between pairs of many-body eigenstates that differ in the values of multiple LIOMs; these matrix elements for large-scale rearrangements play an important part in the low-frequency linear response of the MBL phase. By contrast, in an Anderson insulator, an $n$-site operator can only change $2n$ orbital labels. Third, an MBL system in one dimension contains some density of ``failed'' avalanches, which contribute to the long-time linear response and quench dynamics. In this section, we are concerned with the first two effects (which are not restricted to one dimension, but apply to MBL systems on intermediate timescales in any dimension); for a discussion of Griffiths effects in one-dimensional MBL systems, we refer to Refs.~\cite{kartiek_review, morningstar}.

\emph{Fluctuations in the LIOM description}.---An important but often neglected part of the phenomenology of MBL is that the couplings $J_{ij}, K_{ijk} \ldots$ have broad, approximately log-normal distributions, which broaden as the distance between the l-bits increases. The origin of these broad distributions is similar to that of the broad conductance distributions in localized wires, and can be understood within perturbation theory~\cite{ataf}. The matrix element for an $n$th order perturbative process is generically of the form $\prod_{i \leq n} (M_i / E_i)$, where $M_i, E_i$ are respectively the matrix element and energy denominator at step $i$. Both quantities are drawn from well-behaved distributions; thus, neglecting any possible correlation effects (which is appropriate at strong disorder), one expects the \emph{logarithm} of the overall matrix element to have a normal distribution with variance $\propto n$ by the central limit theorem. This is consistent with what is seen numerically~\cite{vrgop}. The \emph{typical} matrix element falls off with a well-defined localization length despite these fluctuations (since it decays as $\exp(-n/\xi)$, with multiplicative fluctuations of order $\exp(-\sqrt{n})$). However, the existence of these broad distributions has implications for dynamics, as we will discuss below.

\subsubsection{Entanglement growth; quench dynamics; spin echo}

Historically the first nontrivial dynamical effect discovered in the MBL phase was the slow growth of entanglement~\cite{znidaric_many-body_2008, bardarson_unbounded_2012, serbyn_universal_2013}. In an MBL chain initialized in a product state, the bipartite entanglement grows as $S(t) \simeq \xi \log t$. The origin of this logarithmic growth can be explained within the LIOM picture, as follows~\cite{serbyn_universal_2013}. Two LIOMs $i, j$, both initialized in superposition states, become entangled on a timescale $J^{\mathrm{eff}}_{ij}$, because the interactions lead to dephasing (the precession frequency of each spin gets correlated with the state of the other in the z basis). At a time $t$, LIOMs that are a distance $\leq \xi \log t$ from the entanglement cut become entangled with spins on the other side of the cut, so the entanglement grows logarithmically. Also, since the system is only ``visible'' out to a distance $\log t$, the autocorrelation functions of generic local operators after a quench can decay at most as $t^{-\xi \log 2}$~\cite{serbyn_quench}. Local quenches also seem to give rise to logarithmic entanglement spreading~\cite{jed_localquench}.

A more spatially resolved probe of the dynamics in the MBL phase involves the DEER protocol~\cite{serbyn_interferometric_2014}. To motivate this protocol we first consider applying a standard spin echo pulse to a single physical degree of freedom deep in the MBL phase. The physical bit overlaps strongly with a particular LIOM, for which the spin echo is perfect. Thus the spin echo signal saturates at late times with an amplitude that measures the overlap between the physical operator and the LIOM. In what follows we take this overlap to be large and assume we are manipulating l-bits directly. Now we consider flipping a spin at a distance $\ell$ from the original test spin, halfway through the echo sequence. All the couplings \emph{except} the one between the flipped spin and the test spin are echoed out; the overall signal oscillates as $[\langle \cos(i J^{\mathrm{eff}}_{ij} t) \rangle]$, where $J^{\mathrm{eff}}_{ij} \equiv J_{ij} + \sum_k K_{ijk} \tau^z_k + \ldots$, and the average is over both the infinite-temperature initial state and many realizations of the disorder. The log-normal distribution of the effective couplings~\cite{vrgop} makes this DEER amplitude decay as $\log t$ over a temporal range that grows exponentially in time (and is set by the broadening distribution of effective couplings). The original paper~\cite{serbyn_interferometric_2014} explored a more complicated version of this setup, in which one applies a $\pi/2$ pulse to half the system; in that setup the decay of the DEER signal is algebraic in time. 

Another interesting consequence of interaction-induced dephasing is the nature of quantum revivals in the MBL phase~\cite{vpm}.  We consider a quench from an initial product state, and monitor the dynamics of an individual spin.  In a thermal system the initial product state will have non-zero overlap with an exponentially large number of many-body eigenstates; hence, the post-quench dynamics  of any single spin will see rapid dephasing in a thermalizing system. Intuitively,  coherent `revivals' of the initial state are essentially impossible  as they would require the synchronization of an exponentially large number of oscillators. This is true whether we consider the entire product state, or any individual spin. In a localized system --- with or without interactions --- an individual spin has nontrivial overlap with only  $O(\xi)$ l-bits, and therefore an initial product state has overlap with only $O(N)$ many-body eigenstates. In  a non-interacting Anderson insulator, the energy of an l-bit is independent of the states of the other l-bits; therefore, the  dynamics of any single physical  spin is  governed by only $O(\xi)$  frequencies,  and hence coherent revivals will persist to arbitrary long times --- there is neither dephasing nor dissipation.
 In the MBL phase, the many-body energy depends on the states of all the $O(N)$ spins in the  system (e.g., via `Hartree shifts'), and hence eventually one will need an exponentially large number of frequencies to synchronize in order for even a single spin to revive. However, the hierarchy of scales in the MBL system ---  that also governs the slow growth of entanglement --- means that coherent revivals persist upto logarithmically long time scales, i.e. until the system `feels' the Hartree shifts. Note that this does not preclude there being a nonzero long-time average of the spin expectation value, which is simply controlled by the overlap with l-bits. In other words, the MBL phase exhibits dephasing (and hence eventual absence of coherent revivals) but no dissipation. This thought experiment can also be reframed in terms of coupling a distinct `qubit' to the MBL system (as in ~\cite{vpm}), with similar conclusions.  Monitoring single-spin dynamics in this manner thus provides an alternative route to distinguishing between Anderson localization and MBL.

\begin{figure}[tb]
\begin{center}
\includegraphics[width = 0.35\textwidth]{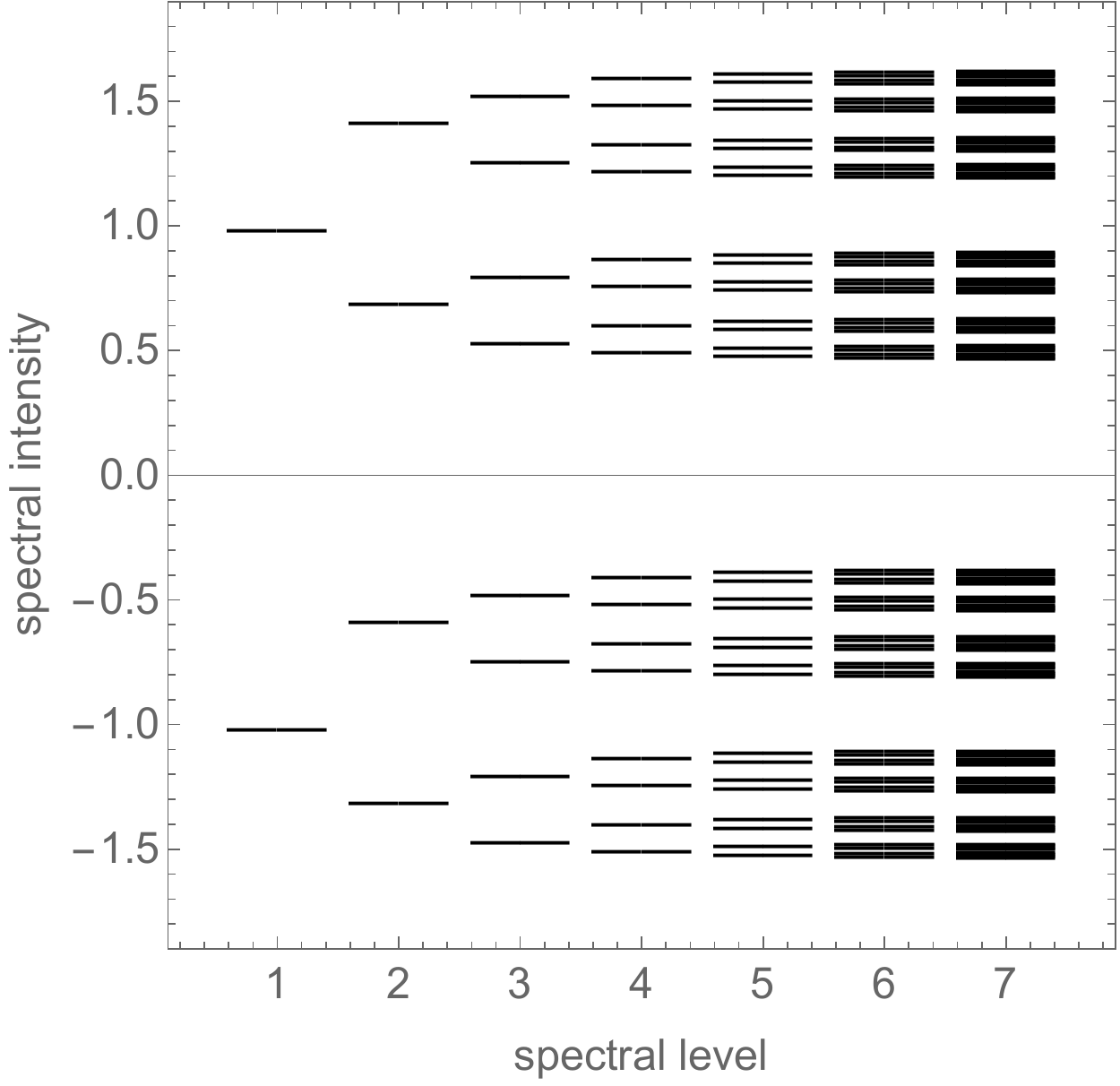}
\includegraphics[width = 0.33\textwidth]{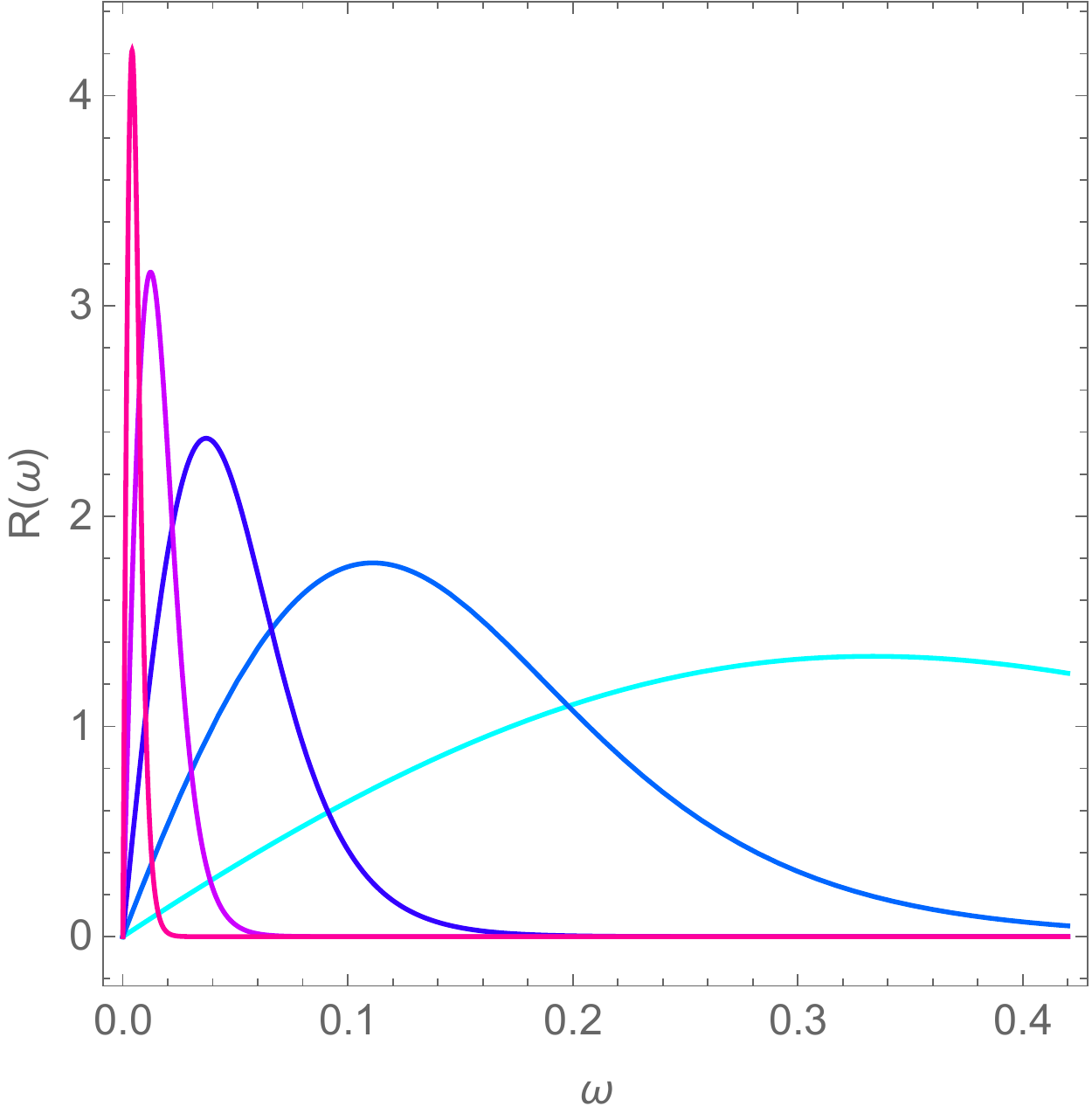}
\caption{\emph{Spectral functions in the MBL phase}. Left: ``spectral tree'' construction of the thermally averaged local spectral function. At higher levels of the tree, the couplings to increasingly distant neighbors are included. The structure of the tree is self-similar, with gaps at all energy scales (of which the smaller ones are resolution-limited). Right: ``Mott'' mechanism for low-frequency divergences in generic spectral functions $R(\omega)$ in the MBL phase. The plots clustered increasingly close to the origin show contributions to $R(\omega)$ from increasingly large-scale rearrangements. There are fewer such rearrangements (so the area under these curves is smaller), but each rearrangement has a smaller associated splitting, so the ultimate result is that spectral weight piles up near zero frequency. There will generically also be a delta-function contribution at $\omega = 0$, which is not shown here.}
\label{spectralfeatures}
\end{center}
\end{figure}

\subsubsection{Thermally averaged spectral functions}

The same mechanism that causes the logarithmic growth of entanglement also manifests itself in the behavior of thermally averaged spectral functions~\cite{Nandkishore14}. One can imagine measuring these, for example, by coupling an MBL system extremely weakly to a bath, waiting for it to equilibrate, and then measuring two-time correlation functions. Again, the local spectral function for a simple local operator (such as a bit-flip) is approximately the same as that for a LIOM: it contains a few large spectral lines corresponding to flipping the l-bits centered near that site, as well as an incoherent background, which we ignore in the following discussion. The large spectral lines can be discussed using LIOMs, since their frequencies correspond to $h^{\mathrm{eff}}_i \equiv h_i + J_{ij} \tau^z_j + \ldots$. In a given many-body eigenstate there will be a sharp delta function at the field $h^{\mathrm{eff}}$. However, the location of this line will vary from eigenstate to eigenstate. The spectral lines form a tree, with the states of farther neighbors having an increasingly weak effect on the frequency. Deep in the one-dimensional MBL phase, the splittings at step $\ell + 1$ on the tree are parametrically weaker than those at step $\ell$, so that deep in the spectral tree there are relatively few crossings. The resulting thermally averaged spectral function is not a continuum, but is instead a zero-measure fractal~\cite{Nandkishore14, vrgop} (Fig.~\ref{spectralfeatures}). 

These local spectral features are, of course, washed out by the spatial/disorder average, since different sites will have their spectral lines at different frequencies. The remaining spectral structure in the averaged spectral function is due to the distinctive properties of local spectral functions near zero frequency in localized systems, which survive the disorder average.

\subsubsection{Mott pairs and low-frequency linear response}

The phenomena discussed above were distinctive to thermal averages, or to quenches from nonequilibrium initial states. However, a different class of slow dynamical properties can be seen in linear response, even for a single eigenstate. These processes have to do with many-particle rearrangements, and dominate the behavior of spectral functions near zero frequency~\cite{Gopalakrishnan15}. We will now estimate the density of these rearrangements and their contribution to response. Here, and subsequently in this review, we will make extensive use of a standard resonance-counting approach: a perturbative resonance occurs when the matrix element of a perturbation between two energy levels exceeds the energy denominator between them. We estimate matrix elements using the properties of the localized phase, and energy denominators from the observation that the density of states near the middle of the spectrum is uniform. 

The intuition behind our estimate of linear response is as follows: in any system there will be some low density of resonant pairs of states, connected by high-order perturbative processes. For instance, an $n$-spin rearrangement will take place at $n$th order in perturbation theory, and will be suppressed by a factor $\sim \exp(-n/\zeta)$ [we use the symbol $\zeta$ for this Fock-space decay coefficient to distinguish it from a spatial localization length]. If two spin configurations that are connected by an $n$-spin rearrangement are resonant, then they will be hybridized into even and odd ``cats'' which are split by an energy $\sim \exp(-n/\zeta)$. A generic local operator will have a large matrix element connecting the two cat states, and thus have a large spectral line at the splitting between them. 

The number of available $n$-spin rearrangements involving a given spin grows exponentially in $n$ (with a prefactor that depends on the Hilbert space size as well as on the entropy of possible shapes), i.e., as $e^{sn}$. Only a fraction $\exp(-n/\zeta)$ of these possible transitions are resonant, however. Thus to leading order the number of \emph{resonant} lines is therefore $e^{n (s - 1/\zeta)}$. (Note that perturbative stability requires $s\zeta < 1$, as otherwise large-scale resonances will proliferate.) These lines are spread out over a frequency range $e^{-n/\zeta}$. Thus the \emph{density} (in frequency) of $n$th order spectral lines scales as $e^{sn}$. This structure is illustrated in Fig.~\ref{spectralfeatures}. Let us consider the response at a frequency $\omega$. By the reasoning above, the response at $\omega$ will be dominated by the largest possible rearrangement that has splitting $\geq \omega$, which corresponds to $n \sim \zeta \log \omega$. Thus, generic low-frequency response functions scale as $R(\omega) \sim \omega^{-s\zeta}$. This scaling is in addition to the delta-function peak that is generically present, due to the finite long-time saturation value of generic local operators in the MBL phase. As the MBL transition is approached from the MBL side, $s \zeta \rightarrow 1$ so $R(\omega) \sim 1/\omega$, i.e., local autocorrelation functions (and therefore, e.g., noise) all approach a $1/f$ form at the MBL transition~\cite{Gopalakrishnan15, kartiek_review}. 

These signatures are simplest to observe in the frequency domain; however, they also lead to slow real-time relaxation, as $|R(t) - R(\infty)| \sim t^{s \zeta - 1}$ at late times. We also remark that some operators that are odd under time-reversal, such as the conductivity, exhibit a different power law because the matrix elements of these operators vanish as $\omega \rightarrow 0$. Specifically, the contribution to the optical conductivity in the MBL phase from these local resonances scales as $\omega^{2 - s\zeta}$~\cite{Gopalakrishnan15}. In the one-dimensional, random case, it is now believed that this mechanism is sub-leading to that coming from fractal thermal Griffiths regions (we will return to this point below, in Sec.~\ref{sec:intro:MHRG}). In quasiperiodic systems or higher dimensions, the fractal Griffiths regions are absent; however,  this perturbative mechanism still survives at intermediate frequencies.

\subsubsection{Breakdown of linear response}

The above discussion of linear response took place in the framework of the Kubo formula. However, one might question what an equilibrium relation such as the Kubo formula \emph{means} physically for a system that does not thermalize. The appropriate thought experiment is as follows: imagine placing the sample between the plates of a parallel-plate capacitor, and driving an a.c. voltage across the capacitor; now measure the rate at which the sample absorbs energy. For a given (sufficiently high) frequency, one can decompose the sample into a sparse ensemble of two-level systems that are resonant with the drive. These two-level systems can be treated in the rotating-wave approximation, in which case they simply absorb energy from the drive at the Golden Rule rate until they saturate (on a timescale set by the inverse of the drive amplitude, $1/A$). The rate of energy absorption (``Joule heating'') precisely matches the a.c. conductivity discussed above~\cite{gopalakrishnan_regimes_2016}. 

This analysis works so long as $A \ll \omega$. However, as one tries to take the d.c. limit at constant drive amplitude, the rotating-wave approximation on the two-level systems breaks down. Instead, one enters a regime in which two-level systems repeatedly cross one another through a series of Landau-Zener transitions~\cite{ponte2015many, abanin2016theory, kns}. These Landau-Zener transitions cause thermalization on sufficiently long timescales. Although the instantaneous Hamiltonian is many-body localized, the unitary time-evolution operator is not. At very slow driving, the system delocalizes with a diffusion constant set by the dominant Landau-Zener transition rate. There also appears to be a nontrivial intermediate-frequency regime in which energy absorption takes place not via Joule heating but as an anomalous, continuously varying power law of time~\cite{Gopalakrishnan15}. Thus, the MBL phase differs from the thermal phase in that the linear-response limit and the zero-frequency limit do not commute: depending on how exactly one takes these limits, one either recovers the linear-response prediction or falls very far out of the linear response regime.

An alternative, somewhat simpler, way to probe the failure of linear response is to apply a slow local perturbation to the system~\cite{kns}. By analogy with the adiabatic manipulation of gapped systems, one might expect that one can slowly manipulate individual spins without changing the state of the others. This turns out to be quite false: rather, when one manipulates a spin slowly, it undergoes Landau-Zener transitions, leading to charge transfer over a scale $\log(1/\omega)$. For concreteness, let us consider tuning the field on a single spin at a rate $\omega$, starting in an eigenstate. As we tune the field, the many-body spectrum rearranges, and the initial state undergoes many avoided crossings. Avoided crossings with matrix elements $\leq \omega$ can be ignored, since the system goes through them fully diabatically. However, avoided crossings with matrix elements $\geq \omega$ are dangerous, because the system crosses them \emph{adiabatically}, leading to the rearrangement of charge (or magnetization). This rearrangement happens over a distance that scales as $\log(1/\omega)$, as previously noted.

\subsubsection{Bistability, noise, and temperature fluctuations}

Our discussion of transport so far has focused on a.c. probes, as these tend to be the most sensitive to the peculiar dynamical properties of the MBL phase. The linear response d.c. conductivity, by contrast, is simply zero in the MBL phase. However, nonlinear d.c. response does show characteristic signatures of localization. This is particularly the case in systems where the MBL transition is temperature-tuned, i.e., where there is a many-body mobility edge. As we will see in Sec.~\ref{sec:unstableMBL}, this transition might be rounded out into a crossover by nonperturbative effects, but the distinction between these is immaterial for the present discussion. When there is a (sharp or slightly rounded) mobility edge, we expect the d.c. conductivity to jump by some orders of magnitude with a modest change in the temperature. This suggests the following scenario for bistability~\cite{basko_expt}, which has been experimentally observed in indium oxide films~\cite{ovadia2015evidence}. Suppose one begins in the MBL phase at low temperature, and ramps up the voltage $V$ across the system. For small $V$, no current will be generated since the conductivity is effectively zero. However, for large enough voltage, we expect MBL to break down, since particles can move down the potential gradient to regions where they are more energetic and therefore delocalized. This will lead to a thermal state in which there is Joule heating, which maintains the system at a higher steady-state temperature than the leads. When the voltage is now decreased, the system can sustain itself in a higher-temperature conducting state through Joule heating, even when the leads are below the putative transition temperature. Thus the system can exhibit temperature bistability. Note that this bistability is not strictly a probe of MBL, but only of a sufficiently abrupt temperature-dependence of the d.c. conductivity. This bistability is accompanied by intermittency and enhanced noise, which has also been observed~\cite{shahar2019}.

\subsection{Strong randomness picture of MBL transitions}\label{sec:intro:MHRG}

To round out our picture of the MBL phase, we briefly discuss some recent theories of the MBL transition in random systems. These theories \emph{assume} a strong-randomness picture of the transition. This choice is dictated in part by the logic of the avalanche scenario, and also in part by practicality: we have no phenomenology for a system that is neither MBL nor thermal, but we have many natural approaches for addressing a system that is spatially separated into MBL and thermal regions. Therefore, we might as well make a virtue of necessity and work out the possible scenarios for strong-randomness transitions. The aim of the recent theories in this vein~\cite{vha, pvp, dvp, thmdr} is to develop an iterative scheme for coarse-graining such heterogeneous systems. Microscopically, such a scheme would have two components: first, one would construct quasi-LIOMs for locally MBL regions; second, one would couple these regions to nearby thermal regions---which would be modeled as random matrices---using the Golden Rule. A fully microscopic version of this scheme (e.g., such as that outlined in Ref.~\cite{thmdr}) for a realistic physical model has not yet been implemented; however, one can both infer some of the general properties of any such scheme, and construct and analyze tractable toy versions~\cite{zzdh, gvs, morningstar}. 

We first discuss some properties common to all of these RG schemes, pointed out in Ref.~\cite{dumitrescu2018kosterlitz}. First, the localization length gets renormalized upwards by the presence of small thermal blocks: these act as local short-circuits for correlations. There do not seem to be any counterbalancing mechanisms that \emph{shorten} the localization length. Second, when the typical renormalized localization length exceeds a critical value, the system undergoes an avalanche, as inclusions spread into the bulk. These two general observations are sufficient to imply an analogy between the MBL transition and the familiar Kosterlitz-Thouless transitions~\cite{kt}, both of which feature two-parameter scaling. For the MBL transition the two parameters are the thermal fraction $f(\ell)$ (i.e., the fraction of the system, at a scale $\ell$, that has already thermalized) and the typical localization length $\zeta(\ell)$ at that scale. To define these variables precisely one must specify an RG scheme, but their physical meaning is as discussed above. In terms of these variables, the most general RG flow equations one can write are:

\bea\label{eq:KTflows}
\frac{d \zeta^{-1}}{d\ell} &=& - c f \zeta^{-1} + \ldots \nonumber \\
\frac{d f}{d\ell} &=& b f (\zeta - \zeta_c).
\eea
These are precisely the KT equations. The observation that KT flows emerge almost immediately from the physical picture of the MBL phase is quite general, and has important consequences for our view of the MBL transition. First, it implies that finite-size scaling near the transition is controlled by a diverging length-scale $\xi_\pm \sim \exp(-c_\pm / \sqrt{|W - W_c|}$. The significance of this length-scale for the MBL transition remains obscure at present, but a practical consequence is that scaling studies of the MBL transition are expected to be extremely challenging. Second, the entire MBL phase is in some sense ``critical'' despite having a well-defined, short localization length for most correlation functions. We have already seen hints of this criticality in the dynamics, particularly in the continuously varying power laws seen in linear response\footnote{At present there is some confusion~\cite{gvs, dumitrescu2018kosterlitz, morningstar} about whether the near-transition region of the MBL phase and the region deep in the MBL phase have different thermal Griffiths effects. The observation that the MBL phase is critical is orthogonal to this: rather, it has to do with the continuously varying power-laws seen in the dynamics (these parameterize the fixed line through $\zeta$).}.

\begin{figure}[t]
\begin{center}
\includegraphics[width=0.5\columnwidth]{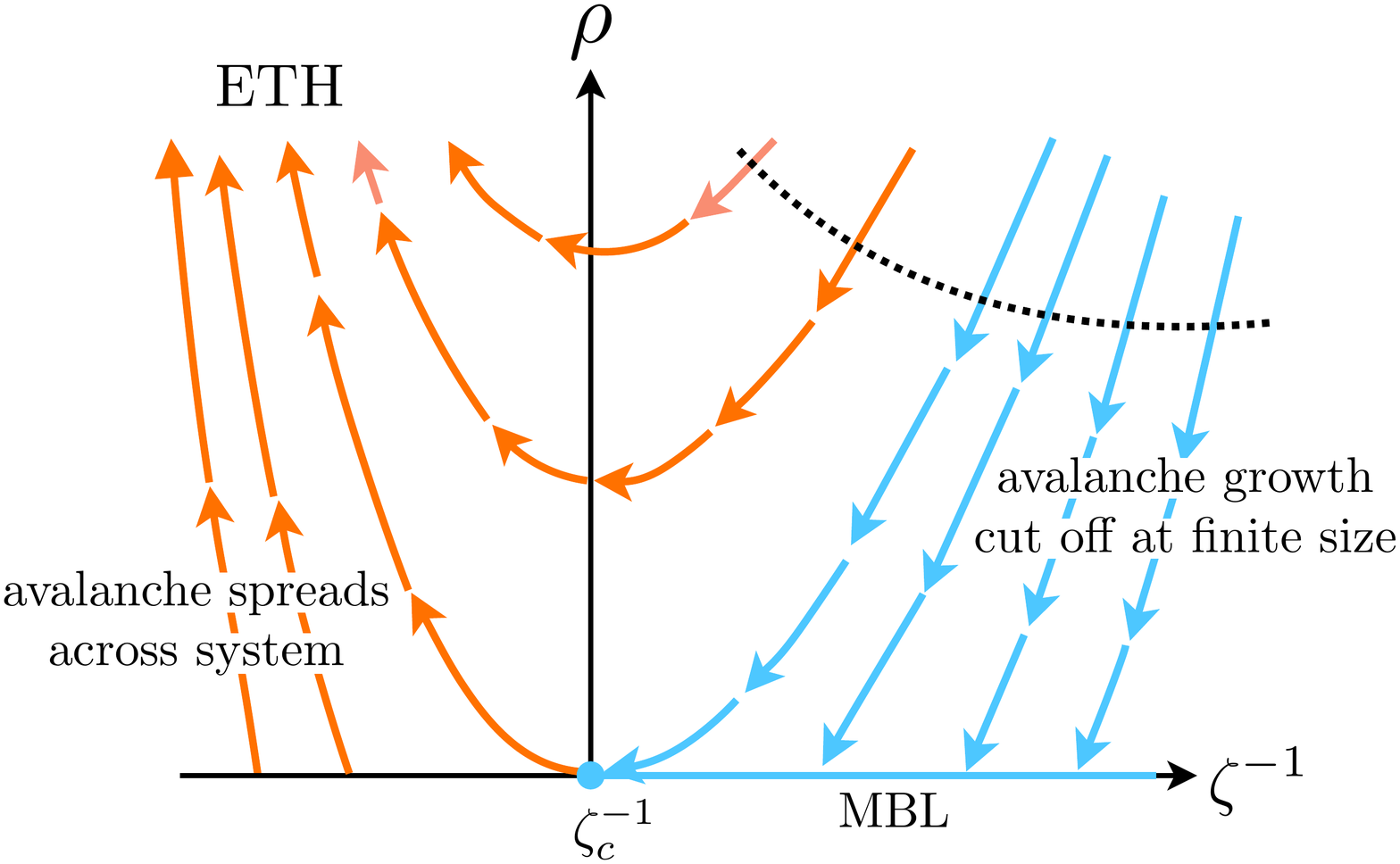}
\end{center}
\caption{\label{fig:KTflowdiag} Kosterlitz-Thouless RG flow proposed to capture the `quantum avalanche' mechanism of the MBL-ETH transition, governed by Eqs. (\ref{eq:KTflows}). When the typical localization length is above a threshold value $\zeta_c$, rare thermal regions seed `avalanches' that spread across the system to drive thermalization. For a sufficiently short localization length, $\zeta < \zeta_c$, avalanches are unable to proliferate and are ineffective on the longest length scales, so that the localized phase is stable. In this picture, the critical point is an endpoint of a line of fixed points characterizing the localized phase, and is hence itself localized, consistent with more microscopic pictures.
}
\end{figure}

Going beyond these general features to address distributions of physical properties (beyond the two scaling variables) requires one to pick a more explicit RG procedure. There are many of these schemes at the moment, which give largely similar phenomenology. 
We briefly review the scheme in Ref.~\cite{morningstar}, because it incorporates the avalanche physics particularly transparently. In this RG procedure, the system is initially divided into blocks that are assigned to be either insulating or thermal. Insulating and thermal blocks alternate in this sequence (since two same phase blocks can be merged into one larger block from the outset). A thermal (T) block is characterized purely by its length $L_T$ (or equivalently its level spacing) and has no internal structure. An insulating (I) block is characterized by both its length $L_I$ and a parameter $d_I$ called the ``deficit,'' which measures how insulating the block is. Concretely, $d_I$ captures how big a thermal block would have to be to thermalize the entire insulating block. One can estimate this as follows: the matrix element across the insulating block is $\exp(-L_I / \zeta_I)$, where $\zeta_I < \zeta_c \equiv 1$ is the critical coupling for the avalanche instability. (We measure lengths in units such that this is true.) For the avalanche to propagate across the insulating block, we need $L_T + L_I = L_I /\zeta_I$, so $d_T \equiv L_T = L_I (1 - 1/\zeta_I)$. 

The RG procedure consists of iteratively coarse-graining this pattern of T and I blocks. One sets a cutoff $\Lambda$. A thermal block is decimated if its length is at the scale $\Lambda$, and an insulating block is decimated if $d_I = \Lambda$; i.e., one finds $\min(L_T, d_I)$ over the entire system and decimates the corresponding block. It is easy to check that under these rules, if a thermal block is at the cutoff, it cannot fully thermalize its insulating neighbors (by definition of $d_I$), whereas if an insulating block is at the cutoff it will be thermalized by its neighbors (which have length $\geq d_I$ by construction). The update rules are as follows: 

(1) When an insulating block is decimated, one constructs a new, larger, thermal block, with length $L^T_{n-1} + L_n^I + L^T_{n + 1}$. Note that the length of the insulating block is always larger than $d_I$. 

(2) When a thermal block is decimated, one constructs a new, larger, insulating block, with length $L^I_{n - 1} + \Lambda + L^I_{n+1}$, and deficit $d^I_{n-1} + d^I_{n+1} - \Lambda$. This captures the fact that the presence of an internal thermal sub-block makes the composite insulating block easier to thermalize. 

These rules are simple to implement numerically on large systems; one can also write an integro-differential equation for the flow of the probability distribution, although this does not have a closed-form solution. The scaling parameters can be extracted as follows. First, the fraction of the system that is effectively thermal at some scale is given by $f = \langle L^T \rangle / \langle L^T + L^I \rangle$. Second, the typical decay length at this scale is given by computing how much correlations decay across the full system (treated at this scale). Each insulating block has a decay $\exp(-L^I_n / \zeta^I_n) \equiv \exp(- l_n)$. The total decay is the product of this over insulating blocks, which we will denote $\exp(-l)$. However, the length of the sample is $L = \sum_I L_I + \sum_T L_T$. Thus the effective decay length at this scale is $\zeta_{\mathrm{eff}} \equiv L / l$. One can numerically extract a two-parameter flow in terms of these parameters $f$ and $\zeta_{\mathrm{eff}}$, which agrees with the phenomenological prediction of KT scaling.

The scheme we have described is in a sense intermediate between, on the one hand, more directly microscopic schemes~\cite{pvp, dvp, thmdr}, which involve beginning with a specific Hamiltonian and constructing resonant clusters of states (so the rules are somewhat more involved than those we have sketched out); and on the other hand the more drastically simplified schemes~\cite{zzdh, gvs} for which closed-form solutions are available, but the relation to the microscopic physics is more obscure. Essentially all of these schemes appear capture the main features of the critical point. However, they appear to make different predictions for low-frequency response in the MBL phase, to which we will now turn. 

\emph{Fractal thermal blocks and low-frequency response}.---We have already discussed one mechanism for low-frequency response in the MBL phase---viz. cat-like resonances. A different mechanism, which the RG schemes naturally capture, is through locally thermal regions in the MBL phase. These absorb at frequencies down to their level spacing (which is also the timescale on which the last spin that was incorporated into the thermal block fluctuates due to its coupling to the thermal block). Thus, there is a contribution to low-frequency response that is proportional to the density of such rare thermal regions. 

A naive estimate of this density of thermal blocks is that it should vanish exponentially in the size of the block. This appears to be an underestimate, however. We can see this from the RG described above: the highest-probability way to create a large T block is not to start out with a large T block, but to have T blocks repeatedly absorb I blocks. Suppose we are concerned with an effective T block of size $L$. The most probable way this was created was through merging two T blocks of size $\zeta L$ with an I block of size $(1 - 2\zeta) L$. However, each of these two constituent T blocks, in turn, was most probably created by merging two still smaller T blocks with a typical I block, and so on. Iterating this procedure gives us a fractal Cantor-set-like distribution of T blocks, with measure $L^{d_f}$, where the logic above suggests that $d_f = \log 2 / (\log 2 + \zeta/(1 - \zeta))$. In particular this fractal dimension goes continuously to zero at the MBL transition. This discussion is oversimplified, as it ignores the flow of $\zeta$ itself under the RG; however, on general grounds we expect the density of thermal regions of size $L$ to vanish as a stretched exponential in $L$.

One of the recent RG schemes~\cite{gvs} has the surprising implication that $d_f = 0$ for at least a finite parameter regime in the localized phase; moreover, this scheme suggests that the probability of a thermal inclusion of size $L$ scales down only \emph{polynomially} with $L$. Numerical solutions~\cite{dumitrescu2018kosterlitz} of the RG schemes in Refs.~\cite{vha,pvp} are consistent with this picture. Polynomial scaling does not seem possible deep in the MBL phase for Hamiltonian systems: Ref.~\cite{jzi} has bounded all spatially averaged correlation functions as decaying slightly faster than a power law (specifically, the bound for the correlation function in an eigenstate is that in general $C(x) \leq C \exp(-\log^2 x)$ at large $x$ for some fixed $C$). However, if the density of $L$-spin thermal blocks decays as a power law of $L$, then one can straightforwardly show that the equal-time spatial energy correlation function in an eigenstate also decays as a power law, contradicting Ref.~\cite{jzi}. Thus, it seems that there must either be two MBL phases (and a hitherto unnoticed transition inside the MBL phase) or a missing ingredient in the solvable RG of Ref.~\cite{gvs}.

We now return to the response functions. Suppose we are interested in response at a frequency $\omega$; this is dominated by inclusions of size $\sim \log \omega$, each of which contributes an amount $1/\omega$ to generic response functions and $\omega$ to the conductivity. Thus, generic response functions in the MBL phase go as $R(\omega) \sim \omega^{-1} \exp(|\log^{d_f} \omega|)$, and the conductivity goes as $\sigma(\omega) \sim \omega \exp(|\log^{d_f} \omega|)$. At sufficiently low frequencies this contribution dominates over the Mott contribution discussed above; however, fractal regions only form on quite large scales, and do not seem to be the dominant effect in the currently numerically accessible regime.

\subsection{Summary}

This section has been a quick overview of the current theoretical understanding of MBL, as well as some of the experiments that motivated this understanding. A central theme is the distinction between ``perturbative MBL'' (i.e., the convergence of perturbation theory around a spatially unentangled state, either up to a high order or to all orders) and ``true MBL'' (i.e., the stability of the perturbative solution against rare-region effects). True MBL seems to exist only under quite restrictive conditions, but perturbative MBL is ubiquitous, as are situations where a highly heterogeneous system has locally convergent regions. This review is concerned with dynamics in the large wedge of model space between perturbative and true MBL. This dynamics is inherently governed by collective processes, and can be analyzed using concepts adopted from the theory of MBL outlined above, as we will discuss in the rest of this review.

\section{Thermal systems near an MBL transition\label{sec:thermoMBL}}

The simplest kind of nearly MBL system is simply a thermal system near an MBL transition. We will focus on the one-dimensional case, where we have the strongest reasons to believe an MBL phase (and thus an MBL transition) does exist. The existing phenomenological models of the MBL transition suggest that it has different character for random and quasiperiodic systems, so we will treat these cases separately.  

\subsection{Disordered 1D systems near the MBL transition}

The phenomenology of disordered systems on the thermal side of the MBL transition has been extensively studied, and is also discussed in depth in Refs.~\cite{gopalakrishnan_griffiths_2016, kartiek_review}. Here, therefore, our treatment will be brief; we will set out the main results and stress a few important recent developments that have occurred since the last review~\cite{kartiek_review}.

\subsubsection{Origin of Griffiths effects and subdiffusion}

In disordered systems, the phenomenological RG schemes as well as numerical analyses indicate that the diffusion constant vanishes within the thermal phase. There is a regime in the thermal phase in which transport is subdiffusive, such that $x \sim t^\beta$, with $\beta \rightarrow 0$ as the MBL transition is approached. Subdiffusion was initially observed numerically~\cite{BarLev_Absence_2015}, and subsequently analyzed as a rare-region effect~\cite{agarwal_anomalous_2015, vha, pvp}. The extent to which the numerical results support the rare-region scenario is still controversial. Here, however, we will assume that the basic picture of Ref.~\cite{drh} is correct and explore its implications; we will revisit the question of numerical support below. We note that there are two compatible but distinct-sounding explanations of subdiffusion based on rare regions. We will present these and then explain how they fit together.

The initial explanation of rare region effects involved coarse-graining the system out to a scale $\ell$ that is large compared with any microscopic scale, and assigning a local control parameter (i.e., distance from the transition) to each region of size $\ell$. The local control parameter has uncorrelated fluctuations from region to region. On these scales the system is thermal in most blocks, so it is appropriate to treat the blocks as resistors in series and add up their resistances. When the system is globally slightly on the thermal side of the transition, any particular region has an appreciable probability of being locally insulating. The least rare insulating blocks are those that are essentially at the critical point, in which the spacetime scaling is $L \sim \zeta \log t$ with $\zeta$ being some finite constant set by the critical theory. These blocks could even be internally thermal, and just disordered enough to fall inside the ``critical fan'' at that length-scale. A string of $N$ critical blocks will occur with probability $p^N$, and have resistance $R \sim \exp(N \ell / \xi)$. Blocks with resistance $R > R_0$ then occur with probability $R_0^{-\eta}$, where $\eta \equiv \xi \log p / \ell$. It is well known that a chain of resistors in which the resistances are distributed according to $P(R) \sim R^{-\eta}$ has a diverging resistance in the thermodynamic limit whenever $\eta < 2$. A more elaborate discussion of this point of view is presented in Refs.~\cite{agarwal_anomalous_2015, gopalakrishnan_griffiths_2016}; we will not dwell on it here.

Although this picture is asymptotically correct (within the assumptions above), it does not describe the relatively small systems that have been numerically studied. For example, numerical evidence suggests that even in the subdiffusive phase, resistances do not add in series\footnote{A. Scardicchio et al., in preparation}. 
The avalanche picture suggests a possible explanation for the numerically observed subdiffusion: the bulk of a system near the MBL transition is locally insulating, but insulating regions are thermalized on long timescales by randomly spaced baths. These baths are Poisson distributed, so the gaps between them are exponentially distributed. Moreover, the decay rate of a typical degree of freedom scales exponentially with its distance to the nearest bath. This picture also yields a power-law distribution of local thermalization times---and therefore subdiffusive transport---as a result of two competing exponentials. According to this picture, however, the spacing between local baths is a nonuniversal microscopic parameter, so the value of the anomalous exponent $\beta$ at the transition is nonuniversal and model-dependent. 

To reconcile these two pictures, we note that the simple rare-bath picture does not incorporate the renormalization of $\zeta$ due to small thermal inclusions. Upon coarse-graining, fluctuations in the density of small thermal inclusions lead to fluctuations of $\zeta$. Near the transition, one begins to see large regions in which $\zeta$ is too small for the region to thermalize internally; thus, the effective baths become increasingly sparse. Thus the density of effective baths vanishes as the MBL transition is approached, consistent with the Griffiths phenomenology.

\subsubsection{Phenomenology of the Griffiths phase}

We now briefly summarize the properties of the Griffiths phase. The discussion here largely recapitulates that in Refs.~\cite{gopalakrishnan_griffiths_2016, kartiek_review}, to which we refer for more detailed arguments and derivations. The basic exponent in the subdiffusive phase is denoted $z$, and is defined as follows: the density of regions that are effectively localized (in the sense, e.g., that generic local autocorrelators remain large) at time $t$ is given by $n(t) \sim t^{-1/z}$. As the MBL transition is approached, we expect that $z \rightarrow \infty$. 

\emph{Autocorrelation functions}.---We first consider the case of a Floquet system with no conservation laws. In this case, generic autocorrelation functions in typical regions should decay faster than a power law. The dominant contribution to spatially averaged autocorrelation functions comes from rare regions; at a time $t$, by construction, the autocorrelation function decays as $C(t) \sim t^{-1/z}$. Typical autocorrelation functions also decay sub-exponentially. One can bound the decay of autocorrelators as follows. A Heisenberg operator grows exponentially slowly in a localized inclusion; therefore, at a time $t$, the operator $O(t)$ has appreciable support only between the nearest inclusions that are ``inert'' at that time, i.e., its support is given by $L(t) \sim t^{1/z}$. Within this region it is a highly entangled (though not purely random~\cite{jonay}) operator. The overlap between such an operator and $O$ is generically exponentially small in $L(t)$, so the autocorrelator will go as $\exp(-t^{1/z})$, i.e., as a stretched exponential. Evidence for this stretched exponential behavior was recently seen in a numerical study~\cite{lezama2019}. 

In systems with conservation laws, a generic local operator has nonzero overlap with the conserved densities, and thus cannot decay faster than diffusively. The leading behavior of these autocorrelation functions follows from that of the density, to which we now turn. (We note in passing that the behavior of operators that do not overlap with the density is addressed in Ref.~\cite{gopalakrishnan_griffiths_2016}. There are two classes of such operators: some, like the current, have sub-leading power-law tails; others, like operators that have no matrix elements within a single sector of the conserved quantity, have no long-time tails and behave as they would in a Floquet system.)

\emph{Diffusion and transport}.---We now consider autocorrelations of the density, $S(x,t) \equiv \langle \rho(x,t) \rho(0,0) \rangle$, or equivalently of its Fourier transform, the dynamical structure factor $S(q,\omega)$. These quantities have two regimes of behavior, a ``local'' regime where $q \gg \omega^\beta$ and a ``transport'' regime $q \ll \omega^\beta$. Here, $\beta$ is an exponent $1/(z + 1)$; the relation between $\beta$ and $z$ can be motivated as follows. Consider a system of size $L$ with a density gradient. The worst bottleneck in this system will take a time $L^z$ for a single particle to cross. However, to equilibrate a macroscopic density imbalance across this bottleneck, one requires $L$ particles to cross, giving a total equilibration time $t(L) \sim L^{z + 1} \sim L^{1/\beta}$. 

When $q \ll \omega^\beta$, the system breaks up into independent regions of size $\omega^{1/\beta}$. Within each of these regions, we expect an effective diffusion constant (and thus, via the Einstein relation, an optical conductivity) $D(\omega) \sim \sigma(\omega) \sim \omega^{1 - 2\beta}$, implying that $S(q, \omega) \sim \omega^{-1 - 2\beta} \sim \omega^{-(3+z)/(1+z)}$. When $q \gg \omega^\beta$, on the other hand, relaxation is primarily local. The dominant contributions to local relaxation come from the inclusions themselves, and have the late-time form $S(0,t) \sim t^{-1/z}$. For large $q$, therefore, we have $S(q, \omega) \sim \omega^{1/z-1}$. Correspondingly, the high-$q$ conductivity goes as $\sigma(\omega) \sim \omega^{1/z + 1}$. The full form of the structure factor beyond these limits is not known. As the MBL transition is approached, $z \rightarrow \infty$ and both regimes of the structure factor go as $S(\omega) \sim 1/\omega$. The fluctuation-dissipation theorem then implies that the noise generated by systems near an MBL transition has a $1/f$ spectrum~\cite{adm2015}. In the previous section we already noted that the same behavior occurs when one approaches the transition from the localized side. 

\emph{Entanglement dynamics}.---
Initially it was believed that bipartite entanglement would grow as $S(t) \sim t^{1/z}$ in the Griffiths phase~\cite{vha}. This (plausible but erroneous) reasoning was based on the intuition that to entangle two regions separated by a weak link, it suffices to send one unit of information through the weak link. This argument in fact describes \emph{operator spreading} (see below), but not entanglement. To see why this reasoning fails for entanglement~\cite{nrh}, first consider placing the entanglement cut at an inclusion. By time $t$ the interactions will have implemented some effective number of gates crossing the entanglement cut (this can be made precise by Trotterizing the dynamics). The amount of entanglement across this cut will correspond to the number of gates applied across it by time $t$. In addition, the entanglement across two neighboring cuts cannot differ by more than one bit, according to a standard theorem in quantum information theory~\cite{nrvh}. The pattern of entanglement across the system can be visualized as a tent, which is pinned down at inclusions (see Fig. 1 of Ref.~\cite{nrh}). 

In the Griffiths phase, we expect that on a given timescale the entanglement has fully spread across inclusions less than a certain size, but is constrained by inclusions above that size. To estimate this critical size, we look for the largest inclusion across which entanglement has equilibrated. Consider a region of the system of size $L$; the worst bottleneck in this region has a characteristic timescale $t_b \sim L^z$. The timescale on which two regions of size $L$ can fully entangle across this bottleneck is $t_E \sim L t_b \sim L^{1+z}$. Since $S(t_E) \sim L$, it follows that $S(t) \sim t^{1/(1+z)} \sim t^\beta$. Thus, entanglement spreading and particle diffusion occur at similar rates in the Griffiths phase. As one goes deep in the thermal phase, particle motion eventually becomes diffusive, but entanglement still spreads with a nontrivial power law between $1/2$ and $1$, as first numerically observed in Ref.~\cite{luitz2016}. Entanglement growth is only purely ballistic when there is no tail of arbitrarily large inclusions in the system. 

We should remark that the discussion above, strictly speaking, concerned the zeroth R\'enyi entropy or Hartley entropy (also known as the logarithm of the bond dimension)~\cite{nrvh}, for which there is an exact mapping to classical surface growth. This is an upper bound on all the other R\'enyi entropies, so the result that entanglement growth is sub-ballistic stands; however, in principle it could grow with an even smaller exponent than $\beta$. Quite recently, it was pointed out that even in clean chaotic systems higher R\'enyi entropies might not grow linearly in time~\cite{rpv_renyi, yichen}. It seems highly implausible that the growth of any R\'enyi entropy would be slower than particle transport, however, so in the Griffiths phase we expect all the R\'enyi entropies to grow as $t^\beta$.

\emph{Operator dynamics}.---To entangle two regions of size $L$ across a weak link, one needs to transmit $L$ bits of information across the weak link; however, quantum information can spread substantially faster~\cite{nrh, debanjan_scrambling}. The moment a single bit of information travels across the weak link, it perturbs the rapid and chaotic dynamics on the other side; thus scrambling (in the sense of the saturation of OTOCs) takes place, at a time $t$, in a region of size $L(t) \sim t^{1/z}$. The shape of the operator front also evolves through various stages, as discussed in Ref.~\cite{nrh}.

\emph{Failure of linear response}.---So far we have been concerned with linear response properties. However, as noted above, a striking feature of the MBL phase is that linear response breaks down in the d.c. limit---for any nonzero drive amplitude $A$, if one fixes $A$ and takes the drive frequency $\omega \rightarrow 0$, the MBL phase is unstable to the proliferation of Landau-Zener resonances. For concreteness consider a Hamiltonian MBL system with a conserved charge, such as the random-field XXZ spin chain. If one drives the system with a slow a.c. field (i.e., a probe that couples to the conserved charge) of fixed amplitude, it will eventually thermalize and will have a \emph{nonzero} d.c. conductivity, which is a nonanalytic function of the drive amplitude (and is thus not interpretable as a linear response coefficient). To get linear response coefficients, instead, one must drive the system at fixed $\omega$ taking $A \rightarrow 0$, and only then take the $\omega \rightarrow 0$ limit. A similar caveat applies to the Griffiths phase: as the drive frequency is decreased, increasingly many inclusions cease to be insulating, so that $z$ renormalizes down and eventually the system heats up. 

\subsubsection{Is the Griffiths phase thermal?}

The thermal Griffiths phase was initially introduced in an entirely classical, linear context, by considering resistor networks with a broad distribution of resistances~\cite{agarwal_anomalous_2015}. However, as we have seen, the actual behavior of MBL inclusions is more complicated than that of linear resistors with high resistance. Similarly, if one quenches a system into the Griffiths phase, there is never a proper ``hydrodynamic'' regime in which all parts of the system can be taken to be near local equilibrium: deep inside large inclusions the system remains far from equilibrium, and these inclusions dominate the response. Thus it is natural to ask whether the Griffiths phase is truly thermal.

\emph{Off-diagonal ETH}.---From the narrow point of view of ETH as defined in the introduction, the Griffiths phase we have described is clearly thermal since all observables eventually relax to equilibrium. However, ETH also contains a less firmly established conjecture about the structure of \emph{off-diagonal} matrix elements of operators~\cite{srednicki_chaos_1994}: off-diagonal ETH says that these matrix elements take the form

\beq\label{odeth}
\langle m | O | n \rangle = \exp(-S/2) \sqrt{f_T(\omega)} r_{mn}.
\eeq
Here, $S$ is the entropy of states in the appropriate microcanonical shell, $r_{mn}$ is taken to be a unit-variance Gaussian random variable that is uncorrelated across different pairs, so $\langle r_{mn} r_{pq} \propto \delta_{mp} \delta_{nq}$ (for a system without time-reversal symmetry), and $f_E(\omega)$ is the spectral function at the appropriate temperature (and/or chemical potential). Essentially, off-diagonal ETH as stated above implies that the matrix elements of local operators are like those of random matrices \emph{except} for the minimal structure they need to make the linear response functions work out. 

This simple form of off-diagonal ETH turns out to be unphysical, because it implies a simple factorization of all four-point correlation functions in terms of two-point correlation functions (using Wick's theorem on products of $r$'s). 
We know, however, that the OTOC (which is a four-point function) and the conventional spectral function (a two-point function) have distinct physical content: for example, in a Floquet system with no conservation laws, the two-point function decays locally while the OTOC spreads to fill in the light-cone. 
Recently, it was pointed out~\cite{foini_kurchan, chalker_eth} that recovering sensible expressions for the OTOC requires one to posit, in addition to Eq.~\eqref{odeth}, a certain structure of four-point correlations of matrix elements. For two operators $X$ and $Y$ at a distance $\ell$ in a system with local interactions but no conservation laws, one requires the following form for the connected four-point correlators of matrix elements~\cite{chalker_eth}: 

\bea
G(\omega_1, \omega_2, \omega_3) & = & N^{-1} \left \langle \sum_{\alpha\beta\gamma\delta} X_{\alpha\beta} Y_{\beta\gamma} X_{\gamma\delta} Y_{\delta\alpha}  \delta(\omega_1 - \Delta_{\alpha\beta}) \delta(\omega_2 - \Delta_{\beta\gamma}) \delta(\omega_3 - \Delta_{\gamma\alpha}) \right\rangle  \\ & & \xrightarrow{\ell \rightarrow \infty} f(\omega_2) f(\omega_3) K_\ell (\omega_1 + \omega_3). \nonumber
\eea
Here, $K_\ell(\omega_1 + \omega_3)$ has to do with the spreading of OTOCs, and is sensitive to the exact nature of the dynamics, and $N$ is the Hilbert space dimension.
Whether locality and causality impose still other constraints on the correlations of matrix elements remains an open question. 

\emph{Numerical results}.---Numerical studies of many-body matrix elements at intermediate disorder (i.e., in the putative Griffiths regime) suggest that the matrix elements are not just correlated but might also deviate strongly from Gaussianity. The matrix-element distribution extracted from exact diagonalization develops power-law tails in the subdiffusive regime~\cite{lbl, lbl2}. 
In addition, if one treats the matrix elements of a local operator as amplitudes of a fictitious wavefunction, the corresponding inverse participation ratio seems to scale as an anomalous power of the Hilbert space dimension~\cite{spa_criterion, serbynmoore, monthus2016many, PhysRevB.96.104201}. 
These observations are consistent with one another; both suggest that the matrix elements between eigenstates at a fixed energy difference are very unevenly distributed, with many anomalously small values and a few very large ones. 
A further piece of evidence for this (as pointed out in Ref.~\cite{serbynmoore}) is that the level statistics in the subdiffusive regime are intermediate between Wigner-Dyson and Poisson; the variance of levels in an energy window also exhibits anomalous scaling~\cite{corentin}. 
Anomalous level statistics are consistent with anomalous matrix-element distributions, since eigenstates that are essentially uncoupled will not repel. 
Finally, equal-time correlators in individual eigenstates also seem to have fat-tailed distributions, even when the averages are consistent with ETH~\cite{colmenarez2019statistics}.
From a random-matrix perspective many of these results are what one might expect. The distribution of matrix elements deep in the thermal phase is a Gaussian, while deep in the MBL phase it is essentially bimodal, with most matrix elements being close to zero and a small number being of order unity. A natural way to interpolate between Gaussian and bimodal distributions is by going through an intermediate family of L\'evy-stable distributions; as the stability parameter is tuned, these distributions develop increasing amounts of weight at very small and very large values.
However, at present our understanding of the microscopic origin of these phenomena is primitive. It is unclear how many of them survive to the thermodynamic limit, and how many are consistent with the Griffiths phenomenology. 

Finally, an aspect of the numerical evidence that might be in tension with the Griffiths interpretation of the subdiffusive phase is the sheer size of the subdiffusive region in parameter space~\cite{BarLev_Absence_2015, agarwal_anomalous_2015, luitz2016, znidaric_diffusive_2016, khait2016}. Although the largest-size numerics clearly shows a diffusive regime at small disorder~\cite{znidaric_diffusive_2016}, as expected under the Griffiths scenario, even there, subdiffusion sets in very far from the putative MBL transition. This, together with the other anomalies we have noted above, might indicate that there is further structure to the observed subdiffusive regime than just the Griffiths phenomenology associated with the MBL critical point. An extreme possibility is that there might be two different mechanisms for subdiffusion, one of them having nothing to do with Griffiths effects. It is also possible, however, that the numerical observations correspond to a very broad critical fan associated with the MBL transition: recall that an inclusion just needs to be internally \emph{critical}, i.e., it needs to be at the nearest point on the critical fan at the relevant size, and if the fan is broad, an inclusion that is large enough to appreciably slow down the dynamics can form even at weak disorder. 

\subsection{Quasiperiodic 1D systems near the MBL transition}

We now turn from random to quasiperiodic systems. Quasiperiodic systems can exhibit single-particle localization, with localization lengths that are short enough to ensure stability against avalanches in one dimension, \emph{even} if such avalanches occur. A result like that of Ref.~\cite{jzi} has not yet been proved for the quasiperiodic case, as the statistics of correlations between many-body energy levels is delicate and not mathematically well controlled; thus it is not even proven---even to all orders in perturbation theory---whether typical regions actually localize. However, no clear mechanism has been suggested that would destabilize such a quasiperiodic MBL phase, so for our present purposes we will assume that this phase exists. 

\subsubsection{Absence of Griffiths effects}

Quasiperiodic systems are ``hyperuniform'' in the sense that they do not have appreciable long-wavelength fluctuations~\cite{crowley2018quantum}. This hyperuniform property persists under coarse-graining. Consider a segment of a generic quasiperiodic spin chain with local interactions, and suppose we want to find another segment where each of the sites and bonds is within $\epsilon$ of the initial segment. We can find an approximant~\footnote{{An approximant is a periodic  system, typically with a long period, that locally resembles a quasicrystal. For  a  1D quasicrystal that is generated by adding an  incommensurate modulation with an irrational wavevector $\gamma$ relative an  existing periodic structure (e.g. a lattice), an approximate can be constructed by considering  a {\it periodic} modulation at a wavevector $\lambda=p/q$ that is a rational approximation to $\gamma$.  A larger denominator $q$ corresponds to a better approximation, in that it matches the quasiperiodic structure over a longer lengthscale.}} with denominator $1/\sqrt{\epsilon}$ that meets this criterion. Since the LIOMs are purely a function of the Hamiltonian and not the states, if the Hamiltonian almost repeats, then so must the structure of LIOMs. Therefore, Griffiths effects of the type we discussed in the previous section cannot occur in quasiperiodic systems. 

A distinct type of Griffiths effect has been proposed in the quasiperiodic case, involving rare configurations of the \emph{state} (which is typically random and has no hyperuniform properties)~\cite{lueschencrit16, lev2017transport, bordia2d17} . In the thermal phase, at least, such Griffiths effects cannot occur at late times, by the following logic. Suppose typical regions of the state allow diffusion, but certain rare regions have anomalously low density and do not allow diffusion. It is straightforward to show that the diffusion constant at the interface between rare and typical regions remains finite, so the typical region ``eats'' the rare region on a timescale $\sim \sqrt{L}$, where $L$ is the size of the rare region. Since the rare regions are exponentially rare, it follows that this mechanism can only give rise to stretched exponential relaxation, and not to anomalous power laws. However, on short timescales it is possible that these rare-region effects are visible, if the prefactor for the associated processes is much larger than for diffusion. 

Even in the localized phase, it is not clear how rare configurations would matter. Since LIOMs imply localization, the only way for rare regions to exist in a quasiperiodic system is for the occupation numbers of orbitals to arrange themselves so as to create a localized region even in the absence of any LIOMs in that region. In the worldview of this review---which is skeptical of many-body mobility edges and translation-invariant localization, as discussed in the next section---such a conspiracy seems implausible, and we can treat the existence of LIOMs as a proxy for localization. (This point is discussed further in Ref.~\cite{agrawal2019universality}.)

\subsubsection{Two universality classes of MBL transition}

The exponents extracted from small-system studies of the MBL transition suggest that (at those scales) the apparent correlation length exponent $\nu \approx 1$~\cite{kjall, lla, ksh}. This is inconsistent with the CCFS bounds for random systems, which require that $\nu \geq 2$ for a stable critical point~\cite{ccfs, clo}. Thus, the true critical properties of the random transition must be different from those observed numerically~\cite{ksh}. This need not be the case for quasiperiodic systems, however, since quasiperiodic systems are only constrained by the weaker exponent inequality (due to Luck) that $\nu \geq 1$~\cite{luck1993classification}. 

There are therefore three logical possibilities for the coarse-grained behavior of the MBL transition in quasiperiodic systems near the critical point (again, assuming this critical point exists~\cite{vznidarivc2018interaction}). The first possibility is that thermal inclusions are simply irrelevant to the critical physics: instead, the phase transition occurs because typical regions undergo an instability of the type first considered in Ref.~\cite{agkl}, at which resonances proliferate everywhere in the system. This scenario has the appealing feature of symmetry: on general grounds, rare regions do not matter on the thermal side (and this is numerically supported~\cite{vznidarivc2018interaction}), so it is tempting to suppose they should not matter on the localized side either. However, there are two basic arguments against this scenario. First, the large-scale numerics of Ref.~\cite{vznidarivc2018interaction} show diffusion for very weak interactions and for disorder considerably above the threshold for single-particle localization. This extreme susceptibility of the single-particle localized case to weak interactions does not seem consistent with a scenario in which the physics of delocalization occurs in typical localization volumes: typical localization volumes are stable, per~\cite{basko_metalinsulator_2006}, until a finite critical interaction strength. 
Second, suppose that one is on the MBL side of the transition, in a typical localized many-body eigenstate. In this state, each degree of freedom experiences an effective potential that is a combination of the quasiperiodic lattice and the random on-site Hartree shifts due to its neighbors. This effective potential is random, and will generically have large-scale spatial fluctuations, which allow for appreciable fluctuations of the effective localization lengths of particles. It seems inevitable that these fluctuations will give rise to rare regions that are anomalously delocalized. Indeed, explorations of the optical conductivity of quasiperiodic MBL systems\footnote{V. Khemani, unpublished}---which is dominated by resonances and rare regions---have found that this observable behaves similarly in quasiperiodic and random MBL systems. 

The two remaining possibilities are those in which rare thermal regions do matter, and the system is able to form internal small baths. No explicit construction of these baths exists, but they provide the simplest mechanism for the results of Refs.~\cite{vznidarivc2018interaction, doggen2} (see also Refs.~\cite{hongyao1, hongyao2}). In the simplest scenario of this type, the quasiperiodic and random instabilities are essentially identical: some set of ``spectator'' spins facilitates the formation of a small delocalized bath, which then thermalizes the rest of the system via the avalanche mechanism. The distribution of these avalanches is random (since it depends on the random occupations of the initial state), which---taken at face value---would suggest that the critical theory should be the same as that in the random case. The main drawback of this scenario is that it implies a many-body mobility edge. If there exist regions and states in which some particles are delocalized over $L_1$ sites, then one cannot construct l-bits in such regions that are localized to fewer than $L_1$ sites. However, the properties of the l-bits do not depend on the state and are purely quasiperiodic; therefore, a breakdown of the l-bit picture in one region immediately implies its breakdown in translations of that region by specific approximants. If MBL occurs only when there is an l-bit construction, this scenario is impossible. 

This brings us to the third scenario, in which the state near the transition consists of a quasiperiodic sequence of thermal and insulating blocks (analogous to the block RG schemes discussed in Sec.~\ref{sec:intro:MHRG})~\cite{agrawal2019universality}. As one lowers the disorder potential, the l-bit construction breaks down in some regions of the potential (e.g., because it is possible to construct avalanches there) while remaining well-defined in others. In this scenario, the critical point is nonrandom (and in fact even has $\nu = 1$~\cite{agrawal2019universality}), but the transition is still driven by the point at which avalanches spread through the system. 

\subsubsection{Quasiperiodic MBL in higher dimensions}

In dimensions greater than one, any avalanche will asymptotically spread through the system. Whether such avalanches can still get started at sufficiently large values of the quasiperiodic potential (or at sufficiently weak hopping) is unclear, however. Moreover, this question might not have a universal answer: different classes of quasiperiodic potential might give different behavior. A demonstration that MBL exists for \emph{any} class of two-dimensional quasiperiodic potential would, however, be extremely significant: many of the exotic phases of driven matter that require MBL to be stable only exist in two or more dimensions. 

\section{Unstable MBL: disordered systems that are constrained to thermalize\label{sec:unstableMBL}}
We turn now to a distinct class of disordered systems: those where \emph{prima facie} the disorder is strong enough for localization, but where for various reasons the system cannot enter a fully-MBL phase characterized by a full set of LIOMs. Instead, the system thermalizes, but at strong disorder the physics of MBL often remains relevant and can both constrain the resulting phase diagram and reveal itself via unusual transport phenomena. 

There are two main classes of such protected thermal systems, that may be distinguished based on whether the eventual thermalization is driven  by `typical' or `rare' features of the disorder distribution. In the former we place systems where the underlying global symmetry structure inevitably leads to an extensive degeneracy that is incompatible with MBL in the restricted sense of having a full set of LIOMs, as well as systems with sufficiently slowly decaying power law interactions. Among the latter we include systems where rare thermal regions drive an `avalanche' instability  to the thermal phase, such as putative MBL phases in $d>1$ or with rapid power-law (or stretched exponential) decay.

\subsection{Symmetry-breaking  and Localization-Protected Order}
We  begin  our discussion by first clarifying what we mean  by `symmetry breaking' in the MBL context. In clean systems, a simple argument due to Peierls precludes the existence of discrete symmetry breaking at non-zero energy density. Recall that the key feature of symmetry breaking is that  a system of size $L$ prepared in one of the distinct broken-symmetry states takes exponentially long to tunnel into its symmetry-related counterparts. At any nonzero energy density, however, the system will host a finite density of domain wall excitations on top of the broken symmetry; in the clean system, these are mobile, and their motion across the system facilitates tunneling between distinct broken-symmetry states, washing out the long-range correlations in the spins. Since the domain walls are interacting point particles in 1D, in the presence of disorder they can become (many-body) localized, and therefore are no longer effective in tunneling between distinct broken symmetry sectors.  Therefore, we can build a highly excited broken-symmetry state of the infinite system by inserting a finite density of such frozen domain walls on top of one of the broken symmetry ground states $\ket{\Psi_0}$ (that we assume, for simplicity, have all the spins aligned). This results in an eigenstate-dependent  `fingerprint pattern' characterizing the discrete broken symmetry; however, each such pattern is exactly degenerate  with the finite set of states related to it by adding the same set of domain walls on top of one  of the other infinite-volume ground states related to $\ket{\Psi_0}$ by the action of the global symmetry. Formally in a finite system the symmetry is of course not broken and the eigenstates are reallty cat states, i.e. are  superpositions of macroscopic broken-symmetry states; the above statements are to be understood in analogy with similar statements about the finite-size ground states of systems that break discrete symmetries, and become precise in the $L\rightarrow \infty$ limit.
 While the local order parameter of any patch of the system fluctuates wildly from eigenstate to eigenstate, i.e. is exponentially sensitive to the energy, the broken symmetry can be diagnosed by a suitable  infinite-temperature generalization of the Edwards-Anderson order parameter familiar from spin glass theory, or by examining spin-spin autocorrelation functions in the time domain.  This is an example of \emph{eigenstate order}~\cite{ppvreview}, introduced in Ref.\cite{huse_localization_2013}, where it was  termed \emph{localization protected order} to emphasize its reliance on MBL to evade thermalization (see also Refs.~\cite{bauer_area_2013,bahri_localization_2015}) . The potential to use MBL to ``protect'' unconventional non-equilibrium phases  is one reason for broader interest in the field. However, as we will see, the complex interplay  of MBL with the preservation and breaking of continuous and non-Abelian symmetry has tempered the initial optimistic pronouncements on the new possibilities that it could open.

\subsection{Systems with continuous nonabelian symmetries}
\subsubsection{Local Symmetry Action and Non-Abelian Symmetry}

We begin by discussing how symmetries are realized in  fully-MBL systems in one dimension. We do not consider the case of putative partially-localized systems with a many-body mobility edge, or MBL in higher dimensions, which are potentially unstable for reasons discussed in the next section. We  also comment that if we are willing to forego a description in terms of LIOMs, there is an additional intriguing possibility: namely, that the system realizes a `quantum critical glass' state that is not MBL but  is nevertheless not fully thermal. Such quantum critical glass phases were proposed as stable excited-state phases of spin chains built from non-Abelian anyons~\cite{QCGPRL}; however, these models have constrained Hilbert spaces  that do not admit a local tensor product structure. This automatically rules out a conventional description in terms of LIOMs. It remains an open question where the combination of disorder and interactions can lead to \emph{emergent}  Hilbert-space constraints, and we do not explore this further.  We mention that a distinct aspect of  thermalization in constrained Hilbert spaces has recently witnessed a  resurgence motivated by experiments on  strongly-interacting trapped Rydberg gases; however, much of this has focused on proximity to integrability in a disorder-free setting rather than links to MBL and so lies  outside the scope of this review.

With these preliminaries in place, we now sketch why continuous non-Abelian symmetries obstruct localization, following the elegant presentation of Ref~\cite{PVMBLSYM}. Although the argument is very general, the intuition behind it is familiar from elementary quantum mechanics: the excited states of systems with non-Abelian symmetries, such as the hydrogen atom and a large spin-$s$ rotor, are generically highly degenerate. If a system with these symmetries were fully localized, one could disentangle it into separate blocks, each with the symmetry acting on it. At high temperatures, most of these blocks would be in excited states and the spectrum would therefore have massive degeneracies, making the system unstable to perturbations.

More formally, since we are considering fully-MBL systems  we  may leverage the l-bit description; let  $\{n_1, n_2, \ldots, n_N\}$ be a complete  set  of l-bit labels, and take $\xi$ to the localization length.   A natural way to incorporate  global symmetries in the MBL setting is to first demand that the single-site Hilbert space $\mathcal{V}$ transforms under some linear representation (either reducible or irreducible)  of the global symmetry group $G$~\footnote{More precisely, to define a representation we must specify both a vector space $\mathcal{V}$ and a linear map $\phi:G\rightarrow GL(\mathcal{V})$ where $GL(\mathcal{V})$ is the group of linear maps on $\mathcal{V}$. However we shall follow the common practice of  referring to the representation  simply by  the Hilbert space on which it acts  $\mathcal{V}$, as in `spin-$1/2$ representation of  $SU(2)$'.}. The $N$-site Hilbert space is then given by the tensor product  $\mathcal{H} = \mathcal{V}^{\otimes N}$. We further require that the representation is faithful (i.e., distinct elements $g\in G$ are mapped to distinct elements when acting on $\mathcal{V}$).  A $G$-preserving fully-MBL phase is is then defined to be one where the l-bit operators commute with the  symmetry generators $g\in  G$;  therefore, each l-bit string $\{n_\alpha\}$ labels a (possibly non-degenerate) multiplet of states $\mathcal{V}_{n_1, n_2, \ldots, n_N}$ that together form a representation  of $G$. Now, let us assume  that  at least one of the l-bit strings $\{n_\alpha^0\}$ labels a non-degenerate eigenstate,  that therefore must  transform  trivially under $G$: in other words, $\mathcal{V}_0  = \mathcal{V}_{n_1^0, n_2^0, \ldots, n_N^0}$ is a dimension-one (singlet) representation of $G$. Changing an l-bit $n^0_\alpha$ creates a local excitation, but the resulting eigenstate transforms in a different representation $\mathcal{V}_{n_\alpha} \equiv \mathcal{V}_{n_1^0, n_2^0, \ldots, n_\alpha,\ldots n_N^0}$ of $G$; this is  either an irreducible representation (or \emph{irrep}) or can  be reduced by adding generic local perturbations. Since we only flipped a single l-bit, it follows that the $\mathcal{D}_{n_\alpha}= \dim \mathcal{V}_{n_\alpha} \equiv \mathcal{V}_{n_1^0, n_2^0, \ldots, n_\alpha,\ldots n_N^0}$ eigenstates in this representation and the original singlet eigenstate all only differ locally in the vicinity  of $r_\alpha$. We now repeat this procedure to excite a different l-bit $\beta$, viz. $n^0_\beta\rightarrow n_\beta$, and let $\mathcal{V}_{n_\alpha n_\beta}$  represent the corresponding representation.

As long as the two l-bits are sufficiently far apart, $|r_\beta -r_\alpha |\gg\xi$, we may argue that the symmetry action factorizes on the two l-bits, i.e. that the  representation on the two l-bits is just the direct product of those  on the individual l-bits, $\mathcal{V}_{n_\alpha n_\beta}=\mathcal{V}_{n_\alpha}\otimes\mathcal{V}_{n_\beta}$. (As  an exact statement, this relies on l-bits  that are  strictly local, but the more general case must be related to this by ``dressing'' with a finite-depth  quasi-local unitary transformation.) Iterating this argument for an extensive set of l-bits $\{\alpha_1, \alpha_2, \ldots,\alpha_p\}$ with $p\sim O(N)$, and $|r_{\alpha_{i+1}} - r_{\alpha_{i}}| \gg  \xi$ shows that the symmetry action factorizes on these l-bits, $\mathcal{V}_{\alpha_1, \alpha_2, \ldots,\alpha_p} =\mathcal{V}_{n_{\alpha_1}}\otimes \mathcal{V}_{n_{\alpha_2}}\otimes\ldots\otimes \mathcal{V}_{n_{\alpha_p}}$. In other words, $G$ has \emph{local symmetry action} on the l-bits.

The fact that MBL systems exhibit local symmetry action is innocuous if the symmetry group $G$ has purely one-dimensional irreps, i.e., if it is Abelian. However, it has profound consequences when $G$ has higher-dimensional irreps. Recall that the Hilbert space is simply a tensor product of the onsite Hilbert spaces, $\mathcal{H} = \otimes_\alpha\mathcal{V}_\alpha$; since it is a faithful representation of $G$, it follows that all irreps of $G$ are contained in $\mathcal{H}$ for $N$ sufficiently large (but finite). Therefore, as $N\rightarrow\infty$, the Hilbert space has (at least) a finite density of multidimensional irreps of $G$. Since the action of $G$ factorizes on the l-bits, it follows that  must  then be a finite density of {local} degeneracies corresponding to these multidmensonal irreps. In other words, simply specifying an l-bit string does \emph{not} fully identify an eigenstate in a $G$-symmetric phase when $G$ is non-Abelian: a finite fraction of the l-bits correspond to the multidimensional irreps, each of which requires an additional label to distinguish states within the degenerate multiplet realizing the irrep. A more formal version of this argument is provided in Ref.~\cite{PVMBLSYM}. Note the crucial role played by local symmetry action: without this, one cannot ascribe a \emph{local} degeneracy to the irrep. 

Such local degeneracies are generically unstable and require fine-tuning to preserve. While symmetry forbids matrix elements between symmetry-unrelated states, since the majority of these states are at finite energy density, it is unavoidable that a finite fraction of these degenerate states are exponentially close in energy to other states to which they can couple by generic symmetry-preserving perturbations. Such perturbations can have one of two consequences for a system with an l-bit description. First,  strong enough disorder can drive spontaneous symmetry breaking in the eigenstates so that they are only invariant under a discrete Abelian subgroup of the full non-Abelian symmetry group. For this to be consistent with MBL, the symmetry that is broken must be discrete since the spontaneous breaking of a \emph{continuous} global symmetry is believed to be incompatible with MBL~\footnote{This is because on general grounds the localization length is expected to diverge at long length scales for at  least a  finite fraction of the spectrum due to Goldstone mode physics~\cite{GurarieChalker}, and  it has been argued that embedding a finite density of such extended states  in the  many-body spectrum destabilizes MBL~\cite{NandkishorePotter}.}.  We now consider three examples that each demonstrate different ways  in which non-Abelian symmetry constrains non-equilibrium/eigenstate phase structure,

\subsubsection{Quantum Clock Models and Broken-Symmetry Paramagnets}
\begin{figure}
\begin{center}
\includegraphics[width=0.5\columnwidth]{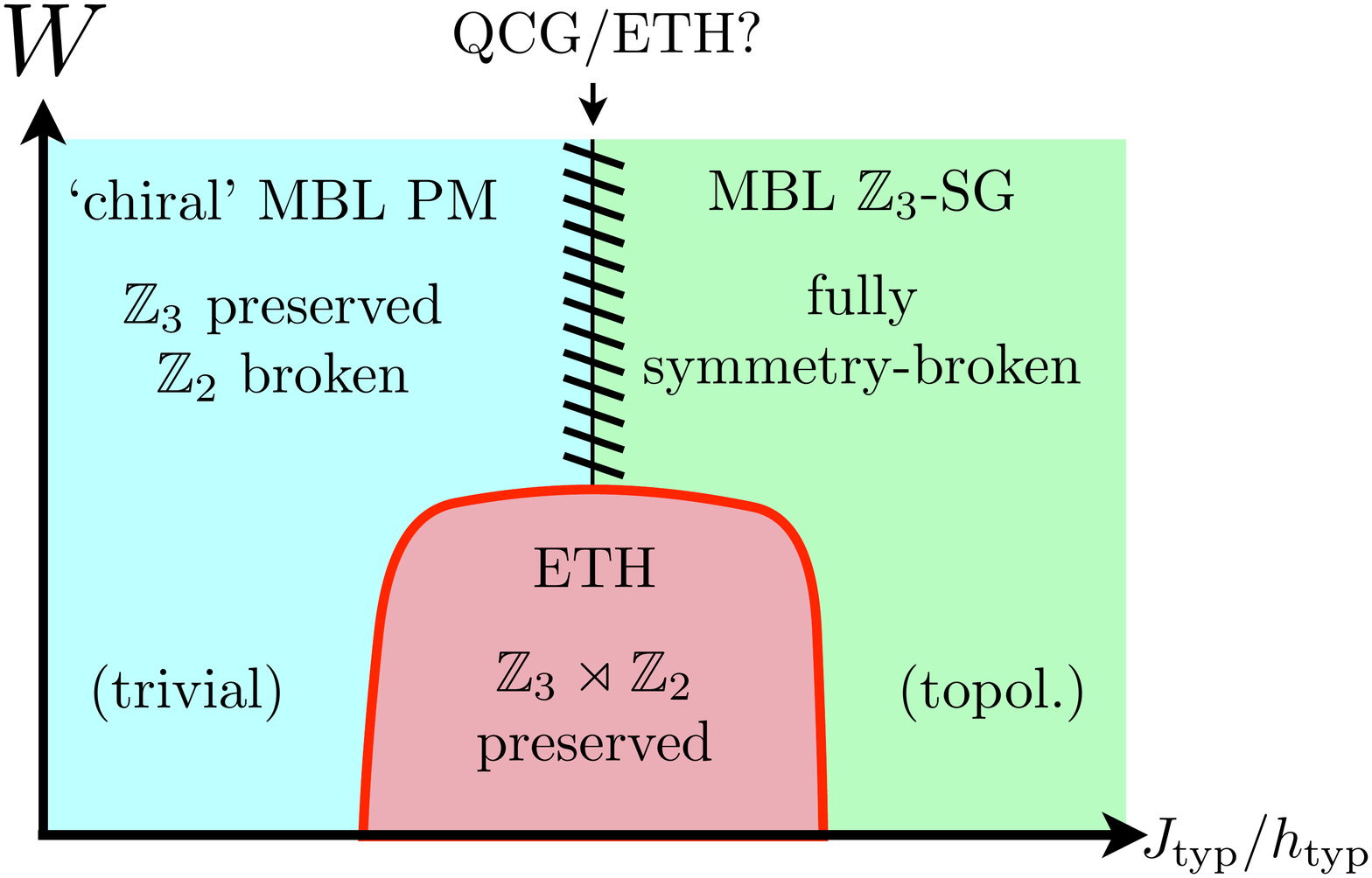}
\end{center}
\caption{\label{fig:Pottsphasediag} Schematic eigenstate phase diagram of the 1D quantum clock model, characterized by non-Abelian $D_3 \cong \mathbb{Z}_3 \rtimes \mathbb{Z}_2$ global symmetry, as a function of the tuning parameter $J_{\text{typ}}/h_{\text{typ}}$ and disorder strength $W$. Here, $J_{\text{typ}}, h_{\text{typ}}$ are \emph{typical} values of the random bond/field terms, which generically have broad distributions. For strong disorder, the model has two many-body localized phases, neither of which preserves the $D_3$ symmetry fully. For $h_{\text{typ}} \ll J_{\text{typ}}$, the eigenstate phase spontaneously breaks the $\mathbb{Z}_3$ symmetry in a glassy manner; this automatically breaks the $\mathbb{Z}_2$ symmetry owing to the semidirect product structure. On the other hand, for  $h_{\text{typ}} \gg J_{\text{typ}}$, the system is in an MBL phase that is a $\mathbb{Z}_3$ paramagnet. However, in highly excited states, this phase breaks the residual $\mathbb{Z}_2$ symmetry, spontaneously generating `chirality' in the system. Numerical studies are inconclusive as to whether there is a direct transition between these MBL phases, or whether there  is an intervening sliver of `quantum critical glass' or ETH phase extending to strong disorder (dashed lines). For weaker disorder, an intervening ETH phase is robustly present in such simulations. A generalized Jordan-Wigner transformation maps the chiral paramagnet/MBL spin glass respectively into trivial/topological phases of an interacting disordered parafermion chain.
}
\end{figure}

As our first example, we first discuss the random-bond quantum Potts/clock chains that provide an example of MBL-enforced symmetry breaking~\cite{friedmanclock}. These models have a   global $D_n$ symmetry that is non-Abelian for $n\geq 3$.  $D_n = \mathbb{Z}_n\rtimes \mathbb{Z}_2$ is the group of symmetries of a regular $n$-gon: $\mathbb{Z}_n$ rotation by $2\pi/n$, and $\mathbb{Z}_2$ reflection (sometimes termed a `chiral' symmetry), whose non-commutativity is indicated by the semidirect product ($\rtimes$). The Hamiltonian for an $L$-site system is given by
\begin{equation}
 \label{eq:Hamclock} H =  -\sum_{j=1}^{L-1}J_j {\hat{\sigma}}^{\dagger}_j {\hat{\sigma}}^{{\dagger}}_{j+1} -  \sum_{j=1}^Lh_j {\hat{\tau}}^{{\dagger}}_j    + {h.c.},
 \end{equation}
where $J_j$, $h_j$ are real and random (for generic complex-valued couplings, the $\mathbb{Z}_2$  is absent and the symmetry group is Abelian). The operators commute on different sites and satisfy 
${\hat{\sigma}}_j^n = {\hat{\tau}}_j^n=1, ~ {\hat{\sigma}}_j {\hat{\tau}}_j = \omega {\hat{\tau}}_j {\hat{\sigma}}_j$.
on a single site, where $\omega = e^{\frac{2 \pi i}n}$.  ${\hat{\sigma}}_j, {\hat{\tau}}_j$ are the natural generalizations of the Pauli $\sigma^z, \sigma^z$ operators of the quantum Ising model;  it is convenient to refer to as `weight' and `shift' operators respectively. For weak disorder, we expect that  highly excited states of the model simply satisfy ETH, and therefore preserve all the symmetries. At strong disorder, the situation is more complex; numerical studies of model \ref{eq:Hamclock} with $n=3$ reveal that it has (at least) two distinct phases in this limit. One of these is a $\mathbb{Z}_3$ spin glass phase that breaks the threefold rotation symmetry; owing to the semidirect product structure, this automatically also breaks the $Z_2$ symmetry~\footnote{This maps via a duality transformation onto a `topological' phase of a parafermionic chain with zero-energy boundary states, providing an example of a system where MBL can be used to stabilize  topologically-protected end states at finite energy density in an intrinsically interacting system.} The second phase that is unambiguously identified in numerics is an unusual `chiral paramagnet'. Here, the system preserves the $\mathbb{Z}_3$ symmetry, but the eigenstates spontaneously break the symmetry between the three distinct  eigenstates of the shift operator $\hat{\tau}_j$   in the $\hat{\sigma}$-basis, effectively generating a chirality or handedness under $\mathbb{Z}_2$, diagnosed by $\hat{\mathcal{J}}_j = \frac{1}{i\sqrt{3}}(\hat{\tau}_j-\hat{\tau}_j^\dagger)$. As in all cases of eigenstate order, in both cases the symmetry is broken in a spatially glassy fashion, and can only be detected by examining order parameters of the Edwards-Anderson type. Besides these two broken symmetry phases, Ref.~\cite{Prakash} argued for the existence also of a quantum critical glass  phase ; while at present there remains as yet no definitive evidence for this phase, it remains an intriguing possibility as a route to a non-ergodic yet non-localized phase. Fig.~\ref{fig:Pottsphasediag} summarizes this phase structure.

\subsubsection{Random-bond XXZ models and symmetry-breaking at the MBL transition}
A second scenario~\cite{XXZpaper} is exemplified by the random-bond XXZ spin-$1/2$ chain, described by
\begin{equation}
 \label{eq:HamXXZ} H =  -\sum_{j=1}^{L-1}J_j (S_j^x S_{j+1}^x + S_j^y S_{j+1}^y + \Delta S_j^z S_{j+1}^z)
 \end{equation}
where as before we choose $J_i>0$ to be random. For $\Delta\neq 1$ the symmetry of the model $U(1) \rtimes \mathbb{Z}_2$, consisting of $U(1)$ rotations about the $z$-axis in spin space, and a global Ising flip $S_z\mapsto-S_z$; similarly to the $D_n$ example generic $U(1)$ rotations by an angle $\theta \not\in \pi \mathbb{Z}$ fails to commute with the Ising flip, since the latter rotates the axis of $U(1)$ symmetry. However unlike in the ${D}_n$ case, there is no way to preserve the $\mathbb{Z}_2$ symmetry and remain localized, since this would require breaking the continuous $U(1)$ symmetry (which as we have noted above is impossible). For $\Delta=0$, the model maps via a Jordan-Wigner transformation to a free-fermion chain with random hopping; in the taxonomy of localization problems and symmetry-protected topological phases~\cite{AZclasses,PeriodicTableKitaev,TenfoldWay} this is an example of what is variously termed class AIII/bipartite random hopping/particle-hole symmetric localization, characterized by a sublattice symmetry\footnote{This is variously  termed a `chiral' or `particle-hole' symmetry, and depending on details --- e.g. if the fermionic hopping terms are real or complex --- can also be related to the BDI topological class.} and the divergence of the single-particle localization length at zero energy.

The ground-state properties of the random-bond XXZ chain are known to be controlled by an infinite-randomness fixed point of the `random singlet' type, accessed via a strong-disorder renormalization group approach (SDRG)~\cite{fisher_random_1994}. For the $\Delta=0$  `XX' chain--- corresponding to free theory in the fermionic language --- it is possible to generalize the SDRG to tackle excited eigenstates~\cite{HuangMoore}. Such generalizations were first introduced to study the excited states of the random transverse-field Ising chain~\cite{pekker_hilbert-glass_2014}. The procedure consists of iterating the following steps: (i) identify the term $H_\text{max}$ in the Hamiltonian with the strongest coupling; (ii) solve $H_\text{max}$ exactly in isolation, i.e. ignoring all the other terms $H - H_\text{max}$; (iii) we then decimate the degrees of freedom involved in  $H_\text{max}$ as follows. We first choose one of the energy levels of $H_\text{max}$; this will either be a singlet or else consist of a symmetry-enforced multiplet. If it is  a  singlet, we freeze the  degrees of freedom  involved  in the  singlet  and compute the couplings between the remaining degrees of freedom mediated by virtual fluctuations of $H_\text{max}$; usually going to second order in pertrubation theory is sufficient. If it is instead a multiplet, we replace it by an effective spin, and compute the couplings of this effective spin with the remaining degrees of freedom (usually, first-order perturbation theory). These processes can also be recast as sequential Schrieffer-Wolff rotations~\cite{pekker_hilbert-glass_2014}. Iterating this procedure produces a flow in the space of Hamiltonians, that can be either numerically studied or analytically recast as a set of integro-differential flow equations for  the \emph{distributions} of couplings. For many problems, this approximate scheme may be self-consistently justified, in that the quality of the approximation is controlled by the breadth of the coupling distributions, which in turn become broader as the RG flows: in other words, its predictions are asymptotically exact.
The original SDRG procedure for understanding ground states always targeted the lowest energy level of $H_\text{max}$ at each step; Ref.~\cite{pekker_hilbert-glass_2014} argued that one could instead target excited states by choosing other levels of  $H_\text{max}$.

\begin{figure}
\begin{center}
\includegraphics[width=0.5\columnwidth]{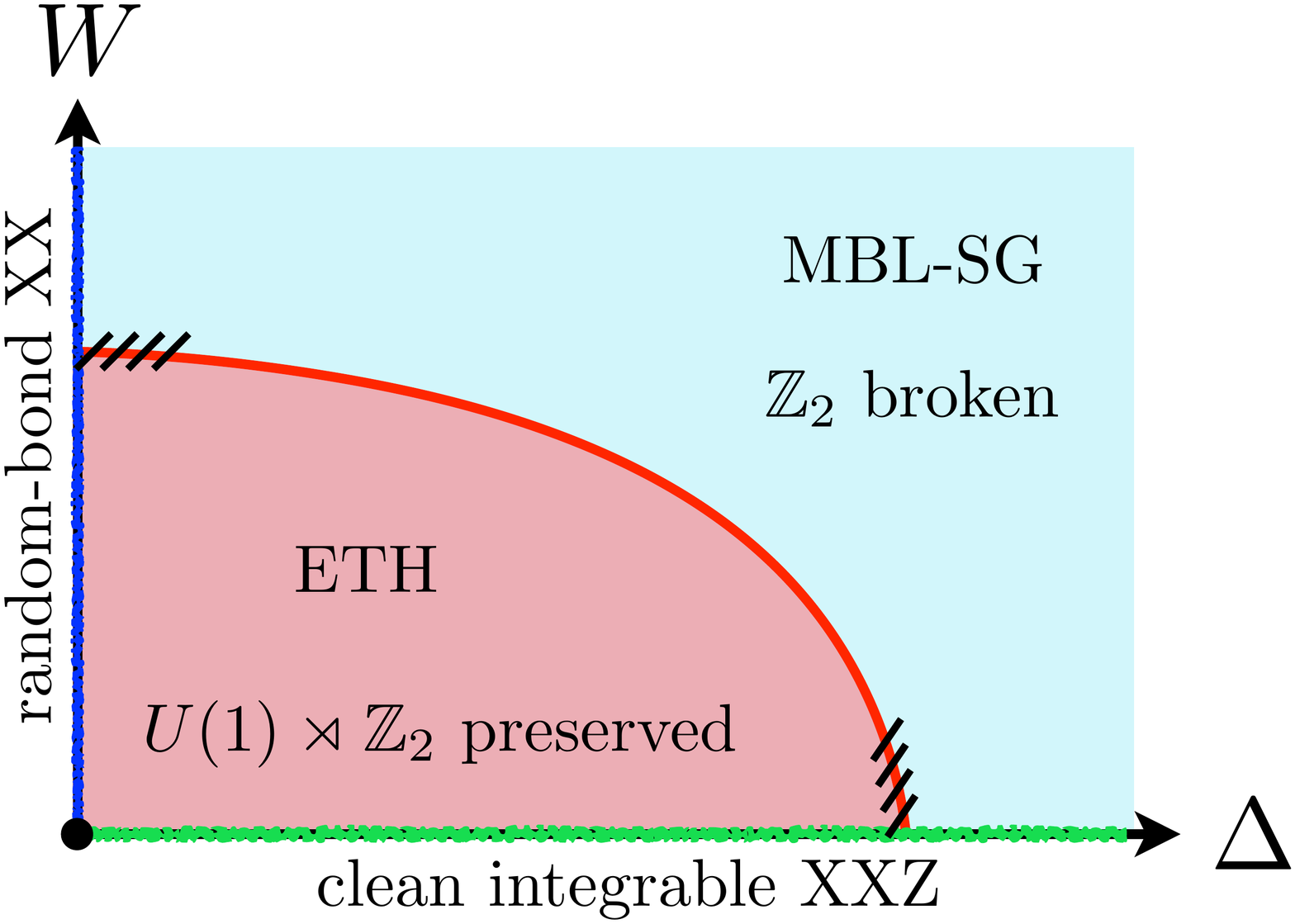}
\end{center}
\caption{\label{fig:XXZphasediag} Schematic eigenstate phase diagram of the random-bond XXZ  chain, characterized by  non-Abelian $U(1)\rtimes \mathbb{Z}_2$ global symmetry, as a function of interaction strength ($\Delta$) and disorder ($W$). Absent disorder (green line), the  problem reduces  to the clean  XXZ chain, which does noot  thermalize owing to being integrable in the conventional, Bethe-ansatz  sense.  For $\Delta=0$ (blue  line),  the  problem corresponds to the  random-bond  XX chain, whose properties are  controlled  by  a   random-singlet fixed point, with extensively degenerate excited  states; this again is non-thermalizing as it can  be mapped to a free-fermion problem via a Jorrdan-Wigner transformation. Away from these special lines, we find that for weak disorder and $\Delta \neq 0$ the system thermalizes (red shaded region); this can be intuitively understood by perturbing in the extensively degenerate manifold of excited states iin the $\Delta\rightarrow 0$ limit. For large disorder, however, we find a broken-symmetry MBL spin-glass phase where the $\mathbb{Z}_2$ part of the symmetry is broken (blue shaded  region). The eigenstate transition between these  phases is characterized both by the arrest of equilibration and the emergence of spin-glass order (and its consequent broken symmetry) at the onset of many-body localization. The hash marks indicate that the terminii of the  transition near the two special lines  corresponding to integrable and free systems is not fully understood, although numerical simulations on small finite systems of $L\lesssim 18$ spins indicate the behavior depicted.}
\end{figure}

When applied to the XX chain, as noted in Ref.~\cite{HuangMoore} this generalization naturally produces an extensive set of degeneracies at any finite energy density --- basically, for any $\Delta$ the spectrum of a individual XXZ bond term consists of two singlets and a doublet, but upon replacing the doublet with an effective spin we find that for $\Delta=0$ this decouples from the rest of the system. A typical finite energy-density excited state will have a finite  density of such decoupled effective spins-$1/2$. This provides a microscopic example of how an extensive degeneracy is  enforced by  non-Abelian symmetry. In both ground states and excited states for $\Delta=0$, the divergence of the single-particle localization length is manifested in the logarithmic scaling of the entanglement entropy of any finite subregion with its size, and in the usual `infinite randomness' length-time relation $l^\psi \sim |\ln t|$ with $\psi=1/2$, corresponding heuristically to a dynamical critical exponent, $z=\infty$. This suggests that the interacting chain is a candidate for `marginal MBL'~\cite{NandkishorePotter}. Indeed, a different generalization of the SDRG adapted to computing time evolution rather than eigenstates~\cite{vosk_many-body_2013} suggested that an initial product state of the N\'eel form, viz, $\ket{\uparrow\downarrow\uparrow\downarrow\ldots}$ would show growth of entanglement and correlations distinct from those of the fully-MBL phase, but that could be captured by random-singlet type arguments. However, the dynamical SDRG is highly state-dependent, and therefore it is natural to ask if these features extend to less special initial states, or indeed if they characterize the entire spectrum. The natural route to this is to incorporate interactions into the excited-state SDRG for the XX chain~\cite{XXZpaper}. This produces $S^z-S^z$ interactions between the previously decoupled effective spins generated by the non-singlet decimations. These interactions
do not naturally renormalize down under the RG, and as other terms get decimated they come to dominate the physics. This suggesting that an effective description of the system at long length scales takes the form of a simple classical Ising chain, $H_{\text{eff}} \sim J_i S^z_i S^z_{i+1}$, suggesting that the system spontaneously breaks the $\mathbb{Z}_2$ $S^z_i \mapsto -S^z_i$ symmetry. In the fermionic language, this is equivalent to breaking the sublattice/chiral/particle-hole symmetry. This leaves only the $U(1)$ spin-rotation symmetry about the $S^z$-axis (fermion number conservation), which is Abelian; for sufficiently strong disorder, the system enters an MBL phase. For weaker disorder, the system remains thermal.  This scenario is borne out by numerics on small system sizes, and is summarized  in Fig.~\ref{fig:XXZphasediag}.

Both these examples illustrate a special feature when MBL emerges in systems with a non-Abelian symmetry group. Recall that the standard thermodynamic arguments that demand that the lower critical dimension $d_\ell=2$ for finite-temperature discrete symmetry breaking remain operational in a system that satisfies ETH. Therefore, in the isolated setting at finite energy density  the symmetry can only broken when the system enters the MBL phase. Conversely, since localization is inconsistent with the full non-Abelian symmetry, the MBL phase must break this down to an Abelian subgroup. Therefore, the onset of localization must coincide with symmetry breaking --- potentially providing a window into the MBL phase transition using spin-glass-type order parameters that are likely more readily accessible than other quantities. (This is also  true of the glassy ordered phases  of the quantum clock models).

\subsubsection{Extended non-ergodic dynamical regime in the random-bond Heisenberg chain~\label{sec:unstableMBL:nonabelian:Heisenberg}}
Having disposed of cases where an MBL phase becomes possible with symmetry breaking, it is natural to ask about the case where any route to reduce the symmetry group to an Abelian one would involve continuous symmetry breaking. In such situations, the system necessarily thermalizes. However, for sufficiently strong disorder thermalization could nevertheless be characterized by an extremely long prethermal regime where the system shows many hallmarks of non-ergodic behavior, even though  it satisfies ETH on the longest timescales.  This is exemplified by the physics of the $SU(2)$-symmetric random-bond Heisenberg chain, 
\begin{equation}
 \label{eq:HamHeis} H =  -\sum_{j=1}^{L-1}J_j \vec{S}_j\cdot \vec{S}_{j+1}.
 \end{equation}
This is a canonical example of a model whose low-energy properties can be accessed by means of SDRG, and with a random-singlet ground state~\cite{PhysRevLett.43.1434,PhysRevB.22.1305,fisher_random_1994}. However,  generalizing the SDRG approach to target excited states, we find that although the effective Hamiltonian remains self-similar,  large effective `superspins' with $S\gg 1$ are generated with finite probability. As a consequence,  on the longest length/time scales a reasonable effective description is in terms of a {\it classical} random-bond Heisenberg model, which thermalizes~\cite{OganesyanPalHuse_classical}. Another way to see this is by numerically implementing the RG scheme; we find that the system flows to strong but not infinite randomness, meaning that it is no longer an asymptotically exact procedure: in other words, the SDRG breaks down when targeting excited states. A similar conclusion was also drawn from numerical studies of random-bond spin-1 chains with anisotropies that only preserved $\pi$-spin rotations about the three orthogonal spin axes; upon switching off these anisotropies, the localization-and-symmetry-protected topological `Haldane phase' of the spin chains gives way to thermalization~\cite{chandran_many-body_2014}. A more formal approach to the Heisenberg chain  instead makes a formal link to random  chains of $SU(2)_k$ anyons~\cite{QCGPRL}. In these models, for any finite $k$ the SDRG is controlled since the largest effective `spin' that it can generate has $S=k/2$. It is possible to define a crossover scale $\xi_k$ such that in an eigenstate the entanglement entropy of a subsystem of size $L\gg \xi_k$ scales logarithmically. While the Hilbert space, algebraic structure, Hamiltonians, and other properties of the $SU(2)_k$ models for finite $k$ are in fact distinct from those of $SU(2)$, in the $k\rightarrow \infty$ limit the two become asymptotically equivalent, at least at the level of the SDRG. By taking this limit, it is possible to see that both $\xi_k$ and the typical size of the effective spin diverges at late stages of the RG diverge, indicating the breakdown of the technique, and presumably the onset of thermalization.

Though the Heisenberg chain is an ergodic system and does indeed thermalize, its approach to equilibrium is extremely slow, with an extended regime where the dynamics `look' non-ergodic (see also Ref.~\cite{agarwal1fnoise}). As argued by  Protopopov et. al.~\cite{ProtopopovSU2}, the slow dynamics is best viewed from the perspective of the SDRG and its broad distributions  and critical correlations. Ref.~\cite{ProtopopovSU2}  examined the  non-ergodic dynamics of the Heisenberg chain by first using the SDRG to approximate the excited eigenstates as a tree tensor network. Each tree is characterized by its geometry and the choice of total block spins at each node. SDRG eigenstates are non-thermal:  for instance, they have logarithmic entanglement intermediate between the area (volume) law scaling for fully MBL (fully ergodic) systems. The origin of this logarithmic entanglement is especially simple to understand in two limits.  Specializing to half-system cut in a system of size $L$, the first limit is to  consider an  eigenstate with `random singlet' structure, with $\sim \log L$ low-energy singlets each with bounded entanglement traversing the entanglement cut. The ground state of the random Heisenberg antiferromagnet is of this form. The second limit is to imagine  the two components of the bipartition each form a large `superspin' with $S\sim L/2 $, with a single `singlet' forming across the cut, thereby also yielding $S_{\text{ent}} \sim \log L$. Realistic highly-excited tree states will be somewhere intermediate between these limits: there can be several large-spin singlets across the cut, but the number of such singlets is bounded by the depth $d< \log_2 L$ of the tree. Therefore the entanglement is upper-bounded by $S_\text{ent}(L/2)< c\log_2^2L$ where $c$ is a positive constant.

However, that such trees only form  an approximate eigenbasis follows from the fact that when written in this basis the Heisenberg chain Hamiltonian \eqref{eq:HamHeis} is not fully diagonal; the off-diagonal terms are resonances that involve a local rearrangement of the tree that characterizes the  approximate eigenstate, i.e. indicates a local breakdown of the SDRG approximation. It can be demonstrated~\cite{ProtopopovSU2} that the correction to the eigenstate entanglement grows logarithmically with the number of such tree states admixed by the perturbation. Accordingly, an exponentially large number of tree states  is required to recover the volume-law scaling of ergodic eigenstates --- put differently, we require an extensive number of hot spots to cooperate in order to drive thermalization.  There are two key $W$-dependent length scales that control resonances: first, they do not appear at on on length scales $L<L_1(W)$; second, although isolated resonances appear on scales $L> L_1(W)$, they are ineffective in proliferating to drive thermalization until we consider scales $L>L_2(W)$, when individual resonances start overlapping. Both length scales remain finite for any $W<\infty$, consistent with eventual thermalization.   However, even for very strong disorder,  $L_2(W) \gtrsim 300$, which lies beyond the range for most `synthetic' platform such as trapped ultracold atoms and ions where truly isolated  systems can be realized. Therefore, for experimentally accessible length and time scales, the SDRG approximation is extremely accurate, and predicts an extended non-ergodic regime.  Further investigations of the dynamics of this regime are clearly warranted, as well as extensions to other symmetry groups. The use of SDRG as a window into the dynamics of such systems is a promising technical tool, but it remains to be seen if  numerical techniques can similarly benefit from making  a link to the tree approximation.

\subsection{MBL in higher-dimensional systems with short-range interactions}

We now turn to systems that are destabilized, not by symmetry, but by spatial dimensionality or interaction range. We first consider systems with short-range interactions in dimensions $d > 1$. At the perturbative level such systems exhibit full MBL when strongly disordered; however, the nonperturbative avalanche instability potentially thermalizes the system on very long timescales. In what follows (and in the rest of this section) we assume that avalanches and related instabilities---such as the ``hot bubble'' instability for MBL systems with putative mobility edges---exist, and set out the consequences for dynamics in the nearly MBL regime. 

To estimate the density of avalanches and thus the rate at which the system thermalizes, we follow Ref.~\cite{gh2019}. Our present aims are a little different, however, and the notation is correspondingly also different. In what follows we consider a generic disordered local Hamiltonian with the disorder bandwidth set to unity, and a nearest-neighbor hopping term of strength $J \ll 1$. 

When $J \ll 1$, the bottleneck for creating an inclusion is finding many spins with energies that are within $J$ of one another. For short-range interactions in generic dimensions, the density of inclusions of size $\ell$ is $n(\ell) \sim J^\ell$. For an inclusion to be able to destroy MBL, it must be able to absorb at least some of the typical spins that surround it. The bath that such an inclusion generates is ``narrow-bandwidth'' (Sec.~\ref{sec:multicomponent:slowbath}) since it can only easily absorb energy at the scale $J$ (or at best $J \ell$). To couple into such a bath, a typical spin needs to undergo a rearrangement that flips $\sim |\log J|$ other spins; the matrix element for this is $\sim J^{\log J} \sim \exp(- (\log J)^2)$ (omitting prefactors that do not affect our conclusions). To act as a bath, the inclusion must therefore have a minimum size of $\ell \sim (\log J)^2$; it follows that $\log[n(\ell (J))] \sim -|\log J|^3$. 

We now consider the long-time dynamics of a system with this sparse network of baths. A typical degree of freedom is a distance $1/n(\ell)^{1/d}$ from the nearest bath, so to leading order its decay rate is $\Gamma \sim J^{1/n(\ell)^{1/d}} \sim \exp[\exp(-|\log J|^3)]$.
Note that this decay rate is \emph{much} faster than the double exponential in $J$ (following from a naive construction of  resonances) that is often claimed. One can go beyond the typical decay rate and estimate the functional form of the decay as well, as follows. At a time $t$, the fraction of the system that has relaxed is the fraction that is close enough to an inclusion. A spin at a distance $q$ from the inclusion decays at a rate $e^{-q |\log J|^2}$. At a time $t$, a region of of size $(\log t / |\log J|^2)^d$ is close enough to an inclusion to relax. Thus there is a large intermediate time window in which local autocorrelators decay with the functional form $C(t) \sim A - (\log t / |\log J|^2)^d$. 

\subsubsection{Weakly interacting, weakly localized Anderson insulators}
A special limit of the higher-dimensional MBL problem is one-dimensional MBL with interactions that fall off as an exponential with a sufficiently slow decay constant. 
A natural context in which this model appears is that of weakly interacting, weakly localized Anderson insulators. If one writes the model in the basis of localized orbitals, the interactions generate two-particle ``hopping'' processes between orbitals, which fall off exponentially in the localization length $\xi$. We consider the limit of $\xi \gg 1$ and $U \ll 1/\xi^3$; here the localized phase is stable against perturbative rearrangements at any order. Once again, we begin by finding the optimal critical inclusion. Here, this consists of a string of sites spaced by a distance $\exp(-r/\xi) \simeq U$. The density of an inclusion of size $\ell$ is given by $U^\ell$. Typical degrees of freedom are incorporated into the inclusion through large-scale rearrangements of $\sim |\log U|$ spins. This case therefore ends up behaving much like the high-dimensional short-range example, but with $U$ replacing $J$.

\subsection{MBL in systems with power-law interactions\label{sec:unstable:powerlaw}}

We now turn to systems with power-law interactions, which fall off as $1/r^\alpha$ with distance, in a $d$-dimensional system. There are three cases: (i)~rapidly decaying power laws, $\alpha > 2d$, for which MBL is perturbatively stable, but is destroyed by avalanches in a way that parallels the short-range case; (ii)~slowly decaying power laws, $0 < \alpha < 2d$, for which there is a \emph{perturbative} instability; and (iii)~``confining'' models with $\alpha < 0$, which typically arise as low-energy descriptions associated with gauge theories (in one dimension, specifically, the Schwinger model~\cite{nandkishore_sondhi, brenes2018, akhtar2018}). We will discuss the first two cases below, and return to the last case when we discuss disorder-free localization in Sec.~\ref{sec:unstable:disorderfree}.

\subsubsection{Rapidly-decaying power-law interactions}

When the interaction power law $\alpha > 2d$, MBL is perturbatively stable, so to destabilize it we need to construct an avalanche, as in the short-range case above. A major difference between power-law and short-range interactions is that in the power-law case, an inclusion of size $\ell$ is parametrically less expensive to create. Once the inclusion has grown to size $q$, it has level spacing $2^{-q}$, and can resonantly absorb any spin at a distance $R$ such that $(J/R^\alpha)^2 > 2^{-q}$. The square in this expression comes from the Golden Rule. The number of such spins is $\sim 2^{d q / (2\alpha)}$. When $q$ is large enough this number is always greater than $1/J$, so the avalanche runs away. The critical size $\ell$ at which this happens is $\ell \sim 2 \alpha |\log J| / d$, and the corresponding density of such inclusions scales as $n(J) \sim \exp(-|\log J|^2)$. 

Again unlike the short-range case, a typical spin couples to these inclusions via a few-spin process, i.e., via a three-spin process in which energy is borrowed from the inclusion. Since all matrix elements are power-law, the relaxation rate scales as $\log \Gamma \sim -|\log J|^2$, i.e., it vanishes barely faster than a power-law in the disorder strength. These decay rates, unlike those in the short-range case, should be readily experimentally accessible. Moreover, an analysis parallel to that in the short-range case indicates that the behavior of the autocorrelation function at time $t$ behaves as $C(t) \sim \mathrm{const.} - t^s$, where $s$ is a positive power law. Eventually, this crosses over to a stretched exponential relaxation (which might be subleading to hydrodynamic power laws in systems with conservation laws).

\subsubsection{Slowly decaying power-law interactions}

We now turn to the case $\alpha < 2d$, for which MBL is perturbatively unstable to few-spin resonances. There are two sub-cases here: when $\alpha < d$, essentially every spin is involved in a resonance~\cite{levitov1989absence, levitov1990delocalization, aleiner2011localization}; by contrast, when $d < \alpha < 2d$, a sparse resonant network of spins forms~\cite{burin1998, burin2006, ylg, burin2015, burin2015b, PhysRevB.93.245427, tikhonov2018}, and acts as a narrow-bandwidth bath for typical spins. 

\emph{Case 1: $\alpha < d$}.---We start with a typical degree of freedom and try to find perturbative resonances at increasing distance scales, as in previous examples. At a distance $R$, the matrix element falls off as $R^{-\alpha}$, but the typical number of spins at that scale is $R^d$, so the characteristic energy denominator scales as $R^{-d} \ll R^{-\alpha}$. Perturbative resonances therefore proliferate\footnote{At first sight one might think the number of spins is $R^{-(d-1)}$. To see why this is not so, let us suppose the interaction at scale $R_0$ is $J_0 \sim 1/R_0^\alpha$. The distance at which the coupling decreases to $J_0/2$ is set by $R_0' = 2^{1/\alpha} R_0$. Thus the distance over which the coupling scale stays approximately the same is $|R_0' - R_0| \propto R_0$. The region is therefore an annulus with all of its dimensions proportional to $R_0$. By contrast, in a localized system with short-range interactions, one finds that (defined analogously) $R_0' = R_0 + \xi$.}. 

Precisely at the critical value $d = \alpha$, it appears~\cite{levitov1990delocalization} that the density of resonances grows logarithmically rather than algebraically, so any spin still asymptotically finds a resonance. In one dimension, for noninteracting fermions, the critical properties of this problem [called the ``power-law random banded matrix'' (PRBM) ensemble] have been extensively studied~\cite{mirlin1996transition, cuevas2001anomalously}. Beyond these critical properties, there is an interesting short-time transient in the autocorrelation function, which was experimentally observed in systems of nitrogen-vacancy centers and explained in Ref.~\cite{kucsko2}. We ask for the probability that the system has not found a resonance by time $t$ (and thus by distance scale $R(t) \sim t^{1/\alpha}$). This is a product of probabilities that no resonances were found at each distance scale less than $R(t)$, which takes the form (using $\alpha = d$): 

\beq
C(t) = \prod_{1 < r < R(t)} \left( 1 - \frac{\mathcal{C} r^{d-1} dr \times J}{r^d} \right) \sim \exp\left(\mathcal{C} J \int_1^{R(t)} \frac{dR}{R} \right) \sim t^{-\mathcal{C} J / d}.
\eeq
This behavior applies at timescales shorter than a crossover timescale (which diverges rapidly as $J \rightarrow 0$) on which typical spins relax; beyond this timescale, most levels are broadened and resonance-counting arguments cannot be applied. 

In the presence of interactions, however, this critical behavior is likely washed out, as the two-spin resonances we will discuss next provide the dominant relaxation channel. 

\emph{Case 2: $d < \alpha < 2d$}.---In this intermediate case, a typical spin no longer finds a resonant partner by the criterion above. Nevertheless, one can form slightly more complex resonant objects as follows. At any distance scale $R$, there will be a density of resonances of characteristic size $R$, and the density of these scales as $n_R \sim R^{d - \alpha}$ as above. Each of these resonances has a matrix element of the same order of magnitude $\sim J/R^\alpha$; thus, if they are within a distance $R$ of each other, they will hybridize with high probability. Now consider one such resonant pair; it has $R^d n_R$ resonances within a radius $R^d$. Thus the total number of resonances within the accessible distance scales as $R^{2d - \alpha}$, and when $R$ is big enough a percolating network of resonances forms~\cite{burin1998, burin2006, ylg}. A major difference between this case and the previous case is that here the resonances are still \emph{sparse} (consisting of a few resonances on the critical scale $R_c \sim J^{-1/(2d - \alpha)}$): a typical spin does not participate in the network. Once again, the resulting bath is a narrow-bandwidth bath, of bandwidth $J^{2d / (2d - \alpha)}$, and the decay rate into it can be addressed as for the case of more rapidly decaying power laws. 

\subsubsection{Mixed power laws and the Fermi glass}

So far, we have considered simple cases where all the power laws coincide. However, many situations involve Hamiltonians in which the hopping and interaction terms decay differently. For instance, electrons on a lattice hop locally but interact via long-range Coulomb interactions. Likewise, in systems of nitrogen-vacancy centers~\cite{ylg} it is natural for the flip-flop interactions between spins to fall off with a power-law $1/R^6$ even though the diagonal ($\sigma^z \sigma^z$) interactions between them fall off as $1/R^3$. We take systems of tightly Anderson-localized electrons interacting via the Coulomb interaction as an illustrative example (this example dates back to work by Anderson and coworkers in the 1970s~\cite{fleishman}). Here, most electrons are essentially inert, but a small fraction form short-range resonances. These resonances are distributed randomly and dilutely across the sample; being literally charge dipoles, they interact via the dipole-dipole interaction. In three dimensions this system falls under case 1 above---so a typical dipole is able to engage in resonant energy transfer with others---while in two dimensions it falls under case 2---a resonant network made of pairs of dipoles (i.e., pairs of resonant pairs of electronic sites) forms. These iterated constructions of resonant objects are generic for mixed power laws. 

Note that in either of these cases, the dipoles only transport energy; moreover, at strong disorder, they form a bath with a narrow bandwidth set by the considerations in previous sections. Charge transport occurs through incoherent hopping mediated by the narrow-bandwidth bath, as already appreciated by Fleishman and Anderson~\cite{fleishman}. 

\subsection{Low-temperature systems with hot bubbles}

We now turn to a different class of instability, which is due to rare fluctuations of some conserved density (such as the energy or particle density). 
Recall that the criterion for perturbative MBL (Sec.~\ref{sec:intro:mblpert}) involved comparing the exponential scaling of typical matrix elements to the density of states. The latter depends on the temperature and chemical potential. 
Even when states at infinite temperature are perturbatively unstable to thermalization, states at nonzero but low temperatures (where the many-body density of states is far lower) will remain perturbatively stable. 
Thus, at the perturbative level, there are many-body ``mobility edges,'' i.e., critical temperatures and densities at which a system becomes unstable to thermalization.
These edges show up in two basic varieties: (i)~the ``regular'' mobility edge scenario, first considered by~\cite{basko_metalinsulator_2006}, in which states with high energy density are mobile but those with low energy density are static; and (ii)~the ``inverted'' mobility edge scenario that occurs in symmetry-protected systems as well as in translation-invariant systems. 

\emph{Regular mobility edge}.---We now lay out the argument of Ref.~\cite{drhms} against the existence of many-body mobility edges (MBMEs). This argument begins by assuming the existence of MBMEs and showing that the assumption is inconsistent. Consider cutting the system up into a large number of mesoscale grains, by removing some fraction of bonds from the Hamiltonian. This gives us a modified Hamiltonian $H'$. The strategy is to construct eigenstates of $H$ perturbatively beginning from those of $H'$. If the grains are large enough, the energy density of a state under $H$ is essentially the same as under $H'$, and we will ignore the distinction. Moreover, eigenstates of $H'$ are product states over grains; such product states have sharp energy densities under $H'$ as well as under $H$. We consider eigenstates of $H'$ that are globally well below the putative MBME, which occurs at some energy density $E_0$. However, statistically, essentially all of these states will have some grains in which $E > E_0$. By assumption, these eigenstates will be locally thermal (which we assume means random-matrix-like). Indeed, a typical eigenstate of $H'$ in a sufficiently large volume will contain arbitrarily long strings of such locally ``hot'' states. 

Starting from these grain-product states, we attempt to construct eigenstates of $H$ by adding in the perturbative coupling between grains. We consider an initial eigenstate with a string of hot grains, stretching from grain $i$ to grain $i + N$, surrounded by cold (typical) grains on both sides. We then argue that this eigenstate is \emph{resonantly} connected to many eigenstates in which the string is translated by one site, i.e., the hot grains stretch from $i + 1$ to $i + N + 1$. The argument for this is simple to state at the intuitive level: first, the hot string acts as a bath and can therefore thermalize the grain immediately to its right (it is resonantly connected to a slightly less hot but slightly larger thermal string). By the same logic, the \emph{translated} hot string can thermalize the grain immediately to its left, and therefore is also resonantly connected to the same state with a slightly larger but cooler string. Therefore, the hot string and its translation are resonantly connected to one another, implying that the hot string can move by an inchworm-like process of sequentially expanding and contracting. However, if the hot string is mobile, the many-body eigenstates of $H$ (which contain a finite density of hot strings) cannot be localized, and there cannot be a true many-body mobility edge.

This discussion has only addressed \emph{typical} eigenstates below the putative MBME. One might wonder if these results extend to \emph{all} states, or if, instead, rare states with no hot bubble configurations can remain localized. Since (according to the argument above) the typical states form a bath, perturbative considerations suggest that any state at the same energy can hybridize with the bath, and therefore all states are delocalized. For a detailed discussion of other potential caveats see Ref.~\cite{drhms}. Although the argument against MBMEs is not rigorous, no compelling counterargument exists at present. The numerical evidence suggests that MBMEs exist; however, it is derived primarily from exact diagonalization of small systems and from DMRG-like methods that target low-entanglement states~\cite{lla, PhysRevB.97.104406, pietracaprina2018}, and is thus potentially consistent with the presence of hot bubbles in the asymptotic large-size limit. However, a quantitative understanding of finite-size effects in this case is lacking.

The rate-limiting step for this mechanism is for a hot bubble to reach the point of interest. In one dimension a hot bubble of size $L$ has a level spacing $\exp(-s(T_h) L)$, and its bandwidth is of order unity; moreover, the speed of the hot bubble does not depend in any obvious way on $L$. Meanwhile, the matrix element for a rearrangement of typical degrees of freedom outside the hot bubble is $e^{-1/\xi_{\mathrm{typ.}}}$. So the criterion for the hot bubble to be able to grow is that $L > 1/(s(T_h) \xi_{\mathrm{typ.}})$. The probability of such a hot bubble then goes as $\exp(-T_c / (T \xi_{\mathrm{typ.}}))$. Thus we expect this mechanism to lead to activated transport in the low-temperature limit. 

\emph{Inverted mobility edge}.---In some systems, such as those exhibiting ``translation-invariant MBL,'' the mobile states are near the edges of the spectrum rather than in the middle. For instance, the mobile states might be those at very low particle density (as in translation-invariant MBL; see below). If particles are mobile beneath a critical density then this case works much like the previous one: an anomalous region acts as a bath and thermalizes the regions around it (heating up slightly in the process), then reconstitutes itself in a different location. The main difference between this case and the previous one concerns finite-size effects: cold regions have a small density of states, for a cold region to act as a good bath it must be extremely large. 
However, the present discussion leaves open the following possibility: the characteristic localization length of excitations could diverge at zero temperature, while obeying the criterion that for regions at any finite temperature $T$ (and corresponding entropy density $s(T)$), the localization length always obeys the avalanche stability criterion $\zeta(T) s(T) < 1$. There are no explicit instances of this behavior, but it is unclear whether this scenario can occur and how it would fit in with our understanding of MBL. 

\emph{MBL in the continuum}.---We now turn to the question of whether MBL is possible in the continuum, in one- or two-dimensional systems. In these systems all single-particle {orbitals} are localized, but the energy-dependent localization length $\xi(E)$ diverges as $E \rightarrow \infty$. In one dimension, $\xi(E) \sim E$, whereas in two dimensions $\xi(E) \sim \exp(E)$. Clearly these systems do not exhibit MBL at sufficiently high temperatures; therefore they are susceptible, at all temperatures, to the hot-bubble argument described above. However, there are also other, more perturbative, channels that lead to delocalization. 

Following~\cite{continuumMBL} we can regard these systems as consisting of two types of single-particle orbitals: (i) typical (cold) {orbitals} with energies comparable to the temperature $T$ and thermal occupation factors and localization lengths of order unity, and (ii) atypical (hot) {orbitals} with energies $E \gg T$ and occupation factors that are suppressed as $\exp(-E/T)$. The potential delocalization mechanism involves a process where a hot electron transitions incoherently between two high-energy {orbitals} by exchanging some energy with the cold sector. (When one restricts one's attention to processes that act entirely within the cold sector, one does not find delocalization~\cite{aleiner2010finite, nandkishore_continuum, michal2016finite}.) 

\emph{One dimension}.---We consider this first in one dimension. Here, the density of states at energy $E$ goes as $\rho(E) \sim 1/\sqrt{E}$, and $\xi(E) \sim E$. One starts with an initial occupied hot orbital at energy $\sim E$, and then exchanges energy with a cold orbital with energy $\sim T$. This can change the energy of the hot orbital by $\sim T$; hot orbitals with an appreciable matrix element for such a transition are those within $T$ in energy and $\xi(E)$ in real space of the initial orbital. Meanwhile, $\xi(E)$ cold orbitals can contribute to this transition. Thus the density of states for (hot + cold) $\rightarrow$ (hot + cold) transitions scales as $1/E \times 1/\sqrt{E} \sim E^{-3/2}$. We now turn to the matrix element, which is given by the overlap of four wavefunctions, 

\beq
V_{ijkl} = \int dx_1 dx_2 \psi^h_i(x_1) \psi^c_j(x_2) V(x_1 - x_2) [\psi^h_k(x_1) \psi^c_l(x_2) - \psi^h_k(x_2) \psi^c_l(x_1)].
\eeq
We expect the first (Hartree) term to be the dominant one, since the second (Fock) term involves an overlap between two wavefunctions with strongly different oscillation amplitudes. To estimate it we first observe that the integrand is only nonzero in a region of size $\sim \xi_c$. Within this region the integrand oscillates rapidly on a scale $E$. Combining the suppression of the integral due to the rapid oscillations ($\sim 1/\sqrt{E}$) and due to the normalization of the hot wavefunctions ($\sim 1/E$) we find that the matrix element \emph{also} typically vanishes as $E^{-3/2}$. Thus, to lowest order in perturbation theory, MBL is perturbatively marginally stable in one dimension in the continuum, when the disorder is white-noise correlated. 

\emph{Two dimensions}.---We now turn to two dimensions, where $\xi(E) \sim \exp(cE)$ and the density of states at high energies is constant. A hot orbital at energy $E$ is connected to $\sim \xi(E)^2$ other hot orbitals, and each such connection could involve $\xi(E)^2$ local rearrangements in the cold sector. Thus the typical level spacing for this process scales as $\sim 1/\xi(E)^4$. Meanwhile, following the same logic as in the one dimensional case, the matrix element for any such transition scales down as $1/\xi(E)^3$. Asymptotically, therefore, orbitals above a critical energy $E_c$ are mobile, and act as a bath for typical degrees of freedom. Since $\xi(E) \sim \exp(cE)$, and $\xi_c \sim T$, it follows that $E_c \sim \log T$. Thus, transport in the two-dimensional continuum is marginally super-activated, with the conductivity scaling at low temperatures as $\log \sigma(T) \sim - \log T / T$. 

In both of these cases, at sufficiently low temperatures this superactivated scaling will cross over to simple activated scaling but with a much smaller prefactor, as rare mobile thermal regions begin to dominate over typical regions. 

\subsection{Translation-invariant systems\label{sec:unstable:disorderfree}}

The question of whether MBL can occur without disorder dates back, in some form, at least as far as the work of Ref.~\cite{kagan2}. The intuition for why this might be possible is similar to that behind the classical glass transition, jamming, or other related phenomena: starting from a typical finite-temperature state of a strongly interacting system, few-particle moves do not, in general, allow one to explore all of the accessible configuration space. Instead, a particle rattles in a cage formed by its neighbors. Because of Anderson localization, quantum mechanics makes this caging phenomenon more severe, since a single quantum particle can be strictly localized even at finite temperature in a background formed by the other particles in the system. 

Slow dynamics in strongly interacting, clean systems had been extensively studied even before the revival of interest in MBL~\cite{eckstein_thermalization_2009, bernier2011, carleo2012}. The first works to revisit this question in the context of MBL were Refs.~\cite{grover_quantum_2014, drh2014, schiulaz_ideal_2014}. Refs.~\cite{grover_quantum_2014, schiulaz_ideal_2014} both considered two-species models of heavy and light particles. 

In the model of Ref.~\cite{grover_quantum_2014}, the heavy and light particles move in the continuum, and heavy particles attract light particles. This work conjectured the existence of a phase in which the heavy particles eventually delocalize, but the light particles are never able to escape from the heavy particles. Thus (according to this conjecture) if one were to perform a projective measurement of all the heavy particles, the remaining light-particle wavefunction would have low (i.e., area-law) entanglement, even though the global wavefunction is volume-law entangled. At present this conjecture is not widely accepted, especially for particles in the continuum: there are many potential mechanisms that would destabilize the proposed state, few of which have been ruled out. 

Ref.~\cite{schiulaz_ideal_2014} considers two species of fermions $c$ and $d$ on a one-dimensional lattice, subject to the Hamiltonian
\beq
H = -J \sum\nolimits_i (c^\dagger_i c_{i+1} + \mathrm{h.c.}) (1 - d^\dagger_{i+1/2} d_{i+1/2}) + \lambda \sum\nolimits_i (d^\dagger_{i -1/2} d_{i + 1/2} + \mathrm{h.c.})
\eeq
We begin by setting $\lambda = 0$, and considering a typical high-temperature state of the system. In this state, $d$ particles are distributed randomly, and constitute broken bonds. The $c$ particles are thus confined to segments; their eigenstates are the appropriate Slater determinants within each segment. These eigenstates are manifestly localized. Now we perturbatively consider the hopping of $d$ particles. Typically, this will not be a resonant process: the energies of the $c$ particles depend on the position of the barrier, so moving the barrier will generically change the energy. However, some perturbative moves will be resonant: for instance, if moving a barrier just interchanges two segment lengths. Ref.~\cite{schiulaz_ideal_2014} estimates the number of perturbative resonances, and finds that these tend to become sparser at higher orders in perturbation theory. Thus, at the perturbative level this model appears to exhibit MBL. However, it is straightforward to construct anomalous regions with a low density of barriers, which thermalize. Thus, one expects this to be a special case of MBL with a mobility edge, as already pointed out in Ref.~\cite{de2015can}. The mobile regions should diffuse, and in the absence of quenched disorder it seems that the diffusion constant should stay finite. 

The model considered contemporaneously in Ref.~\cite{drh2014} is similar in spirit, but has some important differences. This model consists of \emph{bosons} on a lattice, subject to a generalized version of the Bose-Hubbard Hamiltonian:
\beq
H = \sum_i [ n_i^q + g_1 (b^\dagger_i + b_i)^2 ] + g_2 \sum_{\langle ij \rangle} (b^\dagger_i b_j + \mathrm{h.c.}).
\eeq
The second term breaks particle number conservation, whereas the last term (which couples different sites on the lattice) is to be regarded as a small perturbation. The authors choose $q > 2$ to decrease the number of possible local resonances. Since the system consists of bosons and thus has a spectrum that is unbounded above, the infinite-temperature ensemble is not well-defined. Nevertheless, one can imagine working at very high temperatures, where a typical initial state has a large number variance from site to site. The argument for localization now proceeds as in the previous example: a resonant process where two neighboring sites exchange their particle number state involves simultaneously moving many particles so it can only occur at very high orders in perturbation theory. However, lower-order processes give rise to site-dependent self-energies, which are in general much larger than the matrix element for resonant hopping. Thus, at the perturbative level the system is localized. 

As in the previous example, rare cold regions act as baths and prevent full localization~\cite{de2015can}. However, in the present model a notion of asymptotic localization at high temperature can be proved: specifically, the thermal conductivity is upper bounded at high temperature ($\beta \rightarrow 0$) by
\beq
\lim_{\beta \rightarrow 0} \beta^{-n} \kappa(\beta) = 0, \quad \text{for all } n > 0.
\eeq
Subsequent work on lattice models with a finite-dimensional on-site Hilbert space has seen anomalous diffusion on intermediate timescales~\cite{yao_disorderfreembl}, but with an eventual crossover to diffusion in the long-time limit~\cite{papic2015many}. 

We briefly note some other models that have been studied in this context. Ref.~\cite{yao_disorderfreembl} considers a two-leg spin ladder (with legs indexed by $0, 1$), governed by a Hamiltonian of the form

\beq
H = H_0 + \epsilon H_1 + J_z \sum_{i = 1}^N \sigma^z_{i0} \sigma^z_{i 1}.
\eeq
Here, $H_0, H_1$ are generic translation-invariant local Hamiltonians. In the limit $\epsilon = 0$, a generic eigenstate of this model consists of a $\sigma^z$-basis product state on leg $1$, which acts as binary disorder for leg $0$ and localizes it when $J_z$ is large enough. When $J_z\neq 0 $ this model is very similar to the two-species fermionic model discussed above. However, this way of writing the model makes the following point particularly transparent: if one initializes $H_1$ in the state $|+x\rangle^{\otimes N}$ for $\epsilon = 0$, this simply amounts to taking an average over the binary disorder. Thus, quantities such as correlation functions appear MBL even though neither the Hamiltonian nor the initial state is random. This feature is also shared by the model in Ref.~\cite{knolle1, knolle2}, which can be rewritten via a duality transformation to have a similar form to the $\epsilon = 0$ limit of the model discussed above. We expect that all of these models ultimately thermalize away from special fine-tuned points.

\emph{Gauge theories}.---In one dimension, models with extensively many local operators that commute with the Hamiltonian are fine-tuned. However, one general way such Hamiltonians could arise is as lattice gauge theories~\cite{knolle3, brenes2018}. Lattice gauge theories generically have local operators on each link that, e.g., implement Gauss's law on that link when they are in their ground state. The physical subspace of the Hilbert space of the gauge theory is usually taken to be the projection of the full Hilbert space of the lattice model onto a particular sector of these charges (which is typically very simple and not localized). The observation in these works, rather, is that MBL occurs in \emph{typical} sectors of the lattice gauge theory. The localized phases in these models can have unusual spectral properties, explored in Ref.~\cite{nonfermiglasses}.

\subsection{Can there be an avoided MBL critical point?}

In many of the settings discussed here, there is a sharp perturbative MBL transition that is rounded out, asymptotically, because of a nonperturbative instability of the putatively localized phase. Similar situations occur elsewhere: for example, noninteracting Weyl semimetals subject to weak disorder~\cite{fradkin1986, goswami2011, syzranov, pixley_weyl, pixley_weyl2, buchhold2018}. Here, weak disorder is perturbatively irrelevant at the Weyl point, and there is a critical value of the disorder strength at which the semimetal undergoes a phase transition into a diffusive metal. Nonperturbative rare region effects round out this transition, destabilizing the semimetal for arbitrarily weak disorder (but see Ref.~\cite{buchhold2018} for a contrary view). Nevertheless, the avoided perturbative transition manifests itself in the critical behavior of the density of states away from zero energy. In this case, replacing randomness with quasiperiodicity \emph{restores} the sharp phase transition~\cite{pwhg}. 

It is natural to ask whether some generalization of this idea applies to MBL. The distinction between this situation and MBL is that, in dynamics at least, the MBL phase is a critical \emph{phase} with continuously varying dynamical exponents, and the critical point is an endpoint of this phase. Thus the critical behavior at the MBL transition is the same as the behavior in the MBL phase, and there is no distinct quantum critical ``fan.'' 

\subsection{Implications for Localization-Protected Order}
We have already noted that the prospect of using localization to protect unusual non-equilibrium topological and broken-symmetry phases even in the presence of strong interactions is one motivation to study MBL. However, the considerations in this section suggest that many obvious routes to engineering such non-equilibrium may be obstructed by considerations due to dimensionality, symmetry, or some combination thereof. However, there may yet be possible ways to achieve very long prethermal regimes that nevertheless allow for coherent manipulations. Ref.~\cite{SPRVReview} provides a detailed summary of the situation, and we refer the reader to it for further details.

\section{Multiple-component systems\label{sec:multicomponent}}

In this section we consider systems in which some degrees of freedom are MBL and others are thermal. Such systems can be divided into two categories: those in which MBL and thermal degrees of freedom are \emph{spatially} segregated and those in which they are interspersed. This section focuses on the latter case (of ``bulk baths''). The physics of ``boundary baths'' is that of very large thermal inclusions, in the limit where their level spacing goes to zero; this was addressed in the introduction and in the previous section.

\subsection{MBL coupled to a large Markovian bath\label{sec:multicomponent:largebath}}

We first consider the simplest type of bath, which is large (so that there is no quantum back-action), infinite-temperature (so that uphill and downhill transitions are equally likely), and Markovian (i.e., it has zero correlation time and therefore an infinite bandwidth, so it can exchange arbitrary amounts of energy with the system). 

\subsubsection{Generic coupling}

The most generic coupling between an MBL system and a bath allows for processes where LIOMs flip assisted by the bath. The typical rate for such incoherent rearrangements is set by the system-bath coupling (which we assume not to be strongly disordered); for a Markovian high-temperature bath, the dominant processes involve either flipping single LIOMs or locally rearranging them, so the lowest-order process allowed by symmetry is the dominant one. Take the characteristic rate for such a rearrangement to be $\Gamma$; adapting NMR terminology, we can call this a generalized $T_1$ time. Then Ref.~\cite{Nandkishore14} argued that the corresponding $T_2$ time at which local superpositions dephase (i.e., the rate at which spin echo signals decay) is parametrically faster. For specificity we define the $T_2$ time via the decay of the correlator~\cite{shantsev} 

\beq
C_\phi(t) \equiv \left\langle \exp\left( i \int_0^{2t} dt' \mathrm{sign}(t - t') h_{\mathrm{eff.}}(t') \right) \right\rangle. 
\eeq
where $h_{\mathrm{eff.}}(t)$ is the splitting of a given LIOM at time $t$, i.e., its effective field $h^{\mathrm{eff}}_i = h_i + \sum_j J_{ij} \tau^z_j + \ldots$. Thus the $T_2$ time for a given LIOM depends on the states of all the LIOMs near it. Each of these states relaxes on a timescale $T_1$. Thus, the rate $T_2^{-1} = n T_1^{-1}$, where $n$ is the number of neighboring couplings that the spin can ``resolve'' on a timescale $T_2$; these are the couplings such that $\exp(n/\xi) \sim T_2$. Solving these equations to leading order at large $T_1$, we find that $T_2 = T_1 / \log T_1$ in one dimension. 

Analogous results hold for higher dimensions, where $T_2 \simeq T_1 / \log^d T_1$, and for rapidly decaying power law interactions (decaying with power law $\alpha$), for which $T_2 \simeq (T_1)^{1/(1 + \alpha)}$. 

\subsubsection{Lineshape}

Above, we estimated the linewidth of the local decoherence process, set by $1/T_2$. We now turn to its lineshape, which in general is non-Lorentzian. This non-Lorentzian behavior comes from the fact that on timescales short compared with a given coupling $J_{ij}$, that coupling does not affect the slope of the exponential decay of the correlator $C_\phi(t)$. This observation is worked out explicitly, e.g., in Refs.~\cite{shantsev, gopalakrishnan2014mean}. Thus, the phase correlation function that spin echo measures goes as 

\beq\label{phasecorr}
C_\phi(t) \sim \exp[- t \Gamma_1 n(t)],
\eeq
where $n(t)$ is self-consistently determined as above. Thus, there is a temporal regime in which $C_2(t) \sim \exp(- t \log^d t)$ (for short-range interactions) and as $C_2(t) \sim \exp(-t^{1 + \gamma})$ for power-law interactions. Correspondingly the lineshape is not precisely Lorentzian: it interpolates between a Lorentzian of width $1/T_2$ at low frequencies and one of width $1/T_1$ at high frequencies.

\subsubsection{Special cases}

\emph{Pure dephasing}.---The actual system-bath coupling that is relevant to cold-atom experiments is a little different: the system-bath coupling primarily acts by continuously ``measuring'' the system in the computational basis. This process decoheres on-site superpositions on a timescale of order unity; however, when the system is strongly localized, the rate of $T_1$ processes is suppressed by a factor of $(J/E_i)^2$, where $J$ is the hopping and $E_i$ is the nearest-neighbor energy denominator. We take the disorder to be distributed uniformly in the interval $(-1,1)$. Then the probability that the system will have decayed at time $t$ goes as $\int d\zeta \exp[-J^2 \gamma t/(J^2 + \zeta^2)]$. At time $t$, degrees of freedom will have decayed if $J^2 \gamma / (J^2 + \zeta^2) < 1/t$; at large $t$, this implies $\zeta \sim 1/\sqrt{t}$. Thus, upon averaging over this distribution one finds that the autocorrelation function of a p-bit in the presence of a bath decays as a stretched exponential, $C(t) \sim \exp(-\sqrt{t})$. This effect occurs also for an Anderson insulator, and leads to an even stronger separation between $T_1$ and $T_2$ than that noted previously~\cite{fischer_dynamics_2016, levi2016robustness}.

\emph{Particle loss}.---Another special case of experimental relevance is that of particle loss. This case is special because a path that couples to the system merely by absorbing particles does not act as a decoherence channel in the noninteracting limit. In particular, it does not affect the experimentally measured imbalance between even and odd sites, since even and odd sites have the same loss rate. Interactions change this picture, even at the Hartree level: the effective potential landscape each particle experiences is noisy, because of particle loss, and this noise leads to delocalization~\cite{fischer_dynamics_2016, lueschenopen, mbl_loss}. 

\subsubsection{Steady states in damped driven systems}

An MBL system coupled to a bath will thermalize. However, if it is also irradiated with light, it will reach a nonequilibrium steady state in which the absorption of energy from the radiation balances the rate of energy loss to the bath, and maintains an inhomogeneous steady state distribution~\cite{lenarcic}. The properties of this steady state distribution can be worked out straightforwardly in terms of LIOMs: each LIOM can be treated as a damped driven two-level system, which is describable, e.g., using Bloch equations. The resulting steady state has a spatially inhomogeneous temperature profile, corresponding to the fact that each LIOM is at a different temperature\footnote{Recall that any reduced density matrix for a two-level system can be assigned a temperature.}. LIOMs that are resonant with the radiation frequency absorb strongly and heat up to infinite temperature, while those that are far off resonance couple weakly to the radiation and remain close to the bath temperature. These spatial temperature fluctuations act as a diagnostic of localization, since a thermal system would instead absorb radiation relatively uniformly. In interacting systems, one expects that the local temperature will evolve stochastically, since the local transition frequency of the LIOM is dependent on the states of neighboring LIOMs. Thus LIOMs can wander in and out of resonance with the drive. The consequences of this effect for steady state dynamics have not yet been fully explored.

\subsection{MBL coupled to a slow bath\label{sec:multicomponent:slowbath}}

In the previous section we assumed that the bath was Markovian. There were two assumptions here: first, that the bath is sufficiently large compared with the system that the system exerts no back-action on the bath; and second, that the bath has a negligible memory time. We now relax the second assumption while maintaining the first. Since we are considering infinite-temperature baths, we can equivalently think of this situation as one in which the system is coupled to classical noise with a generic temporal correlation function~\cite{crow_2014}. We take the system-noise coupling strength to be spatially uniform at first; we will subsequently comment on random coupling. 

We refer to Ref.~\cite{nandkishore_general} for a more detailed exposition of some of the points presented here.

\subsubsection{Weak coupling to a narrow-bandwidth bath}

We first consider the case in which the system-bath coupling $g$ is small compared with the bandwidth of the noise $W$, which in turn is much smaller than the characteristic energy scale of the system (which we take to be unity). We also assume that the power spectrum of the noise falls off faster than a power law of $\omega$ (although a simple generalization exists for the case of a power law faster than $1/\omega^2$). To leading order in $g$ there are two channels by which the system can typically find an energy denominator of order $W$: either a single particle can hop a distance $\sim 1/W$, or one can rearrange $\sim (1/s) \log W$ particles. In an interacting system, the latter process is parametrically faster, with a matrix element $\sim \exp(-\log W / (s\zeta))$ which in turn implies a rate $\Gamma \sim g^2 W^{2/(s\zeta) - 1}$~\cite{gopalakrishnan2014mean}. 

Now let us consider the autocorrelation function. At time $t$, some fraction of spins with nearby small denominators will have relaxed. Relaxation via a narrow-bandwidth bath via an $n$-spin process would scale as $g^2 \exp(-2n/\zeta)/W = 1/t$, so $n \sim (\zeta/2) \log t$. At $n$th order, the typical denominator for this process is $e^{-sn} \sim t^{-s\zeta}$. Combining these expressions, we find that the autocorrelation function $C(t) \sim t^{-s \zeta}$ on timescales short compared with the characteristic decay rate. 
These characteristic rates are valid even when $g > W$, since the low-order processes are off-shell and for the on-shell processes the rate is sufficiently strongly suppressed that the Golden Rule applies. However, details of the short-time behavior of the autocorrelation function will be modified when $g > W$. 

\subsubsection{Strong coupling to narrow-bandwidth classical noise}

The Golden Rule analysis sketched above remains valid for classical noise even when the system-noise coupling is much larger than the bandwidth of the noise, provided that $W \ll g \ll 1$. In this regime the criterion for the validity of the Golden Rule is whether the actual rate $g^2 W^{2/(s\zeta) - 1} < W$. When this condition is satisfied, the Golden Rule still continues to apply. On the other hand, when $g$ is large enough, the Golden Rule ceases to apply. Instead, the system exchanges energy with the bath through Landau-Zener transitions, with a rate limited by the correlation time of the bath, which (in this simple model) goes as $1/W$. 

\subsubsection{Large-bandwidth slow baths and spectral diffusion}

We have considered, so far, a bath with a single characteristic timescale, so the bandwidth $W$ and the correlation time of the bath $\tau$ are interchangeable concepts. We now briefly consider noise that has a large bandwidth but a slow relaxation time; such noise can be generated by physical systems in which relaxation is slow compared with the characteristic dynamics, e.g., systems that are nearly localized. In the context of MBL, this type of noise was first discussed in the context of an attempt to construct a mean-field theory of the MBL transition~\cite{gopalakrishnan2014mean}. However, a different natural setting in which it occurs is that of a system coupled to a spin bath undergoing spectral diffusion. 

This case combines features from some of the previous cases. When the system is sufficiently weakly coupled to the bath, the system cannot resolve the fact that the bath is slowly fluctuating, and simply sees it as a large-bandwidth bath. For this answer to be internally consistent, we require that the Golden Rule decay rate $\gamma$ must satisfy $\gamma \tau \ll 1$. In the opposite limit, the system sees a bath of reduced bandwidth that depends on how much spectral diffusion has taken place on the timescale it takes the system to relax; thus the relaxation rate and the effective bandwidth are related and must be determined self-consistently together. On a timescale $1/\gamma$, the effective bandwidth is $\sim 1/(\gamma^{1/2} \tau^{3/2})$. Computing $\gamma$ for noise of this effective bandwidth, using the appropriate previous case, and self-consistently determining the bandwidth, allows one to solve for the decay rate in this regime.

\subsubsection{Pure dephasing and transient subdiffusion}

We can make these general considerations more concrete by considering a specific solvable case. In this limit, the system is a strongly localized Anderson insulator (with a localization length $\xi \ll 1$) and the noise couples to the on-site potential:

\beq
H = \sum_i (\epsilon_i + W_i(t)) c^\dagger_i c_i + J (c^\dagger_i c_{i+1} + h.c.).
\eeq
where the noise is Gaussian and spatially uncorrelated, has strength $W$ at all sites, and has a general two-time correlation function $\langle W_i(t) W_i(0) \rangle = C(t)$. We work perturbatively in $J$, which is taken to be the smallest scale in the problem~\cite{gik}. To leading order, the dynamics consists of incoherent classical hopping~\cite{alp}, with an amplitude on a given bond set by 

\beq
\Gamma_{i, i+1} \equiv \Gamma(\omega = \epsilon_i - \epsilon_{i+1}) = 2J^2 \int_{0}^{\infty} dt \cos[(\epsilon_i - \epsilon_{i+1}) t] |C^\phi(t)|^2,
\eeq
where 

\beq
C^\phi(t) \equiv \left \langle \exp\left( - \int_0^t dt' (t - t') C(t) \right) \right\rangle
\eeq
is the correlation function of the on-site phase\footnote{Note that this is subtly different from the phase-correlation function in Eq.~\eqref{phasecorr}---the two correspond respectively to Ramsey interferometry and spin echo---but we will be cavalier and use the same terminology for both.}. One can now regard the system as a chain of resistors, each with resistance set by $1/\Gamma_{i, i+1}$. Transport in this resistor network depends strongly on the \emph{high-frequency} asymptotics of the noise and the statistics of large deviations of the disorder. If we take the noise to be generated by a local quantum model, then $\Gamma(\omega)$ falls off at least exponentially in $\omega$ at large frequencies. The nature of transport then depends on the tail of the disorder distribution. If this tail falls off faster than $\Gamma(\omega)$, then the probability of finding an extremely slow bond is negligible and there is a well-defined diffusion constant. On the other hand, if the disorder distribution is itself fat-tailed (e.g., $P(|\epsilon_i - \epsilon_{i+1}|) \sim 1/|\epsilon_i - \epsilon_{i+1}|^\gamma$) then this model has a vanishing diffusion constant and exhibits subdiffusive transport. This subdiffusive transport has also been analyzed numerically in Refs.~\cite{bonca1, bonca2, bonca3} and a generalized Einstein relation has been constructed for the subdiffusive regime~\cite{bonca3}. 

It is unclear under what conditions subdiffusion persists in the long-time limit. When a link is effectively ``blocked'' (because the detuning across it is too high), one must consider processes that tunnel virtually through the link, to find a farther site that is less detuned. This is analogous to variable-range hopping~\cite{ahl} but with bandwidth replacing temperature. Even in noninteracting models, this variable-range hopping process can lead to a crossover to diffusion at late times, as it allows particles to bypass the worst bottlenecks. In interacting models, the number of possible relaxation channels grows exponentially, so there is generically always a crossover to diffusion at sufficiently late times.

\subsection{Quantum baths and back-action\label{sec:multicomponent:backbath}}

We now turn to baths that cannot be treated as classical noise, because they are strongly coupled to the system and this changes their properties. The strong system-bath coupling invalidates the simplest Golden-Rule treatments, in terms of the naive microscopic variables; however, one might still be able to use the Golden Rule with appropriately renormalized couplings. We first consider the case of large, narrow-bandwidth baths, then turn to ``baths'' that are comparable in size to the system, and finally to baths that are much smaller than the system. 

\subsubsection{Large slow baths: the ``frozen core''}

We first consider the case of a large slowly fluctuating bath. For concreteness, we take the system to live on the $x$ axis, and the bath to live in the whole XY plane. Thus the Hamiltonian is of the general form
\beq
H = H_S + W_b H_b + g O_S O_b,
\eeq
where $O_S, O_b$ are generic norm-1 local operators, and $H_S, H_b$ are generic local Hamiltonians. $H_S$ is taken to be in the MBL phase, whereas $H_b$ is either translation-invariant or weakly disordered, and would on its own be thermalizing. We are concerned with the case $W_b \ll g \ll 1$. This situation is generic in electron spin resonance~\cite{frozencore, frozencore2}: the hyperfine coupling between the electron and nuclear spins is much stronger than the dipolar coupling between nuclear spins, while the electron spins themselves interact strongly. Suppose the electron spin is frozen on some timescale. It then exerts a field on nearby nuclear spins that far exceeds the bandwidth of the nuclear Hamiltonian; thus it ``pulls them out'' of the nuclear band, so that they are now far detuned from all the other nuclear spins and are therefore no longer able to resonantly exchange energy with any other spins. These spins cease to function as a ``bath,'' and are best regarded as instead being incorporated into the system. This effect attenuates with distance---as a power law in the electron-spin case, and exponentially in the model outlined above---so nuclear spins far from the electron (or bath degrees of freedom far from the system) are weakly perturbed and continue to act as a bath. 

Thus the boundary between the ``system'' and the ``bath'' in this instance is deep inside the region that we initially took to be the bath. Any spin that is coupled to the system more strongly than $W_b$ is pulled out of the bath and frozen; the first layer of spins remaining in the bath are those that were coupled to the system with couplings that are of strength $\sim W_b$. In the present setup, this depleted bath is still infinite in extent, and can now be treated as a standard narrow-bandwidth bath, for which the Golden Rule is safe: in order for the system to undergo a transition, it must undergo a large-scale rearrangement of $s \log W_b$ spins. The decay rate in this case therefore scales as 

\beq
\Gamma \sim W_b^{1 + 2/(s\zeta)} \ll W_b, 
\eeq
so the Golden Rule is internally consistent. Autocorrelation functions also behave as in the narrow-bandwidth case (Sec.\ref{sec:multicomponent:slowbath}): decay sets in on a timescale $1/W_b$, then decays as a power law in time until the timescale $1/\Gamma$.

A more formal way to treat the strongly coupled bath degrees of freedom would be to use a canonical transformation that ``dresses'' system degrees of freedom with a distortion of the bath; this approach is carried out, e.g., in Ref.~\cite{banerjee_altman}, for the problem of noninteracting electrons coupled to a marginally localized phonon bath.

\subsubsection{The MBL proximity effect}

We can modify the previous discussion as follows: instead of taking the bath to be much larger than the system, we can take them to be the same size. This leads to a model of a two-leg ladder with a localized large-bandwidth leg coupled to a small-bandwidth but (in the absence of inter-leg coupling) thermalizing leg~\cite{mblprox1}. A concrete Hamiltonian of this type that has been studied is as follows: 

\beq
H = H_1 + W_b H_2 + g n_1 n_2,
\eeq
where $H_1$ is a generic MBL Hamiltonian, $H_2$ is a generic local thermal Hamiltonian, and the two are coupled via a density-density interaction. We further assume that the total particle number on either leg is conserved.  
When the coupling along the rungs of the ladder is weak, a straightforward perturbative calculation shows that the system globally thermalizes~\cite{mblprox2}. However, when the coupling along the rungs is strong, the degrees of freedom on the small-bandwidth leg experience strong random fields ($\gg W_b$); these fields localize them, suggesting that the composite system has an MBL phase in this regime~\cite{mblprox1, mblprox2}. 

However, we note that there can be no LIOM description in this limit: in a typical state there will be arbitrarily long segments in which the MBL leg has no particles. The ETH leg is therefore \emph{thermal} in these segments. The situation is thus analogous to a many-body mobility edge. Following the arguments in that case, one expects that this void-bath combination will be mobile, and will therefore thermalize the entire system on very long timescales.

\emph{Experimental implementation}.---The setup described above, in which a disordered system is coupled to a clean system, was explored experimentally in Ref.~\cite{rubio2018} in the quantum gas microscope setup. This is accomplished by trapping two Zeeman sub-levels of rubidium atoms in an optical lattice, and turning on a disorder potential that only couples to one species. One then varies the ratio of atoms in each species and observes the consequences for the relaxation of both species. The total number of atoms is $\sim 100$. When all the atoms are in the ``dirty'' species, there is no bath, and one can map out the phase diagram of the isolated system. On the other hand, when the population ratio of clean to dirty atoms is large, the bath is ``big'' and the bath delocalizes the system (even when, for those parameters, the dirty component on its own would be localized). As the ratio of atoms in the clean species is decreased, the bath becomes increasingly less effective at thermalizing the dirty species, and eventually ceases to delocalize the dirty atoms at all. On the other hand, the dynamics of the clean species does not seem to be sensitive to the number of clean atoms. In the large-system limit, we expect that both of these results cannot be true: either the clean system will eventually be localized by the proximity effect, or the dirty system will eventually thermalize. There is no reason to expect a mixed scenario with coexisting localized and delocalized degrees of freedom to be stable, although it would be very interesting if such a scenario were found to be possible. 

\subsubsection{Zero-dimensional bulk baths}

We now turn to the case where the bath is small but coupled globally to the system~\cite{huse2015localized, plhc}. A simple version of this model takes the bath to be coupled separately to $N$ separate LIOMs~\cite{plhc}:

\beq
\hat{H} = \sum_{i = 1}^N \left[ \Delta_i \hat{\tau}^z_i + \lambda \left(\hat{A}_i \hat{\tau}^z_i + \hat{B}_i \hat{\tau}^x_i \right) \right],
\eeq
where we have put hats over operators to distinguish them from coupling constants. The detunings $\Delta_i$ are taken from a distribution of unit width, and the $2N$ matrices $\hat{A}_i, \hat{B}_i$ are random matrices from the Gaussian orthogonal ensemble that each act on a ``bath'' Hilbert space of dimension $m$. The random matrices are normalized so their bandwidth (or equivalently their operator norm) is set to unity, thus the typical matrix element in the bath is of size $1/\sqrt{m}$. Ref.~\cite{plhc} analyzes the case $m = 2$, in which case the ``bath'' is a single spin $\sigma$. We can write down a simple Hamiltonian for the $N + 1$ spins as follows (we return to suppressing hats):

\beq
H = \sum_{i = 1}^N \left[ \Delta_i \tau_i^z + \lambda (\epsilon_i \tau^z_i \sigma^z + \delta_i \tau^x_i \sigma^x + \zeta_i \tau^x_i) \right].
\eeq
where $\epsilon_i, \delta_i, \zeta_i$ are all random numbers with unit variance. We now consider perturbing the system in $\lambda$. We anticipate that the system thermalizes when $\lambda > 1/N$, as Ref.~\cite{plhc} showed, and consider $\lambda$ of that order. We now consider second-order processes in $\lambda$ that involve flipping two LIOMs and possibly flipping the central spin as well. The matrix element for these processes is $\sim \lambda^2$ and there are $N^2$ possible processes, so there are typically $O(1)$ resonant flips at this order. A crucial point now is that after this second-order process has happened, the effective field on the central spin $\sum_i \lambda \epsilon_i \tau^z_i$ will have changed by $\sim 1/N \gg 1/N^2$. Therefore, the resonance condition on top of the final configuration is different from that on top of the initial configuration: most of the initial resonances are gone (though of course undoing the resonant hop is still resonant), and new resonances appear. One can thus traverse configuration space by repeating this process indefinitely, so the system ultimately thermalizes. 

One might question why we chose to go to second order. There are $2N$ possible resonant transitions at leading order, each involving one LIOM flip with or/without the central spin flipping. When $\lambda \sim 1/N$ a typical many-body configuration has $O(1)$ resonant spin flips even at first order. However, the shift in the effective field due to these is not much larger than the resonance window, so a single spin flip does not completely change the structure of subsequent resonances, and does not delocalize the system. The key role played by these energy shifts was first emphasized, in a somewhat different context, in Ref.~\cite{gornyi2017spectral}.

A similar problem was also studied for single-particle hopping coupled to a random matrix~\cite{huse2015localized}; however, the results there are very specific to single-particle problems and thus lie outside the scope of this review.

\subsection{Coupled identical MBL chains}

An interesting variant of the physics discussed in the previous sections is that in which the various subsystems are identical: i.e., they all experience the \emph{same} disorder realization. This situation was first explored experimentally~\cite{Bordia16}. It is a natural setup to realize, as the purely 1D experiment consists of tubes that are separated from one another by a strong optical lattice potential. Lowering this optical lattice naturally gives rise to a transverse coupling between tubes; however, all the tubes involved experience identical quasiperiodic potentials.

\subsubsection{Two coupled chains: Mobility emulsions}

Before turning to the (still poorly understood) many-chain problem, we first briefly discuss the case of two coupled identical chains. Our discussion follows that of Ref.~\cite{mobemulsion}. Specifically, we consider a Hamiltonian

\beq
H = \sum\nolimits_{\alpha = 1,2} \sum\nolimits_i \left[ (\sigma^+_{\alpha, i} \sigma^-_{\alpha, i+1} + \mathrm{h.c.}) + \Delta \sigma^z_{\alpha, i} \sigma^z_{\alpha, i + 1} + h_i \sigma^z_{\alpha, i} \right] + J_\perp \sum\nolimits_i (\sigma^+_{1, i} \sigma^-_{2,i} + \mathrm{h.c.}).
\eeq
The on-site fields are drawn from a distribution of width $W$. When $W$ is small the system thermalizes, so we focus instead on $W \gg 1$. For $J_\perp = 0$ the system consists of two decoupled spin chains, each deep in the MBL phase. An important observation is that almost all eigenstates break the reflection symmetry under interchange of the two legs of the ladder. 

This symmetry-breaking is argued to persist even when $J_\perp$ is small and nonzero: moving a single spin between the legs of the ladder is generically going to change the interaction energy, and is therefore off-resonant. Thus, a ``mirror glass'' that spontaneously breaks the mirror symmetry persists for small $J_\perp$. When the density is sufficiently low or $J_\perp$ is sufficiently large, the symmetry is restored, but the system still potentially remains localized. 

A helpful way to conceptualize this problem is to work in the basis of localized single-particle eigenstates of the decoupled chains. Each such eigenstate is doubly degenerate; the pair (which we can identify at strong disorder with a rung) can be empty, singly occupied, or doubly occupied. Empty and doubly occupied rungs are dynamically inert. However, singly occupied rungs have two allowed states per rung. For these states, one can define a pseudospin corresponding to the leg index. In this representation, $J_\perp$ is a constant transverse field, while $\Delta$ generates an Ising spin-spin interaction between neighboring singly occupied rungs. In the limit where every rung is singly occupied, {the random field term cancels between the two rungs;} the model is thus effectively clean, and therefore \emph{thermalizes}. We expect thermalization in this sector to asymptotically ``infect'' the entire system. In the opposite limit, where most rungs are inert, the singly occupied rungs are Poisson distributed so their couplings have strong positional randomness, and localization remains stable. 

As with the MBL proximity effect, the present system is also susceptible to being destabilized by rare thermal configurations. In this case, the thermal configurations are rare strings of consecutive singly occupied rungs, which form a locally thermal region. These locally thermal regions are in principle mobile, and can destroy MBL throughout the system on the longest timescales.

\subsubsection{Many coupled chains}

Although the two-leg ladder is relatively straightforward to analyze, the most intriguing experimental findings are for anisotropic two-dimensional systems, consisting of many identical chains with nearest-neighbor hopping~\cite{Bordia16}. Although some works have explored this setup numerically~\cite{mueller_coupled, rossignolo_coupled}, the dynamics of these systems is not yet well understood. The most striking experimental observation is that the timescale on which a density-wave pattern decays scales either logarithmically or as a very slow power law of $J_\perp$. Assuming a power law, the observed exponents are never more than $1/3$. No clear theoretical mechanism has been proposed for this anomalous dependence; moreover, it is unclear what ingredients are necessary---for instance, these anomalous exponents have not been experimentally studied in the random case. 

We briefly review a few limiting cases. The first is that of very weak interactions (which is not directly related to the experiment~\cite{Bordia16}). Here, one can begin in the noninteracting limit, where the potential is separable; thus, the single-particle eigenstates are tightly localized along the $x$ direction and plane-wave like in the $y$ direction. As a minimal model of this system, consider two clean wires each with bandwidth $J_\perp \ll 1$, and a detuning $W \gg 1$. Interactions between (as well as within) wires rapidly thermalize each wire; however, thermalization between wires is much slower, as hopping a particle changes the energy by $\sim W$, and (by analogy with doublons in a Mott insulator) only happens in this limit at a rate $\sim \exp(-W/J_\perp)$. This extremely sensitive dependence on $J_\perp$ is the opposite of what is experimentally observed. A much less strongly suppressed channel involves rearrangements of $\sim \log (W/J_\perp)$ particles along the $x$ direction, using the narrow-bandwidth bath of transverse excitations in each tube. This mechanism gives a rate $\Gamma \sim U^2 (J_\perp/W)^{2/(s\zeta) - 1}$. On the one hand, this mechanism gives a power-law dependence on $J_\perp$ with a continuous exponent. On the other hand, in the regime where MBL is stable in an individual tube, $s\zeta < 1$, so one cannot explain the very small observed exponents through this mechanism.

In the actual experimental regime, $U \sim W \geq 1 \gg J_\perp$. In this regime, interactions generically frustrate transverse motion, because a particle can move from one leg to the next only if its surroundings on both legs are identical. The corresponding two-leg ladder is in the ``mirror glass'' phase, where it spontaneously breaks mirror symmetry. A naive extension of this result to the many-chain problem would suggest that this case should spontaneously break translational symmetry along the $y$ direction, so that relaxation should be insensitive to small but finite $J_\perp$. This scenario has the opposite problem from the previous one: if the system remains in a glassy state, there is no obvious reason why a small $J_\perp$ should affect relaxation rates. To summarize, at present neither of these limits appears to lead to predictions that match experiment particularly closely, and the actual dynamics of these systems should be regarded as an open question. 

\subsection{Localized charges coupled to thermal spins}\label{sec:multicomponent:spins}

The next variant of the two-species problem we will consider involves localized degrees of freedom coupled to degrees of freedom that are protected against localization. The simplest and most broadly relevant instance of this paradigm is the Hubbard model subject to random chemical potentials. This model retains a global $SU(2)$ symmetry under spin rotations, although the charge degrees of freedom are allowed to localize. Further, in one dimension, it undergoes spin-charge separation in the clean limit. In what follows, we will discuss the dynamics of this model in both the weak-coupling and strong-coupling limits. 

\subsubsection{Hubbard model at small $U$}

We begin with the weak coupling limit, taking $U \ll T$, and also working in one dimension for concreteness. At $U = 0$ one fills orbitals at random; some fraction of orbitals ($\sim T$ at low temperatures and $o(1)$ at high temperatures) will be singly occupied. These singly occupied orbitals are spin-degenerate, so at $U = 0$ there is an exponential degeneracy corresponding to the entire manifold of spin states corresponding to this randomly chosen charge state. When $U \neq 0$, exchange interactions lift this degeneracy. Ignoring charge fluctuations to leading order, one can derive an effective Heisenberg model for the spin sector. At high temperatures and localization lengths $\xi \geq 1$, the Heisenberg model is not strongly random and thermalizes on a timescale comparable to $U$. 

Assuming the Heisenberg model thermalizes efficiently, we can treat the spin excitations as a narrow-bandwidth bath for charge rearrangements. The charge and spin degrees of freedom are coupled because a charge hopping event locally modulates the couplings of the effective Heisenberg model: 

\beq
H = \sum_i J(n_i, n_{i+1}) \mathbf{\sigma}_i {\mathbf \cdot \sigma}_{i + 1}.
\eeq
Thus, one can naturally write the perturbation coupling the charge and spin sectors as $\sim n_i \mathbf{\sigma}_i {\mathbf \cdot \sigma}_{i + 1}$. When $\xi$ is large, $n_i$ is well approximated by its mean value, so \emph{fluctuations} in $n_i$ (which couple to spin) are relatively small and perturbation theory in the spin-charge coupling is legitimate. However, when $\xi$ is small, $n_i$ is essentially a binary quantum variable, and the spin-charge coupling is of the same order of magnitude as the bare spin-spin coupling itself. 

As $T$ is decreased, the bonds in this effective model become increasingly random: in the low-temperature limit, a fraction $\sim T$ of the orbitals are singly occupied; these can be treated as Poisson distributed, so the spacings between singly occupied orbitals follow an exponential distribution that broadens at low $T$, specifically $P(x) \sim \exp(-c T x)$. We also know that $J \sim \exp(-x/\xi)$. Putting these two observations together, one sees that $P(J) \sim J^{- 1 + c' \xi T}$, where $c'$ is some nonuniversal prefactor of order unity. Thus, at sufficiently low $T$ the effective Heisenberg model becomes increasingly nonergodic on short length-scales, and its thermalization time diverges in a super-activated fashion (see Sec.~\ref{sec:unstableMBL:nonabelian:Heisenberg}). How the charge sector thermalizes in that regime is not quantitatively understood at present.

\subsubsection{Hubbard model at large $U$: spin-incoherent Luttinger liquid\label{sec:multicomponent:spins:SILL}}

\begin{figure}[tb]
\begin{center}
\includegraphics[width = .4\textwidth]{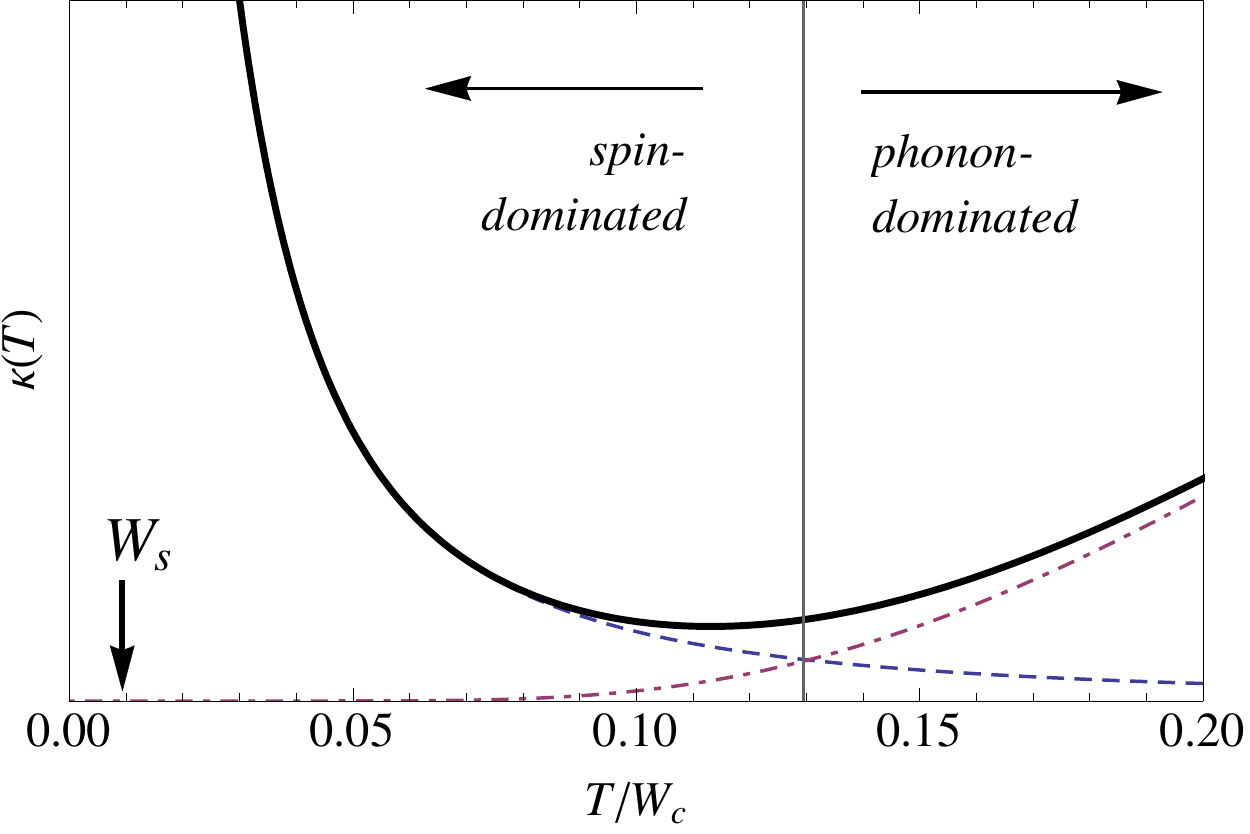}
\includegraphics[width = .4\textwidth]{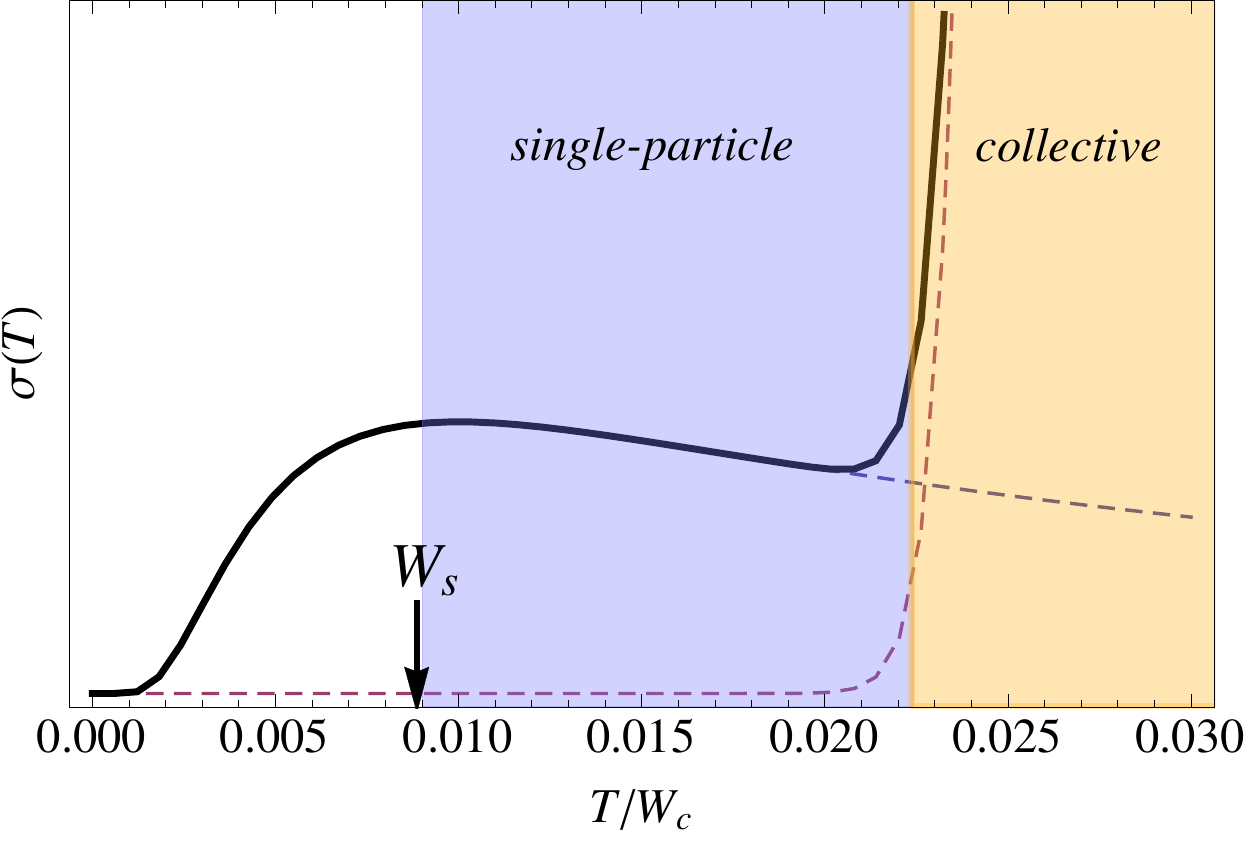}
\caption{Evolution with temperature of the thermal (left) and charge (right) conductivity of the disordered SILL, reproduced from S. A. Parameswaran and S. Gopalakrishnan, \emph{Phys. Rev. B} {\bf 95}, 024201 (2017). Copyright (2017) the American Physical Society; reuse permitted according to APS copyright policies.}
\label{sillfigs}
\end{center}
\end{figure}

Another regime that is tractable by somewhat similar methods is the spin-incoherent regime of the Hubbard model, $t^2/U \ll T \ll t$. This is an example of a spin-incoherent Luttinger liquid (SILL)~\cite{fieteRMP}, in which the charge degrees of freedom are essentially in their ground state but the spin excitations are at infinite temperature. We will address the case of the disordered SILL using general phenomenological arguments, following Ref.~\cite{parameswaran_spin-catalyzed_2017}. In the Hubbard model, in the $U \rightarrow \infty$ limit the system can be regarded as free fermions at twice the density. Adding disorder that couples to charge gives rise to Anderson localization of these free fermions, and (since $T > 0$) the state of the charge sector in the $U \rightarrow \infty$ limit can be regarded heuristically as a Slater determinant of these free fermions. The rest of the argument goes through as before: for finite $U$, superexchange leads to dynamics in the spin sector, with a bandwidth $t^2/U$. Charge fluctuations are suppressed because of the effective fermionic statistics; thus the Heisenberg model will generically be thermalizing in this regime, and the rest of the analysis goes through as it did in the weak-coupling case. 

One can also address the SILL more generally, away from this free-fermion limit. A useful model~\cite{fieteRMP} of the SILL is as a fluctuating charge density wave (CDW) in one dimension. Any disorder pins the CDW. There are two types of excitations above the pinned CDW ground state: Gaussian fluctuations of each particle around its minimum, and instantons, or events where a particle can tunnel between two approximately equal-energy minima. Gaussian fluctuations involve essentially no charge transfer; thus, charge conductivity is dominated by instantons. Meanwhile, the Gaussian fluctuations do carry energy, and can transport energy by hopping between orbitals. Either of these transport channels ultimately involves inelastic processes in which energy is exchanged with the spin sector (which, as above, forms a bath whose bandwidth is set by the superexchange scale, which we denote $W_s$). 

An interesting feature of this regime is the distinction between spin, charge, and energy conductivity. Spin conductivity scales as $W_s/T$, since spin excitations are at very high temperatures. Meanwhile, charge conductivity occurs through the hopping mechanisms we have discussed here. The most interesting case, however, is that of thermal conductivity. At low temperatures, this takes place mostly through the spin sector; the spin diffusion constant goes as $W_s$, and each spin only transports $W_s$ of energy, so the thermal conductivity goes as $\kappa_s \sim W_s^3/T^2$. At higher temperatures, phonons with energy $\geq T$ become an important channel for energy transport. The significance of high-energy phonons in the present problem comes from the narrow-bandwidth nature of the spin bath. In general situations, phonons can locally exchange energy with the bath, and should not be regarded as conserved. However, in the present setup, phonons can efficiently exchange only a small fraction of their energy with the bath. Thus, the dynamics has an emergent, approximate conservation law for phonons with energies that lie in a given energy window: in other words phonons are ``foliated'' according to their energies. A straightforward calculation~\cite{parameswaran_spin-catalyzed_2017} shows that the phonon contribution to the conductivity scales as $\kappa_{ph} \sim T^{-3/2} \exp[-1/(W_s T^3)^{1/4}]$. The total energy conductivity is the sum of the spin and phonon parts. When $W_s \ll 1$, and the disorder is not too large, the thermal conductivity has a finite-temperature minimum, and overall has the form shown in Fig.~\ref{sillfigs}. 

\subsubsection{Numerical studies}

The overall picture of the Hubbard model that we have presented here is supported by detailed numerical studies~\cite{prelovsek2016, mierzejewski2018, mierzejewski2019}, but a few words are in order about how the conclusions of these works relate to ours. At the system sizes available to exact diagonalization, the apparent physics is that charge degrees of freedom are \emph{localized} whereas spin degrees of freedom are delocalized. We expect that the apparent localization of charge is a finite-size effect: on general grounds, any state in which some degrees of freedom are localized and others are actually thermal will be unstable to weak perturbations that mix the sectors. However, these perturbations might only set in for large systems. From this perspective, the key observations of these works are, first, that the spin indeed fails to localize in $SU(2)$ invariant models, and second, that spin transport (even treating the charges as fixed) can be slow and indeed even subdiffusive. The latter result agrees with our discussion above, in the weak coupling case, indicating a broad distribution of hopping matrix elements for spin.

\section{Driven systems, slow heating, and prethermalization \label{sec:prethermal}}

{Thus far, we have focused on instabilities of MBL in Hamiltonian systems, although most of the arguments presented above would extend with few changes to periodically driven Floquet systems, provided that the driving takes place at frequencies comparable to the characteristic energy scales of the system. The main effect of driving in this limit is to eliminate the conservation law for energy, and thus to remove all hydrodynamic modes from the problem. Thus, Floquet systems generically heat up to infinite temperature by absorbing energy from the drive.}

{However, the assumption that Floquet systems simply heat up to infinite temperature misses rich phenomena that occur when the drive frequency is much larger than the system's characteristic energy scale. These phenomena} manifest either as long prethermal dynamical regimes, possible even in clean systems, or as fully non-ergodic stable phases of MBL systems. In this section, we briefly review the standard Magnus expansion approach to periodic driving~\cite{bukov_universal_2015, vandersypen}, and then explain its breakdown in many-body systems. We then explain why a large class of lattice systems experience exponentially slow heating, before discussing both prethermal and localized Floquet systems. In both cases, the prethermal regime/localized Floquet phase can exhibit a variety of interesting phenomena --- including `time crystalline' phases~\cite{khemani2016, ebn, moessner2017equilibration, yao2018time} and `anomalous' topological states impossible in the undriven setting~\cite{rudner2013, rudner2016, adrianpo, rudner2017}. A detailed discussion of these lie outside the scope of this review, but our considerations apply, {\it mutatitis mutandis}, to these phases as well.

\subsection{Magnus expansion and its breakdown in many-body Floquet systems}
Let us consider a periodically driven system, described by a time-dependent  Hamiltonian  $H(t+T) =  H(t)$. A key idea in Floquet theory is that the periodic dynamics are characterized by studying the prooperties single-period unitary time evolution operator or Floquet operator,
\begin{equation}
F = U(T) \equiv  \mathcal{T} e^{-i \int_0^T H(t) dt},
\end{equation}
whose eigenstates, known as Floquet eigenstates, have a particularly simple evolution in time: $F\ket{\psi_\alpha} = e^{i \epsilon_\alpha T} \ket{\psi_\alpha}$, with the Floquet quasi-energy defined on the circle, $\epsilon_\alpha \equiv \epsilon_\alpha + \frac{2\pi}{T}$. It is possible to formally rewrite the Floquet operator in terms of an effective  time-independent `Floquet Hamiltonian', $H_F$, via $F = e^{-iH_F T}$. However, the definition of $H_F$ is ambiguous in general, owing to the branch cut in defining the logarithm of $F$, and the Floquet  Hamiltonian need not even be local. A conventional approach attempts to build a simple Floquet operator using the Magnus expansion,
$H_F =  H_F^{(0)} + H_F^{(1)}+ H_F^{(2)} +\ldots$, where $H_F^{(0)} = \frac{1}{T} \int_0^T H(t) dt$,  $H_F^{(1)} = \frac{1}{2T} \int_0^T dt_1 \int_0^{t_1} dt_2\,  [H_1(t_1), H_2(t_2)],$ etc.  This expansion is controlled in the high-frequency limit $T =  \frac{1}{\nu} \ll 1$, but even in this limit converges only for bounded Hamiltonians. 
For a generic many-body system in of linear size $L$ in $d$  dimensions,  $\|H(t)\|\sim L^d$  so that in the thermodynamic  limit the Magnus expansion fails in general~\footnote{This is sidestepped in non-interacting many-body systems --- which are highly non-generic --- by replacing the many-body unitary evolution operator with the unitary operator describing the  evolution  of \emph{single-particle} states, which has $O(1)$ norm.}. A distinct approach is  instead necessary to incorporate the presence of many-body resonances and their role in heating the system --- to which we now turn.

\subsection{Exponentially slow heating in many-body systems}
We now present an argument for the exponentially slow heating, that is the basis of the proof in Ref.~\cite{adhh}. Consider a generic many-body lattice system $H$, with a typical energy  scale $W$ for a single excitation,  globally driven at a high frequency $\omega \gg W$ by a term of the form $gV\cos \omega t$ where  $\omega = 2\pi/T$ and $V = \sum_j V_j$ is a sum of local terms. Let us consider the  energy absorption rate $dE/dt$; to leading order in $g$ this can be related to the dissipative portion of the linear response function  via $dE/dt  = 2g^2 \omega \sigma(\omega)$, after averaging over a single period. The latter can in turn be expressed in terms of the local operator $V_i$ through $\sigma(\omega) = \sum_{i,j} \sigma_{ij}(\omega)$, where 
\begin{equation}
\sigma_{ij}(\omega) = \frac{1}{2}\int_{-\infty}^\infty dt\, e^{i\omega t} \langle [V_i(t) , V_j(0)] \rangle_\beta,
\end{equation}
where $\langle \ldots \rangle_\beta$ indicates a thermal average (we assume we begin in a thermal state at temperature $T = \beta^{-1}$) and $V_i(t)$ indicates the Heisenberg dynamics of $H$. To see why the heating might be slow for a local system, consider replacing  the global drive $V$ by a single local term $V_i$; in this case, the relevant linear response function is
\begin{equation}
\sigma_{ii}(\omega)  =  \sum_{m,n} e^{-\beta{E_n}} |\bra{m} V_i \ket{n}|^2 \delta(\omega -(E_m - E_n)) 
\end{equation}
Absorbing a single quantum of energy $\omega$ from the drive creates $O(\omega/W)$ local excitations in the system, but since $V_i$ is a single local term and a lattice system has a finite local state space, there is no obvious matrix element capable of creating such a process. 
We can rewrite $\sigma_{ii}(\omega)$  in terms of a $n^{\text{th}}$-order commutator $[[[V,H],H,],\ldots]$, suppressed by $\omega^{2n}$. We can then use  the locality of $H$ to bound  the commutator in the matrix element by $\varepsilon^n n!$ where $\epsilon>0$ is a constant energy scale that is proportional to the norm of a typical local term $h_j$ in $H$ and by the range of the $h_j$ and $V_j$. The worst failure of the series  then occurs at order $n_*\sim \omega/\varepsilon e$, when the factorial dominates the suppression by powers of $\varepsilon/\omega$.  This allows the spectral function to be bounded  by $A(\omega)< e^{-\kappa\omega}$, with $\kappa = 2/\varepsilon e$. The locality of the Hamiltonian and the perturbation allows  cross-terms involving $V_i, V_j$ to be estimated  and similarly bounded. Putting these arguments together leads to a controlled version of the high-frequency expansion that can be made non-perturbative both in the strength of the driving and the interaction, and bounds the heating rate by a frequency-dependent exponential $C e^{-\kappa\omega}$ where $C>0$ is an $O(1)$ constant. This bound was expanded beyond linear response in Ref.~\cite{Abanin2017} at the cost of  slightly weakening the exponential term to $e^{-\kappa \omega/\log^3 \omega}$, although for $d=1$ the exponential bound continues to hold for \emph{strictly} local $H$. 

{The argument above indicates that} generic many-body systems with local Hamiltonians{, driven at high frequency $\omega$}, only heat exponentially {slowly in $\omega$}. This {observation} can also be reframed as a statement about thermalization: until an exponentially long time scale $\tau_* \sim e^{\kappa \omega}$, the system is `prethermal', i.e., far from the infinite-temperature Gibbs state.  In this setting it is possible to also consider the thermalization of  time-independent systems  with local Hamiltonians of the form $H =  H_0  + V+ \mu  N$, where $N$ is a conserved quantity of $H_0$ (i.e. $[H_0, N]=0$) and the scale $\mu$ is large compared to the local energy scales of $H_0, V$. In this setting,  since $V$ is a sum of local terms it only has small matrix elements between configurations with very different values of the conserved charge $\mathcal{N} = \langle  N\rangle $. One can then argue for exponentially slow relaxation to equilibrium of non-equilibrium states with $\mathcal{N}\gg 1$, with $1/\mathcal{N}$  the small parameter. {This mechanism therefore underlies the \emph{generic} phenomenon of slow relaxation in lattice Hamiltonians that have a large separation of scales between their constituent terms.}

\subsection{Prethermal states in driven systems}
Since we have demonstrated that generic driven many-body systems only heat after  an exponentially long time scale $\tau_*$.  it is natural to ask what governs the dynamics in the prethermal regime $t\leq \tau_*$. As argued by Ref.~\cite{adhh} upto this time scale, the prethermal plateau  is  governed by a quasi-conserved effective time-independent Hamiltonian
\begin{equation}
H_* = H_0 + \frac1\omega H_1 +  \frac1{\omega^2} H_2 + \ldots +   \frac1{\omega^n} H_n.
\end{equation}
The construction of $H_*$ proceeds by  unitarily transforming $H(t)$, systematically eliminating the time dependence order-by-order in $T$. On a technical level this is accomplished by using a time-periodic unitary $Q(t+T) = Q(t)$, with $Q(0) = I$  to rewrite the wavefunction via $\ket{\varphi(t)} = Q(t) \ket{\psi(t)}$, where $\ket{\varphi(t)}$ coincides with the original wavefunction $\ket{\psi(t)}$ at stroboscopic times $t_m =  mT$. The evolution  of $\ket{\varphi(t)}$ is governed by a Schrodinger equation with a modified Hamiltonian,
\begin{equation}
i\partial_t \ket{\varphi(t)}  = H_Q(t)  \ket{\varphi(t)}, \,\,\,\,\,\,\,\text{with}\,\,\,\,\,\,\, H_Q(t) = Q^\dagger H(t)Q - iQ^\dagger\partial_tQ.
\end{equation}
It is possible to show that $H_Q$ can be chosen so that all time dependence in eliminate upto an order $T^{n_\text{max}}$. 
 This procedure cannot continue to an arbitrary order, but it can be carried out up to an optimal order  $n_\text{max}^* \sim \omega$ that follows from a similar argument to that for the heating rate above. Truncating at this order yields $H_*$ and the heating timescale $\tau_*$ as above.

Let us consider the dynamics of the system in light of the existence of the time-independent prethermal Hamiltonian $H_*$. Let us assume that $H_*$  is ergodic, and suppose the system is initially prepared in a non-equilibrium state $\ket{\psi}$ and then driven at  high frequencies so that the time scale $\tau_*$ is appreciable. For times $t\lesssim\tau_*$ the system will reach a steady state controlled by ETH as applied to $H_*$. In other words, local  observables $\bra{\psi(t)} O \ket{\psi(t)}$ have thermal values controlled by the density matrix $\rho  \propto e^{-H_*/T_{\text{eff}}}$, with $T_{\text{eff}}$ is  the effective temperature set by the initial energy density of the state. If the latter is less than the infinite temperature predicted by ETH applied to the full driven system, the system does not initially appear absorb energy or to heat up at short for times $t\lesssim \tau_*$. Beyond this timescale the system begins to absorb energy and relax to an infinite-temperature steady state. This can be used to define a variety of interesting phenomena and quasi-phase structure  in the prethermal plateau. A detailed discussion of these is beyond the scope of this review; for examples of the rich possibilities, see e.g. Refs.~\cite{bukovetalprethermalBH,else_prethermal}.

\subsection{Localization protection against heating}
So far, we have discussed how many-body systems in many circumstances heat on an exponentially slow timescale, even if  the Hamiltonian that governs them in the undriven setting is ergodic. Such systems show rich prethermal behaviour, but on the longest time scales they do eventually heat up to a featureless infinite-temperature state. The latter fate is avoided by many-body localized systems under sufficiently fast periodic driving. {For MBL systems, ``sufficiently fast'' turns out to be a much milder condition than in the thermal case, with the critical frequency scaling to zero when the system is deeply localized, as we now discuss.}

Let us  consider driving an MBL Hamiltonian, viz $H(t) =  H_{\text{MBL}} +  g  V(t)$, with $V(t  + T) =  V(t)$ and $\frac{1}{T} \int_0^T dt V(t) = 0$. It is possible to show that the dynamics of the system are controlled by a Floquet Hamiltonian $H_F$ that is itself MBL, as long as the driving frequency is sufficiently high, and the disorder sufficiently strong~\cite{abanin2016theory, Ponte15, ponte2015periodically, Abanin2017}:
\begin{equation}
\frac{g}{\nu} \ll 1, \,\,\,\,\,\,\,\text{and}\,\,\,\,\,\,\, \frac{g^2}{\nu W } \ll 1,
\end{equation}
where $W$ is the strength of the disorder. To arrive at this conclusion we can write the Schrodinger equation for the unitary time evolution operator, $i \frac{d}{dt}U(t) = H(t) U(t)$, and decompose $U(t) =  P(t) e^{-i H_{\text{eff}}t}$ where  $P(t) = P(t+T)$ and satisfies $P^\dagger(t) \left( H(t)  - i \frac{d}{dt}\right) P(t) =H_{\text{eff}}$. We can solve for $P(t)$ iteratively and gradually eliminate time-dependent terms, thereby constructing the Floquet Hamiltonian $H_F$. This perturbation theory converges, with the convergence criteria as above. The formal proof is technical and the reader is referred to Refs.~\cite{adhh,Abanin2017} for details; we note here that the relevant arguments are  modifications of those used to prove the stability of MBL systems. Accordingly, many of the statements regarding MBL systems may be translated also to the Floquet  setting, with the caveat that various arguments will require modification to incorporate the ability of Floquet systems to absorb energy from the drive.

To see why even MBL systems are unstable  to low-frequency driving, it is useful to consider the driven system as  a many-body Landau-Zener problem for the instantaneous eigenstates of $H(t)$. The change in the Hamiltonian with time can lead to level crossings; while both adiabatic and diabatic crossings are innocuous, `interrmediate' crossings that lead to states getting entangled are dangerous and can cause delocalization. At low frequencies there are many diabatic crossings and delocalize the MBL phase leading to thermalization under the drive, as discussed above in Sec.~\ref{sec:intro:mbldynamics}.

A further interesting  direction blends the themes of the above lines of study: it is  possible to imagine a disordered system such that the prethermal Hamiltonian $H_*$ is MBL rather than ergodic, so that the system does not even equilibrate to a prethermal state for times $t\leq \tau_*$ but instead behaves like an MBL system on these time scales. {One concrete example of such a system would be, e.g., a system of many identical coupled disordered channels~\cite{Bordia16}, in which the interchannel hopping is {periodically} modulated in time, with zero mean. In the high-frequency limit, the interchannel hopping would average out and one would have many independent MBL channels, but at finite frequency this cancellation would cease to be operative at sufficiently late times, and the system would thermalize.}

\section{Common dynamical signatures \label{sec:dynamicalsig}}

This section ties together various strands from our previous discussions of the diverse types of nearly MBL dynamics. The basic phenomenon in each of these cases was a \emph{separation of scales} between the characteristic timescale for local interactions and the timescale(s) on which the system thermalizes. This phenomenon was due, in turn, to the fact that local transitions in the MBL phase are detuned away from resonance in most of the system. This detuning is pervasive in strongly disordered systems, as well as in strongly interacting systems at high temperature, where any one particle in effect experiences a random potential due to all the others. Thus the system appears MBL on short or intermediate distance scales. On the longest timescales, however, enough of the available moves are resonant, and these resonant large-scale (or long-distance) moves eventually thermalize the system. 

The phenomena we will discuss here appear in most instances of nearly MBL systems. There is one important exception, however, which is that of an MBL system coupled to a generic, large-bandwidth bath. Here, the crossovers are much more conventional: a typical LIOM has some lifetime for decaying into the bath; on timescales much shorter than this lifetime, the behavior is MBL-like, whereas on timescales that are much longer, it is thermal.

\subsection{Anomalous relaxation and dynamical heterogeneity}

An immediate implication of this picture is that nearly MBL systems are dynamically heterogeneous, like systems near a classical structural glass transition~\cite{biroli2013, berthier2011, berthier2011a}. Three features of the dynamics of glasses are particularly salient to MBL: first, they have a broad distribution of relaxation timescales; second, there is spatial structure (and a characteristic length-scale) to the fast and slow regions; and third, transport is ``facilitated'' by certain configurations of disorder or particles. In classical glasses these phenomena can readily be seen, at least qualitatively, by monitoring videos of the dynamics. They can also be quantitatively captured by the four-point dynamical susceptibility 
\beq
G_4(\mathbf{x},t) \equiv L^{-d} \int d^dx \langle o(\mathbf{x},t) o(\mathbf{y},t) o(\mathbf{x},0) o(\mathbf{y},0) \rangle,
\eeq
where $o(\mathbf{x},t)$ is the excess density of particles at $(\mathbf{x},t)$. A related quantity that is often used in the glass literature is $\chi_4(t) \equiv \int d^d y G_4(\mathbf{y}, t)$. Evidently, $G_4$ tells one how ``frozen'' the dynamics is: thus, for instance, in a glassy equilibrium state like a spin glass, it would have a finite asymptotic value at late times and large separations. In a structural glass, it instead captures the \emph{length-scale} over which dynamics is correlated at a timescale $t$, i.e., the size of the regions in the glass that stay frozen over that timescale. These spatio-temporal correlations, in turn, come from the facilitated nature of dynamics in glasses: whether a particle is mobile or stuck depends on the local configuration surrounding it, and on whether those particles are mobile or stuck, and so on.

The relevance of this to nearly MBL systems is fairly direct, though the connection has remained relatively unexplored. It is most direct in the case of translation-invariant MBL, where the dynamics is precisely of the facilitated type: degrees of freedom can relax when they are in or near a hot bubble, and not otherwise. Thus the dynamics at any given time is slow in the large bulk of the system that lacks bubbles, and fast where the bubbles are present. Likewise, it is obvious why nearly MBL systems in more than one dimension would be dynamically heterogeneous: in these systems relaxation is fast near the thermal inclusions and slow elsewhere. Dynamical heterogeneity is, however, a more general consequence of the arguments in the preceding sections. When the locator expansion breaks down, it first (in the temporal sense) does so \emph{somewhere}. These anomalously thermal degrees of freedom (which might not be contiguous) then form a bath for the rest of the system. Since the bath is sparse and interactions are local, both dynamical heterogeneity and a broad distribution of timescales directly follow.

Even in systems with perturbatively destabilizing power-law interactions [Sec. ] (which are relatively nonlocal and thus unfriendly to spatial heterogeneity) there are atypical degrees of freedom that live on the resonant percolating network, and other typical degrees of freedom that are some distance from this network. Relaxation takes place in multiple well-separated stages, as it radiates out from the rapidly relaxing regions to the slow ones. (In the case of power-law interactions, this typically only gives rise to stretched exponential relaxation, but for local interactions the hierarchy is more pronounced and one typically finds a power-law distribution of rates.)

\subsection{Emergent conservation laws and transport hierarchies}

\subsubsection{Emergent conservation laws}

The dynamical heterogeneity of nearly MBL systems need not be purely spatial; another common type of heterogeneity is the existence of multiple types of excitations that relax on very different timescales. The simplest instance of this phenomenon is the Hubbard model at large $U$, where doubly occupied sites can only dissociate on exponentially long timescales (Sec.~\ref{sec:unstable:powerlaw}); however, a closely related phenomenon is the exponentially slow heating of rapidly driven Floquet systems (Sec.~\ref{sec:prethermal}). Yet another instance is that of localized phonons in the spin-incoherent Luttinger liquid (Sec.~\ref{sec:multicomponent:spins:SILL}), which are approximately conserved because their characteristic energies are much larger than those of the spin bath. In the simple case of the Hubbard model at low filling, one large microscopic ratio $U$ seeds at least three separate dynamical timescales: a fast scale of order unity (in units of the hopping) on which singly occupied sites move, a slower timescale $\sim U$ on which doubly occupied sites move and on which superexchange happens, and a much slower timescale $\sim \exp(U)$ on which doubly occupied sites decay (Sec.~\ref{sec:prethermal}). Related systems, such as the Bose-Hubbard model and the anisotropic XXZ model, have an infinite hierarchy of timescales, as well as an infinity of approximately conserved charges, that come from a single large parameter. These conservation laws lead to a nontrivial, multiple-scale hydrodynamics near the MBL transition, the consequences of which---for instance, for long-time tails~\cite{pomeau1975, gopalakrishnan_griffiths_2016}---have not yet been fully explored.

\subsubsection{Particle vs. energy transport}

When there are multiple conserved charges (exact or approximate) in a nearly MBL system, the corresponding transport coefficients will generally be very different, because the excitations that form the effective bath might not transport all forms of charge. The charges that are transported by the bath will in general diffuse much faster than the others, which have to move via variable-range hopping mediated by the bath. We have illustrated this phenomenon with various examples, which we briefly recapitulate. 

(i)~In the charge-disordered Hubbard model at large $U$, thermalization takes place because there is a symmetry-based obstruction to localizing the spin sector. At moderate disorder (i.e., when the wavepackets of neighboring localized charges have reasonably large overlap), spins form a thermal bath of bandwidth $1/U$, which then facilitates incoherent hopping of the charge. Energy diffusion is dominated by the spin sector. Charge diffusion is parametrically slower than spin or energy diffusion (Sec.~\ref{sec:multicomponent:spins}). At intermediate temperatures $1/U \ll T \ll 1$, the spin \emph{transport} coefficients are suppressed by further factors of $1/(UT)$. 

(ii)~In the electron glass, charges hop locally but energy is transported over long distances through interactions among charge dipoles. These dipoles form a resonant network of charge-neutral excitations, with a bandwidth set by the disorder strength. Charge moves through incoherent local processes that absorb energy from the neutral bath, as described in Sec.~\ref{sec:multicomponent}. 

(iii)~Generically, different types of excitations in the MBL phase have different characteristic localization lengths. Thus, in the thermal Griffiths phase near the MBL transition, inclusions might be more effective at blocking spin transport than energy transport, as numerically seen in Ref.~\cite{varma17, lerose19}. Whether this leads, asymptotically, to a strong quantitative suppression of the diffusion constant, or to the coexistence of energy diffusion and spin subdiffusion, is still an open question. 

These observations suggest that the thermoelectric properties of MBL systems might be an interesting question for future research. Little is known about these properties, even for Anderson insulators, but there have been some recent developments in the noninteracting case~\cite{chiaracane2019quasi}.

\subsubsection{Hierarchies in the dynamics of quantum information}

In addition to particle transport, nearly MBL systems can exhibit a large separation of scales between the distinct rates for quantum information spreading. Some relevant scales are the rate at which a Heisenberg operator grows, as measured by the OTOC, and the rate(s) at which entanglement grows---which are in general different for every distinct R\'enyi entropy. In a chaotic system with no conservation laws, all of these quantities grow linearly with comparable velocities~\cite{mezei, nrvh, nvh, curtvonk}. A single conservation law suffices to introduce considerable fine structure to the dynamics of information~\cite{rpv_renyi}, for instance causing higher R\'enyi entropies to spread diffusively rather than ballistically. A finite density of localized inclusions has more drastic effects: even though operator spreading remains ballistic, entanglement spreads sub-ballistically, and the operator front is parametrically broader than in a clean system. An analytic understanding of these phenomena seems within reach, given recent developments in random unitary circuits (Sec.~\ref{sec:tools:RUCs}), but at present many of these questions remain open. 

\subsection{Fragility of linear response}

Another general feature of nearly MBL states is the fragility of linear response. In the MBL phase itself, as we discussed in Sec.~\ref{sec:intro:mbldynamics}, linear response breaks down for any fixed drive amplitude as the drive frequency is lowered, since low-frequency driving causes the system to delocalize via Landau-Zener transitions. Linear response fails in a similar way in the subdiffusive, thermal phase: as the frequency is lowered, inclusions begin to delocalize through Landau-Zener transitions, and thus cease to act as inclusions. Thus at any finite drive amplitude, the subdiffusive phase becomes diffusive at sufficiently low frequencies. Similar results also hold for other types of nearly MBL states: slow or strong driving destabilizes the nominally MBL regions and thus enhances transport. 

Even if one drives a system sufficiently fast to avoid destabilizing MBL, the dynamics of energy absorption are much more nontrivial than Joule heating. One can regard a nearly MBL system as an ensemble of two-level systems (TLSs) weakly coupled to a thermal bath, with a broad distribution of lifetimes $\rho(\tau)$\footnote{Strictly speaking, the lifetime here is the rate at which the effective splitting of a TLS fluctuates, i.e., a $T_2$ time rather than a $T_1$ time. Hartree shifts due to transitions in TLSs cause some previously resonant, saturated, TLSs to wander out of resonance and other, previously non-resonant, TLSs to come into resonance. This process counteracts saturation~\cite{Faoro2015}.}. The nature of this distribution varies depending on the precise case being considered, but it is typically fat-tailed, approaching $\rho(\tau) \sim 1/\tau$ as one approaches the MBL transition. Suppose the system is driven at a frequency $\omega$ and amplitude $A \ll \omega$ for a time $t \gg 1/A$ (both $\omega$ and $A$ are assumed to be small compared with the characteristic local energy scales). TLSs with $A \tau \gg 1$ undergo many Rabi oscillations before they decay; thus they can absorb one unit of energy every $\tau$ units of time. On the other hand, TLSs with $A \tau \ll 1$ decay before they saturate and thus remain in the linear response regime. The total rate of Joule heating in this regime, in the long-time limit, is (adapting the arguments in Ref.~\cite{gopalakrishnan_regimes_2016} to open systems):
\beq
\Gamma = \frac{A^2 \omega}{T} \int_0^{1/A} d\tau \rho(\tau) + \frac{\omega^2}{T} \int_{1/A}^\infty d \tau \frac{\rho(\tau)}{\tau}.
\eeq
The first term corresponds to the linear-response absorption rate, while the second, nonlinear contribution is the steady state heating rate due to nearly saturated TLSs. This two-part structure of absorption is generic in systems that are nearly MBL; in most (but not all) cases, the $A$-dependence of the limits of the integral leads to nontrivial $A$-dependence in $\Gamma$, and thus to anomalous heating in the steady state.

\subsection{Temporal crossovers in response functions and entanglement dynamics}

We close this section by summarizing the complex set of dynamical crossovers that generically exist in nearly MBL systems. There are three dynamical regimes in a nearly MBL system: (i)~an early-time regime where the system has not yet resolved the processes that will thermalize it; (ii)~an extended, intermediate-time regime in which the system is in the process of being destabilized; and (iii)~a thermal regime at very long times. These ``times'' may be the duration after a quantum quench, or the inverse frequency of an equilibrium response function; they also correspond to \emph{lengths}, defined by the spread of entanglement and quantum information over a given timescale. 

In some sense these three regimes always exist in systems that are described as ``prethermal''; however, in most other contexts the second regime occurs on a particular, well-defined timescale. Thus, if one plots the quantity of interest vs. time, putting time on a \emph{logarithmic} scale, it appears to remain at a ``prethermal plateau'' for a certain time interval, and then fall off a cliff until it reaches the thermal value. In these contexts, regime~(ii) is most naturally regarded, not as an extended temporal regime, but as the time when the observable falls off the prethermal plateau.
In nearly MBL systems, on the other hand, regime~(ii) involves a hierarchy of timescales, and in some cases (e.g., the thermal Griffiths phase) can extend out to infinity. 

We now briefly summarize a few salient features of the early- and intermediate-time regimes in MBL systems. The late-time thermal regime, when the system has fully thermalized, is the same for nearly MBL systems as for any other many-body system, and need not be discussed with a specific reference  to the prior  near-MBL history.

\begin{figure}[tb]
\begin{center}
\includegraphics[width = 0.35\textwidth]{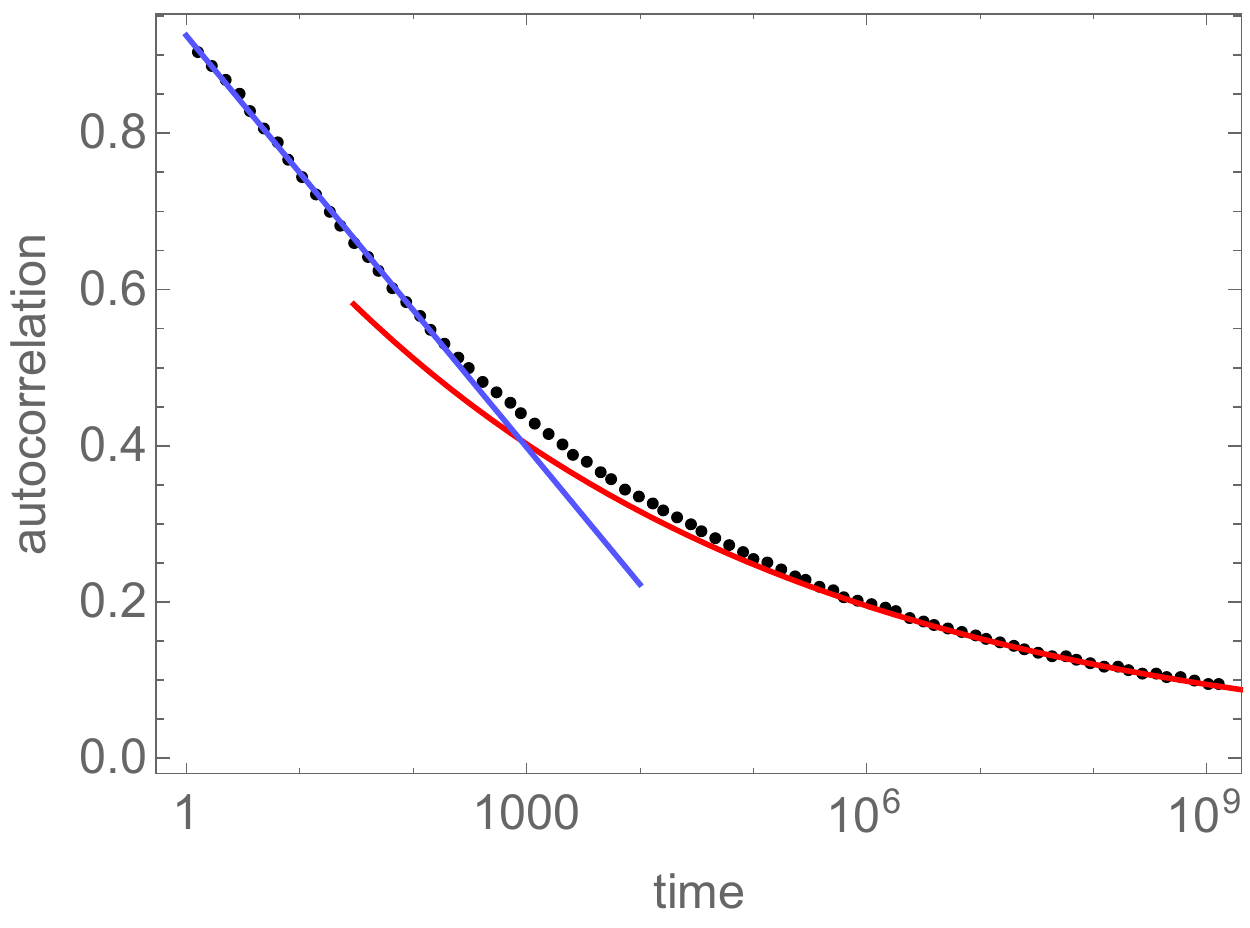}
\includegraphics[width = 0.35\textwidth]{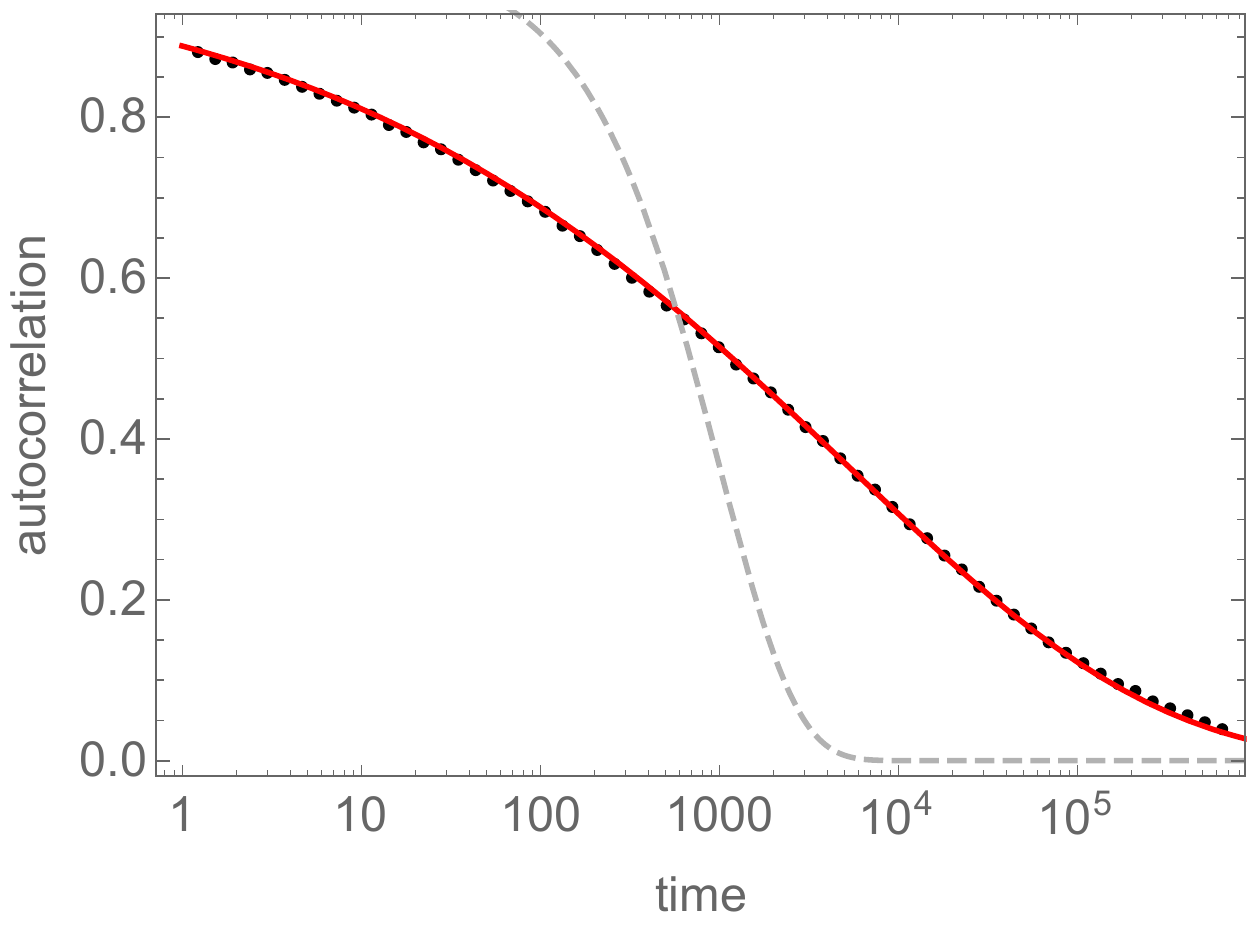}
\caption{Relaxation of local operators or autocorrelation functions in unstable MBL systems with short-range (left) and power-law (right) decay of interactions. In the short-range case, the decay is initially logarithmic (blue) then crosses over to a Griffiths power law (red). In the long-range case, a stretched exponential captures the relaxation at essentially all times. When the interactions are $1/r^\alpha$ the decay goes as $\exp(-\mathrm{const.} \times r^{1/\alpha})$ [here $\alpha = 4$]. For comparison, a simple exponential decay (dashed gray line) would look like a steep cliff on this logarithmic scale.}
\label{prethermalplots}
\end{center}
\end{figure}

\emph{Early-time prethermal behavior}.---At early times, the dynamics is locally MBL except possibly in isolated patches of the system; a ``causal window'' of length $\log t$ (or a finite size sample of size $L \sim \log t$) has a very small probability of hosting a dangerous resonance or thermal inclusion. The behavior in this temporal regime is qualitatively what one would expect in the MBL phase, except that certain stability constraints are absent here. For example, we argued in Sec.~\ref{sec:intro:mbldynamics} that the stability of the MBL phase requires that $\sigma(\omega)/\omega \rightarrow 0$ as $\omega \rightarrow 0$. If the MBL phase is asymptotically unstable, this constraint naturally does not apply. Various signatures of quantum order can presumably be seen in two-time correlation functions in the prethermal regime. 

\emph{Intermediate times}.---As time progresses, an appreciable part of the system begins to feel the instability. Particularly in short-range systems that are unstable to avalanches, this happens over many decades: the ``avalanche'' progresses logarithmically outward in time\footnote{Shivaji Sondhi has remarked that it is really more like a glacier than an avalanche.}, so quantities such as spatially averaged autocorrelation functions decay logarithmically. Even after the bulk of the system has thermalized, rare insulating clusters persist, so the logarithm crosses over into a Griffiths power law, which persists (in this simple case) to arbitrarily late times. Similar phenomena (sometimes with power laws instead of logarithms) occur in all the other examples we have explored. The nature of this temporal crossover is shown, and contrasted with the ``standard'' prethermal scenario, in Fig.~\ref{prethermalplots}. 

To summarize, the nearly MBL systems studied here are similar in exhibiting a broad distribution of relaxation timescales, spatially heterogeneous dynamics, and many other phenomena that have been extensively studied in the context of glasses. These parallels are natural since both nearly MBL systems and actual structural glasses are not thermodynamically distinct from the thermal (i..e, ergodic) phase; thus, many of the natural approaches to discussing them are similar. The extensive existing literature on the glass problem~\cite{berthier2011} suggests many new aspects of MBL dynamics that might be fruitful to study, such as aging dynamics.

\section{Theoretical methods and challenges \label{sec:tools}}

The previous sections have presented a phenomenology of the MBL phase, the MBL transition, and the dynamics of systems that are in various senses ``nearly'' MBL. This phenomenology was built around the somewhat aggressive assumption that resonances and thermal regions proliferate whenever they can. This assumption of maximal infectiousness is a plausible conjecture, for which some numerical evidence exists, and which there is no strong \emph{a priori} reason to doubt: the MBL problem involves motion in a high-dimensional space, where weak localization effects and other known instabilities of the delocalized phase are not expected to be large. Nevertheless, the picture discussed here is largely conjectural, as neither controlled calculations nor dispositive numerical evidence exists either way. The existing numerical methods are severely restricted by system size, and since a nearly MBL system looks MBL on short scales, these methods are biased toward finding an MBL phase even for parameters where it does not truly exist (in the conventional thermodynamic limit). Given the inherent limitations of current numerical methods for quantum dynamics, the true asymptotic regime of MBL might be largely inaccessible.

In this section we review various recent ideas that might overcome this difficulty. Consistent with the spirit of this review, we will focus on methods that address dynamics and the structure of correlation functions; thus, in particular, we will not discuss the various interesting schemes that have been proposed for constructing exact excited eigenstates of MBL systems~\cite{kps, ypc}. We will also focus on methods that are scalable to very large systems (hundreds of sites); thus we will omit some important recent advances that have made exact simulations feasible for spin chains of length $L \geq 30$. These advances are already well described in the reviews~\cite{Luitz16Review, pietracaprina2018}, and require a technical discussion that is outside the scope of this work. 
The approaches we will discuss are all approximate, and the point of our discussion will be to sketch the nature and inherent limitations of each approximation. The discussion naturally divides itself into two parts: first, we will introduce some methods that build on the structure that exists in the MBL phase, and discuss how to extend them to nearly MBL systems; second, we will turn to a discussion of numerical approaches that capture the thermal phase, and capture how thermalization and transport begin to break down in the presence of disorder. 

\subsection{Numerical construction of integrals of motion}

Within the MBL phase itself, numerical studies are stable, and many quantities can be computed with relatively weak finite-size effects. A number of methods perform well here: exact diagonalization has weak finite-size effects down to very low frequencies, while schemes for studying the dynamics based on matrix-product states (MPS's)---in particular, time-evolving block decimation (TEBD)~\cite{vidal_efficient_2003}---remain tractable out to late times, since the growth of entanglement is logarithmically slow. For MPS methods, the computational cost is polynomial in the evolution time; for exact diagonalization, the cost is exponential in system size, but a system of size $L$ is free of finite size effects out to timescales exponential in $L$, so once again the cost scales polynomially with time. One can therefore extract converged results for quantities such as the distribution of local expectation values, autocorrelation functions, etc., as well as aspects of slow dynamics such as the logarithmic growth of entanglement and the power-law temporal decay of dynamical correlation functions. However, even in the MBL phase, some predictions of the theory we have reviewed here remain out of reach: in particular, the stretched-exponential decay of equal-time correlation functions within an eigenstate, and the prediction in Sec. 1.4 for the asymptotic low-frequency limit of the conductivity. These predictions relate to the existence of fractal thermal inclusions, which are expected to be present but for which there is relatively little direct evidence (as one would expect, given that one is looking for sparse fractal clusters in a system of size $L \approx 20$---see, however, Ref.~\cite{herviou}). 

To reach much larger systems on present-day computers, the most natural strategy is to leverage some kind of renormalization-group approach: i.e., to solve the system one length scale at a time, and only keep the most important degrees of freedom from one scale while treating the next. This is the approach advocated in many RG papers~\cite{pvp, dvp, thmdr} on the MBL transition, but the actual implementations of the RG make further approximations that decrease the microscopic accuracy. We now briefly outline a class of methods for actually carrying out this procedure in the MBL phase. Our focus will be on methods that are used to construct LIOMs, since the generalization of the LIOM concept to imperfectly MBL systems is more direct than the corresponding notions for eigenstates. {We note that the strong-disorder RG can provide additional perspectives on dynamics; we refer the reader to  Sec.~\ref{sec:unstableMBL} and references therein for a discussion on the use of SDRG {and on novel methods for identifying perturbative resonances between approximate SDRG eigenstates~\cite{ProtopopovSU2}).}

In principle, exact diagonalization of a small system gives one complete information about that system; however, extracting LIOMs from exact diagonalization data is challenging. Recall that the LIOMs are formally formally defined as follows: for each LIOM, one takes half the states to have eigenvalue $+1$ and the other half to have eigenvalue $-1$ under that LIOM. There are factorially many such assignments of states to LIOMs; however, we seek an assignment that is \emph{local}, which rules out nearly all such mappings\footnote{However, there remains some residual freedom as to how the mappings are chosen, corresponding to local transformations between the variables. For example, instead of the variables $\tau^z_1, \tau^z_2$, one might choose the variables $\tau^z_1, \tau^z_1 \tau^z_2$, and the mapping would remain local.}. Clearly, cycling through the possible mappings looking for a local one is completely infeasible, and we seek a better way to proceed. 

\subsubsection{Direct association of LIOMs to physical spins}

Deep in the localized phase, one expects LIOMs to be close to physical spin operators. For simplicity we consider a model in which the $z$-component of magnetization is conserved (giving us a natural ``axis'' to work with in physical space). The simplest algorithm for associating spins to LIOMs in this regime is along the following lines (this specific approach was implemented in Ref.~\cite{peng2018}): 

(1) Pick a spin $j$. Evaluate $s_j \equiv \langle n | \sigma^z_j | n \rangle$, and sort the eigenstates in descending order by $s_j$. Denote the eigenstate index under $s_j$ as $m_j$ and the corresponding many-body energy eigenstate as $| m_j \rangle$.

(2) Create an operator $\tilde\tau^z_j \equiv \sum_{m_j \leq 2^{L - 1}} | m_j \rangle \langle m_j | - \sum_{m_j > 2^{L-1}} |m_j \rangle \langle m_j |$. 

(3) Compute the trace of this operator with all the $\sigma^z$, i.e., evaluate $T_{ij} \equiv \mathrm{Tr}(\tilde\tau^z_i \sigma^z_j)$. Let $j_0 = \max(|T_{ij}|)$, i.e., find the physical spin operator that overlaps most strongly with $\tilde\tau^z_i$. Define $\tau^z_{j_0} \equiv \tilde\tau^z_i$. This step accounts for cases where we misidentified the center of the operator.  
 
(4) Now consider the $2^{L-1}$-dimensional space of $+1$ eigenvalues of $\tau^z_{j_0}$. Within this manifold, repeat steps (1)-(3) to construct a new operator. Then continue this process until all the conserved operators are found.

It is intuitively clear that deep in the localized phase this procedure will generate local operators; also, there is compelling numerical evidence that the operators generated are in fact local. 

Although this prescription naturally generates the $\tau^z$ operators (and thus suffices to generate the effective Hamiltonian for the MBL phase), it is nontrivial to get the rest of the Pauli algebra from it. The remaining Pauli operators have exclusively off-diagonal matrix elements in the many-body eigenbasis. Multiplying each many-body eigenvector by an arbitrary phase would leave the $\tau^z$ operators unchanged, but scramble the $\tau^{x,y}$ operators, and potentially spoil their locality properties. However, for most purposes it is not crucial to have the rest of this algebra, in which case the algorithm described above efficiently constructs all LIOMs.

\subsubsection{Finding LIOMs by time evolution}

An alternative, and instructive, way to construct LIOMs is through the dynamics of initially local operators~\cite{chandran2015}. The key idea is as follows: we observe that for any operator $O$,

\beq
O_\infty \equiv \lim_{T \rightarrow \infty} \frac{1}{T} \int_{-T}^T dt O(t)
\eeq
commutes with the Hamiltonian (or Floquet unitary), since commuting $O(t)$ with the Hamiltonian simply amounts to a time translation, under which the integral is invariant. Thus the operator $O_\infty$ is an integral of motion by construction, but is generically not local. However, we now show that $O_\infty$ is a quasi-local operator (i.e., local up to exponentially small tails) provided that \emph{some} complete basis of LIOMs exists. The argument runs as follows: if such a complete basis does exist, then one can write the operator $O$ (which is local by assumption) in the LIOM basis as a sum of products of Pauli strings with coefficients that decay exponentially in the string length (here $\alpha = 0,1,2,3$ labels Pauli matrices):

\beq
O = \sum\nolimits_{n \geq 0} \exp(-2n/\xi) \sum\nolimits_{\{\alpha\} \in \{ 0, 1, 2, 3 \}^{2n}} \prod\nolimits_{\{\alpha\}} C^{\alpha_{-n} \alpha_{-n + 1} \ldots \alpha_{2n}} \tau^{\alpha_{-n}}_n \tau^{\alpha_{-n+1}}_{-n + 1} \ldots \tau^{\alpha_n}_n.
\eeq
The expansion of $O$ contains two types of terms: those with the off-diagonal Pauli matrices $\tau^{x,y}$ and those without. Under time evolution, the former terms oscillate and cancel out, provided there are no exact degeneracies in the spectrum, while the latter terms commute with the time evolution operator and do not grow. Therefore, if $O$ was initially quasi-local, so is $O_\infty$. 

One can construct an infinite set of mutually orthogonal [in the Frobenius sense, $\mathrm{Tr}(O_1 O_2) = 0$] conserved operators using this procedure, since time-evolution preserves this inner product. However, the operators generated by this procedure do not obey a Pauli algebra, and thus cannot be used to write a Hamiltonian as simple as the standard l-bit model. Nevertheless, the fact that one can construct these operators using time evolution rather than exact diagonalization means that this construction is compatible with techniques that are more scalable than exact diagonalization, such as Krylov-space methods or TEBD. Moreover, if one truncates the integral at finite times, this method gives useful information even outside the localized phase, as we will discuss below in Sec.~\ref{slowops}. (We note that the procedure of looking for operators that almost commute with the Hamiltonian~\cite{cpls} is effectively a finite-time variant of this procedure.)

\subsubsection{Generalized Schrieffer-Wolff transformations}

A different class of approaches to this problem involve diagonalizing the Hamiltonian through a series of local unitary transformations. The essential concept here is the Schrieffer-Wolff transformation~\cite{swtrans}, which is a way of doing perturbation theory at the level of the Hamiltonian rather than the state. We will outline the idea behind this transformation in a very general way, and then discuss how to implement it. The Schrieffer-Wolff transformation begins with a Hamiltonian that is decomposed as $H = H_0 + \lambda V$, where $H_0$ is ``simple'' (e.g., consists of local fields and other diagonal terms, so that the spectrum has local labels), and $V$ is some arbitrary perturbation. We now perform a unitary transformation $\widetilde{H} = e^{i \lambda S} H e^{-i \lambda S}$. We see that, if $V = - i [S, H_0]$, the transformed Hamiltonian is 

\beq
\widetilde{H} = H_0 + O(\lambda^2) + \mathrm{constant}.
\eeq
The eigenstates of $\widetilde{H}$ are therefore (to leading order) the same as those of $H_0$ (up to a possible global shift). The unitary $e^{iS}$ can also be seen as a transformation rotating the (simple) eigenstates of $H_0$ into the more complicated eigenstates of $H$.
In the present case, we would like $H_0$ to contain all terms that are purely diagonal in the computational basis; since there are a large number of terms in the Hamiltonian that do not commute with $H_0$, it is necessary to perform a large number of these unitary transformations to bring the Hamiltonian to the l-bit form. We now briefly mention some schemes for organizing this sequence. 

\emph{Locator expansion}.---The simplest of these schemes, and the first to be implemented, is the locator expansion done in operator language~\cite{ros2015integrals}. Here, one starts from a particular site, and progressively rotates out couplings between that site and its neighbors; at each state this process generates smaller but longer range couplings, which decay exponentially with distance. At high orders in perturbation theory, there are many possible paths between two configurations, so the bookkeeping is quite involved; for details see Refs.~\cite{ros2015integrals, PhysRevB.93.054201}. 

\emph{Flow equations}.---The Wegner-Wilson flow equation method~\cite{kehrein_book} is an \emph{exact} but numerically intensive technique for carrying out these Schrieffer-Wolff transformations. In this method one implements the rotations \emph{differentially}, leading to a flow of the Hamiltonian along a curve in parameter space parameterized by a number $\beta$. The Hamiltonian parameters as well as the differential transformations evolve along the flow, as captured by the system of differential equations~\cite{pcor}:

\bea
H(\beta) & = & H_0 (\beta) + V(\beta), \nonumber \\
\eta(\beta) & = & [H_0(\beta), V(\beta)], \nonumber \\
\frac{dU}{d\beta} & = & \eta (\beta), \nonumber \\
\frac{dH}{d\beta} & = & [H(\beta), \eta(\beta)].
\eea
This flow is initialized at $\beta = 0$ with the data $U(0) = 1, H(0) = H$. The desired, fully diagonal, representation of the Hamiltonian is arrived at when $\beta \rightarrow \infty$. This system of equations is an exact transformation of any Hamiltonian to the desired diagonal form; however, it is challenging to solve---and is currently restricted to smaller system sizes than full diagonalization---but it has the advantage over the previously mentioned methods that it (a)~generates all the LIOMs at once, and (b)~automatically generates local $\tau^{x,y}$ operators without phase ambiguity. This motivates the search for schemes that are simpler to implement. One such scheme---which exploited the strong-randomness limit to discretize and simplify the flow equations---was applied to noninteracting power-law systems in Ref.~\cite{quito}. An alternative approach, which has been explored in depth in the MBL context, is the method of displacement transformations, which we will now describe.

\emph{Sequential displacement transformations}.---A somewhat more intuitive way of organizing the sequence of rotations was developed in Refs.~\cite{rademaker1, rademaker2}, independently of the flow-equation literature. The approach can be summarized as follows. We begin with a generic fermionic Hamiltonian, which we write in the form 
\beq
H = \left[ \sum\nolimits_\alpha n_\alpha + \sum\nolimits_{\{\alpha\}} f(\{n_\alpha\}) \right] + \sum\nolimits_{\{\alpha\}} X_{\{\alpha\}}, 
\eeq
where $n_\alpha \equiv c^\dagger_\alpha c_\alpha$ and $X_{\{\alpha\}} = n_{\alpha_1} n_{\alpha_2} \ldots c^\dagger_{\alpha_m} c^\dagger_{\alpha_{m+1}} \ldots c_{\alpha_{m+k}}$. Note that $\alpha_{m+1} \ldots \alpha_{m+k}$ must all be different as otherwise the term could have been absorbed into an $n$. The part of the Hamiltonian in square brackets is diagonal and need not be rotated away; our objective is to perform Schrieffer-Wolff transformations to eliminate the remaining terms. 

We organize the off-diagonal perturbations in the Hamiltonian by the total number of fermion operators they contain; this is termed the ``order'' of the term. At each order there are ``classical'' terms (which depend only on the $n$'s) and ``quantum terms'' (the rest). We can rotate away any quantum term $X$ by the displacement transformation 

\beq
\widetilde{H} = e^{-\lambda(X - X^\dagger)} H e^{\lambda(X - X^\dagger)}.
\eeq
This transformation can be regarded as a sort of polaron transformation, in which the bosonic operator $X$ is shifted, and the other operators are ``dressed'' by this shift. To see how this works, we follow Ref.~\cite{rademaker1} and consider a concrete four-site example:

\beq
H = \sum_{i = 1}^4 \xi_i n_i + V_{13} n_1 n_3 + V_{24} n_2 n_4 + V(c^\dagger_1 c^\dagger_3 c_2 c_4 + \mathrm{h.c.}).
\eeq
The operator $X = c^\dagger_1 c^\dagger_3 c_2 c_4$ and we attempt to rotate it away via a displacement transformation. One can check that the displacement transformation eliminates the term $V(X+X^\dagger)$ if one chooses $\lambda$ such that

\beq
\tan(2 \lambda) = -\frac{V}{\xi_1 + \xi_3 + V_{13} - (\xi_2 + \xi_4 + V_{24})}.
\eeq
More generally, the cancellation holds when $\tan(2\lambda)$ matches the ``naive'' first-order perturbation theory shift to the state (recall that we are treating all classical terms as part of $H_0$ so they contribute to energy denominators). Note that $\lambda$ remains finite even if the term being eliminated is resonant. Generically, the shift procedure will generate new terms in the Hamiltonian. It can be shown that these new terms are always at least of the same order as the term that was rotated away, and are typically higher order and have smaller coefficients. By eliminating all terms out to a certain order, one can construct an effective Hamiltonian in which the lowest ``quantum'' terms left are large-scale rearrangements. 

The displacement transformation approach has some superficial similarities with strong-randomness RG; however, it is different in a number of respects. First, it is exact, if one keeps track of the generated terms at all orders. This is true even in the thermal phase, although there the procedure will ultimately generate a very large number of individually small terms, with essentially every coupling being resonant. Second, it is organized in terms of Fock-space distance rather than real-space distance: thus it applies to systems that have long-range interactions. Third, it extends naturally to the case of imperfect MBL, since one can simply cut off the procedure when all remaining couplings are below a certain energy scale (at which, e.g., coupling to the bath becomes important). This is the question we will focus on in the next section. 

\subsection{Extension to slow dynamics near the MBL phase}

Nearly MBL phases look essentially localized at short distances, before crossing over to thermal behavior at much longer scales. Thus, at short length-scales typical regions of the system will be effectively in the localized phase, and will have approximate l-bits, i.e., operators that are very slow to relax. These slow operators will control the dynamics of nearly MBL systems on intermediate timescales, before the eventual crossover to thermal relaxation. In this section we sketch how the ideas from the previous section about constructing l-bits can be used to capture this intermediate-time dynamics. 

\subsubsection{Slow local operators}\label{slowops}

In systems that are not strictly MBL there are no true l-bits; however, there are \emph{approximate} l-bits, which barely relax out to very late times. We should distinguish here between two notions of slow relaxation of an operator: the first, more natural, notion is that the autocorrelation function of the operator has a slow asymptotic decay. However, the asymptotic behavior of operators is a subtle question that cannot readily be answered in finite size numerics. Therefore, we consider a somewhat different notion of slowness: an operator is taken to be slow if its commutator with the Hamiltonian is small, in an appropriate norm. Since commutators are straightforward to compute, this notion of slowness is much more tractable than the natural, asymptotic one. An operator that is slow in this ``short-time'' sense barely decays at all until some long time $T$; beyond that timescale, however, it may decay arbitrarily fast. However, the commutator lower-bounds the eventual thermalization time, since it controls when an operator begins to spread. Also, in the context of nearly MBL systems, we note that there are many approximate LIOMs that almost commute with the Hamiltonian, and that will be picked up by this diagnostic.

Searching for operators that are slow in the short-time sense is further simplified if one quantifies the ``size'' of the commutator using the Frobenius norm~\cite{slowest_operators, pancotti_slow}: $\Vert [O, H] \Vert_F \equiv \sqrt{ \mathrm{Tr}([O,H] [O,H]^\dagger )}$. The Frobenius norm of an $N\times N$ matrix can equivalently be thought of as the Euclidean norm of the ``vectorized'' matrix, i.e., the $N^2$-sized vector of entries in the matrix. The square of the Frobenius norm can be written as a quadratic form and thus efficiently minimized over all operators of a certain size~\cite{slowest_operators}; the minimization problem can also be written as an eigenvalue problem that can be solved by the Lanczos method~\cite{pancotti_slow}. 

In thermal states the slowest operator with support over $M$ sites is well described by the following construction, which is closely parallel to the construction of l-bits through time evolution. Given an arbitrary local operator $A$ we define a slow operator as $\mathcal{S}_A = T^{-1} \int_{-T}^T dt \cos[\pi t / (2 T)] A(t)$. Because of the Lieb-Robinson bound this operator has support only over a region of size $v T$ where $v$ is the Lieb-Robinson speed. To estimate its commutator, we use

\beq
[\mathcal{S}_A, H] = \frac{1}{T} \int_{-T}^T \cos(\pi t/ (2T)) [A(t), H] = \frac{i}{T} \int_{-T}^T \cos(\pi t/ (2T)) \frac{d A(t)}{dt} = \frac{i}{T^2} \int_{-T}^T \sin(\pi t/(2T)) A(t).
\eeq
Thus, the norm of the commutator is suppressed by a factor of $1/T$ when it has support $\sim v T$. As one increases disorder, the Lieb-Robinson velocity becomes increasingly heterogeneous, and eventually becomes peaked at zero as one enters the subdiffusive Griffiths phase. Here the distribution of commutators at a given operator size becomes very broad, with a tail peaked at zero corresponding to operators that begin in localized inclusions. One expects the density of operators with commutator $\leq \lambda$ to scale as $\lambda^{1/z}$, where $z$ is the dynamical exponent introduced in Sec.~\ref{sec:thermoMBL}. Thus the probability distribution will be $P(\lambda) \sim \lambda{-1 + 1/z}$. As one enters the MBL phase $z \rightarrow \infty$ so the slowest operators will follow a $1/f$ distribution, corresponding to the proliferation of LIOMs. These regimes of behavior, as well as a more quantitative analysis of the distribution functions, are numerically explored in Ref.~\cite{pancotti_slow}. 

Many questions remain about the structure of slow operators, however. One of their useful properties is that they give a way of efficiently capturing the dynamical heterogeneity of systems near an MBL transition; this aspect has not yet been fully explored. (It is worth remarking that slow operators are spatially correlated in the Griffiths scenario, since an inclusion contains many slow operators; this aspect of the Griffiths scenario has not yet been tested.) Numerically extracted slow operators might also serve as a starting point for constructing a fluctuating hydrodynamics of systems near the MBL transition. 

\subsubsection{Boltzmann equation for approximate LIOMs}

Instead of constructing approximate LIOMs by time evolution, one could instead construct them by Schrieffer-Wolff transformations, such as the generalized shift method. Suppose one applies this method to systems that are MBL on short scales but asymptotically thermal, e.g., because of large-scale rearrangements, or long-range resonances. The first several levels of transformations will typically be non-resonant, but as the procedure continues an increasing fraction of the eliminated terms will have perturbative resonances, i.e., the shift parameter $\lambda$ will be attracted to $\pi/4$. At the point when an appreciable fraction of displacements are resonant, the effective LIOMs have ceased to be localized; beyond this point, one can treat the system as effectively thermal, and thus treat the residual couplings as collisions in the framework of the Boltzmann equation. The advantage of this approach is that the collision integrals are microscopically derived rather than phenomenological. In analogy with Fermi liquid theory, where one dresses the quasiparticles with interactions and computes the lifetimes of dressed quasiparticles, this approach would both constitute a microscopic kinetic theory of thermalization near the MBL transition, and include a nontrivial collisionless piece due to the diagonal interactions. Thus it would include the phenomena we have sketched out in Secs.~\ref{sec:intro}~\ref{sec:dynamicalsig} and, such as the decoherence of a given LIOM because of the noise from transitions in neighboring LIOMs. This direction has not been actively pursued so far, but might be promising for future work.

\subsection{Numerical methods for the thermal phase}

The repertoire of quantitative methods to describe dynamics in the thermal phase is more limited at present. We expect that the coarse-grained theory of the thermal phase is hydrodynamics; however, the implications of hydrodynamics for questions such as the nature and correlations of many-body eigenstates are still an active research topic, and computing hydrodynamic coefficients in generic quantum models remains challenging. Quantum systems in the thermal phase generically have chaotic dynamics and are not exactly solvable---although, remarkably, both random unitary circuits~\cite{nrvh, nrh, nvh, curtvonk, rpv, kvh, cdc1, cdc2, Lashkari2013, Maldacena2016, ZhouNahum, ZhouChen, ChenZhou} and Sachdev-Ye-Kitaev (SYK)-type models~\cite{symodel, maldacena_syk, polchinski2016} allow certain dynamical quantities to be computed exactly. Even if these models have similar \emph{universal} properties to generic quantum chaotic models, there is no simple microscopic path from one to the other: thus, quantities such as the diffusion constant of a generic quantum model are difficult to compute reliably. Methods that work well in the MBL phase, such as exact diagonalization and TEBD, are severely affected by finite-size and/or finite-time restrictions in the thermal phase. However, understanding the incipient breakdown of thermalization coming from the thermal side requires reliable ways of simulating the coarse-grained dynamics of the thermal phase. Many recent numerical methods have been applied to this problem; in this section we briefly outline a few approaches. We focus on the basic concepts rather than on implementation details, for which we refer to the literature. We group the methods that we will consider into four classes: random circuit methods (which allow for exact solutions but are not microscopic), variational methods for dynamics, self-consistent methods, and specialized linear-response methods. This classification is only rough, since there is not a sharp distinction between, say, variational and self-consistent methods. In each case we are primarily interested in the application to disordered systems for which thermalization begins to fail. 

\subsubsection{Random quantum circuits}~\label{sec:tools:RUCs}

Random quantum circuits can be regarded as an extension to local many-body quantum dynamics of the central principle behind random-matrix theory: viz. to consider a system that has the same symmetries as the system you care about, but is otherwise completely random. ETH can be regarded as a random matrix ansatz for the eigenstates and level correlations of a chaotic many-body Hamiltonian. This ansatz can only hold for timescales that are long compared with an appropriate ``Thouless time,'' set by how long information takes to traverse the system~\cite{vha, PhysRevB.96.104201, cdc2, vsuntajs2019quantum}. The Thouless time is defined as $t_{\mathrm{Th}} \sim L^2$ for a diffusive system, and scales as an appropriate power law $\geq 2$ in the subdiffusive regime. How it scales in Floquet systems with no conservation laws is a more delicate question~\cite{cdc2, gharibyan2018onset, bkp}: however, in the sense that we are using the term here, it is linear at best in system size. On timescales shorter than $t_{\mathrm{Th}}$ the system cannot be modeled as a random matrix because initially local operators still retain spatial structure on these timescales and thus cannot be treated as purely random. 

Random unitary circuits (RUCs) generalize the random-matrix philosophy to systems with spatial locality. The simplest example of an RUC is a system in which random two-site gates (i.e., $q^2 \times q^2$ unitary matrices that are drawn randomly with Haar measure) are applied to randomly chosen bonds of a spin chain with local Hilbert space dimension $q$~\cite{nrvh}. Alternatively, instead of applying the gates randomly, one can apply two-site gates to all the even bonds at once, followed by all the odd bonds, forming a ``brickwork'' pattern of gates. In addition to locality, one can also incorporate conservation laws. RUCs make precise the notion of ``generic'' behavior in local quantum systems. In the simplest and most tractable cases, gates are temporally random---being drawn independently at each time step---and this temporal randomness prevents phenomena like MBL. However, anomalous transport in the thermal phase can easily be studied by applying the gates at bond-dependent rates, or (in the brickwork geometry) applying gates that interpolate between the identity and a Haar-random unitary~\cite{nrh}. 

Many properties of RUCs are analytically tractable in the ``semiclassical'' $q \rightarrow \infty$ limit: one can compute the growth of entanglement and its fluctuations, as well as the spreading of operators. Importantly, some properties can be analytically computed even away from that limit: examples include the circuit-averaged spread of out of time order correlators, as well as the evolution of the circuit-averaged purity $\mathrm{tr} \rho^2$~\cite{nvh, curtvonk}. These calculations are possible via a mapping onto the partition function of a classical Ising model. They can also be generalized to more structured gate sets, such as those with a conservation law~\cite{rpv_renyi}: however, in that case the partition function corresponds to a more complicated classical model. 

Although temporally random circuits do not give rise to MBL, Floquet circuits do, and one can apply a spatially random but temporally periodic set of gates to generate dynamics that is potentially localized. A concrete proposal along these lines is in Ref.~\cite{cdc2}. However, at present the model only admits analytic calculations in the $q \rightarrow \infty$ limit, in which MBL is absent. Whether this model can be studied away from that limit (even at large finite $q$) remains an interesting open question. 

\subsubsection{Variational methods}

Quantum circuits extend the RMT approach to local systems; thus they have both the strengths of RMT (allowing for specific universal predictions) and the limitations (e.g., not being able to address timescales before RMT sets in). Variational methods based on matrix product states or operators take a much more microscopic view. These methods build on TEBD, which describes the time-evolution of MPS's/MPO's provided entanglement is not too large. The major drawback of TEBD is that---for a fixed size of MPS/MPO---the algorithm gives unreliable results at late times, for instance by violating conservation laws or losing the positive-definiteness of the density matrix. Analogous issues arise in field-theoretic approaches to dynamics, and are fixed there by working with conserving approximations~\cite{PhysRevLett.62.961}; the methods below construct similar conserving approximations to time evolution with tensor networks. 

\emph{Time-dependent variational principle (TDVP)}.---Given a Hamiltonian, the TDVP~\cite{leviatan} begins by invoking the Dirac-Frenkel variational principle $\langle \delta \Psi | i \partial_t - H | \Psi \rangle = 0$~\cite{dirac1981principles}, and a generic variational parameterization of the state $|\Psi(\alpha)\rangle$ in terms of some variational parameters $\alpha$, to write down a classical Lagrangian for the $\alpha$:

\beq
\mathcal{L} (\alpha, \adot{\alpha}) = \langle \Psi[\alpha] | i \partial_t - H | \Psi[\alpha] \rangle. 
\eeq
We now derive equations of motion from the classical action in this variational subspace. Equations of motion derived from the variational principle are known to conserve both energy and the norm of the state. In general, such equations of motion will be classically chaotic; thus an appealing feature of the TDVP is that it reduces quantum chaos to classical chaos, which is better understood. For instance, an intriguing empirical relation seems to hold between the Lyapunov spectrum of the classical dynamics and the entanglement growth rate~\cite{hallam}. However, the question of when TDVP gives a reliable approximation to the true quantum dynamics is unclear at present: although the asymptotic behavior is qualitatively correct by construction, the diffusion constants for some models are quite far off~\cite{kloss}. As of now, this approach remains an area of active development and improvement (see, e.g., Ref.~\cite{paeckel2019}). 

Recently, the TDVP approach has been applied to study MBL for relatively large systems $L \approx 100$~\cite{doggen1, doggen2}. Because of the slow growth of entanglement in and near the MBL phase, TDVP is essentially just TEBD out to fairly long times. The picture that emerges from these TDVP studies is consistent with our general theoretical expectations: in particular, a large subdiffusive regime is seen, with exponents that stabilize around $L \approx 50$, and the MBL transition point drifts to larger disorder as system size is increased. This approach was then extended to quasiperiodic spin chains, for which finite-size effects and anomalous power-laws appear to be absent.

\emph{Alternatives to TDVP}.---A conceptual drawback of TDVP is that it employs a variational wavefunction that has clearly incorrect entanglement properties. It would be better, from this point of view, to work with reduced density matrices, which genuinely have low operator entanglement (i.e., can be represented as low-bond dimension MPO's) in the thermal phase. A concrete implementation of this approach is the ``density-matrix truncation'' (DMT) method~\cite{dmt}. In the DMT method, one time evolves the density matrix of the full system as an MPO, using TEBD. If one were to use TEBD naively, errors in the truncation step would quickly make $\rho$ an unphysical density matrix, i.e., it would cease to be positive-definite and unit-trace, and would also cease to conserve energy. In DMT, the physicality of the density matrix is maintained by carefully truncating the density matrix in a way that preserves all local operators that have support on fewer than $n$ sites. This method has been applied with some success to the dynamics of Floquet systems~\cite{dmt2}. 

There is a large space of potential variants of DMT, for instance methods that avoid some of the issues with positivity by working with purifications of the density matrix. 
However, these have not yet been widely applied in the present context. We also note that the TDVP can be regarded as a semiclassical limit of a functional integral over MPS's~\cite{green2016feynman}; thus, instead of going to higher bond dimensions, one might be able to achieve better results by adding ``quantum corrections'' on top of TDVP.

\subsubsection{Self-consistent methods}

\emph{Self-consistent theories of level broadening}.---
The variational methods above aimed to give a full (though approximate) description of the quantum state. A more conventional approach, rooted in field theory, aims instead to derive equations of motion for few-body observables, and to decouple or resum the BBGKY hierarchy~\cite{van2008equilibrium} in some appropriate approximation. (In a sense, this is also what DMT does.) The first detailed exposition of this approach is already present in the seminal work of Ref.~\cite{basko_metalinsulator_2006}. The idea in that work is to note that local spectral functions consist of sharp lines in the MBL phase and broadened ones in the thermal phase; thus, by solving self-consistently for the broadening $\Gamma$, one can find both an MBL phase with $\Gamma = 0$ and a thermal phase with $\Gamma > 0$. This approach, relying as it does on the properties of \emph{local} spectral functions, is similar in spirit to the ``typical-medium theory'' description of the low-temperature dynamics of random systems~\cite{tmt}. This correspondence was used in Ref.~\cite{gopalakrishnan2014mean} to incorporate the effects of Hartree shifts into the self-consistent theory: Hartree shifts cause energy levels to jitter (leading to ``spectral diffusion''~\cite{gornyi2017spectral}) and thus enhance the decay rate of putative LIOMs. 

The underlying idea in these approaches is to consider the dynamics---say, for concreteness, the spectral function $S(\omega)$---of a single degree of freedom (or more generally a local region) in the presence of a bath, which is self-consistently taken to have the same spectral function $S(\omega)$. In the high-temperature, weak coupling limit one can simplify this problem by ignoring the back-action of the bath on the system and approximating it as a classical noise source. The dynamics of the system in the presence of this noise source can then be computed as outlined in Secs.~\ref{sec:multicomponent:largebath}-~\ref{sec:multicomponent:slowbath}; the only new twist is the self-consistency requirement. 

This self-consistent approach has been worked out relatively completely in the lowest-order (self-consistent time-dependent Hartree-Fock) approximation~\cite{wgk} (see also Refs.~\cite{reichman2014, BarLev_Absence_2015, lev2017transport}). The self-consistent Hartree-Fock theory can be derived as a conserving approximation~\cite{wgk}. It gives physically reasonable results---subdiffusion in the random case, and somewhat faster decay in the quasiperiodic case---but is biased toward finding subdiffusion even in regimes where it probably does not exist. To see why this happens, let us simplify and restrict ourselves to Hartree terms. At the Hartree level, the system consists of single particles moving in noise generated by the time-dependent potential due to other particles. If a system is started far from equilibrium, any given particle initially feels large-amplitude noise as the other particles near it oscillate at different frequencies. However, at late times the density profile becomes stationary and the noise amplitude dies out. The Hartree approximation predicts slow dynamics in this limit because the single particle states are all localized absent noise. This slowdown is an artifact: in fact, even at late times, the autocorrelation function of the density oscillates and is not quiescent. Thus the Hartree method becomes unreliable whenever the system is near a steady state. However, one sees an abrupt dynamical slowdown at much earlier times, suggesting that this method is capturing some sort of dynamical crossover from delocalized to localized behavior. The physics of this crossover is as follows: when the single particle localization length is long, each orbital ``feels'' noise due to many others, and this noise contains many frequencies. This leads to transitions, which in turn generate more noise, and the system delocalizes. By contrast, when the single-particle localization length is short, each orbital only experiences noise at a few frequencies, which are usually not resonant with any transitions, so the localized state remains stable. 

As this discussion would suggest, the Hartree and Hartree-Fock theories do not correctly describe linear response on top of a steady state. To describe linear response, or to cure the deficiencies of the Hartree-Fock theory, one must incorporate collisions in (e.g.) the self-consistent Born approximation (SCBA)~\cite{reichman2014}. Implementing this within the field-theory framework is numerically intensive, however, and does not seem to yield large advantages over exact diagonalization. 

We also note some related field-theoretic attempts, which (unlike any of the work we have surveyed so far) attempt to exploit the known low-temperature properties of interacting electron gases as a starting point for studying MBL~\cite{liao2017response, mbl_boltzmann}. These techniques also seem to lead to descriptions of the thermal phase in terms of self-consistent noise~\cite{liao2017response}. However, these methods involve many subtleties specific to low temperatures and are outside the scope of the present review.

\emph{Cluster truncated Wigner approximation}.---Another self-consistent approach, with a somewhat different starting point, is the cluster truncated Wigner approximation (CTWA)~\cite{ctwa}. This approach works as follows: one divides the system into non-overlapping clusters, each of size $\ell$, then constructs a basis of $4^\ell$ operators within each cluster (here, as elsewhere in this review, we assume spin-$1/2$ systems unless otherwise specified). The basis operators are taken to be orthonormal under the Frobenius inner product $(A | B) \equiv \mathrm{Tr} (A^\dagger B)$. We denote the basis operators in cluster $j$ as $X^\alpha_j$, $\alpha = 0, 1, \ldots 4^{\ell} - 1$, with $X_i^0 \equiv \mathbb{I}$. A generic local Hamiltonian can be decomposed into terms that act within a cluster and terms that couple adjacent clusters. Terms that act within a cluster are \emph{linear} in terms of the basis operators, while terms that couple different clusters are products of basis operators (one on each cluster). The resulting Hamiltonian has the form

\beq\label{ctwa}
H = \sum_i^\alpha B_{i\alpha} X^\alpha_i + \sum_{\langle ij \rangle} C{i\alpha j\beta} X^\alpha_i X^\beta_j.
\eeq
We now introduce a Schwinger boson representation for the operators, considerably simplifying what follows. Let us define basis states $|d\rangle$ on a cluster, where $|d\rangle$ is the binary representation of the number $d \in \{ 1, 2, \ldots 2^\ell \}$. So for example $|2\rangle = |\downarrow \downarrow \ldots \uparrow \downarrow \rangle$. We can represent $X_\alpha$ in terms of Schwinger bosons as $T^{pq}_\alpha b^\dagger_p b_q$, where $T^{pq}_\alpha = \langle p | X_\alpha | q \rangle$. Plugging this into Eq.~(\ref{ctwa}) gives a bosonic Hamiltonian with quadratic (intra-cluster) and quartic (inter-cluster) terms. 

The truncated Wigner approximation (TWA) is a well-established method for treating such bosonic Hamiltonians. It consists of replacing the bosonic operators with $c$-numbers, which leads to a time-dependent Gross-Pitaevskii equation, and initializing each bosonic variable in an initial state drawn randomly from an appropriate Gaussian probability distribution. One then evolves the Gross-Pitaevskii equation and averages over initial conditions. The classical nonlinear equations of motion will generally be chaotic, and thus lead to thermalization. In this sense there are parallels between CTWA and methods like TDVP (though a major difference is that TDVP is deterministic while CTWA is inherently stochastic). The techniques can potentially be combined, e.g., by using a restricted MPO basis of operators for the CTWA.

\subsubsection{Methods for linear response and local operators}

The methods we discussed above were all, in principle, well suited to studying systems that are far from equilibrium. Thus, they relied on knowing, or at least approximating, the global state. However, many of the key questions regarding MBL are addressable at the level of linear response, and linear response is often much simpler than the full dynamics of the state. The disparate methods described below attempt to make use of this simplification to reliably extract the dynamics of disordered spin chains.

\emph{Light-cone methods}.---Many questions about the dynamics of many-body systems---e.g., transport and autocorrelation functions in equilibrium---can be phrased in terms of the Heisenberg evolution of initially local operators. The operator can be represented as a matrix-product operator (MPO), which can be regarded equivalently as a state in a doubled Hilbert space, if one ``flips'' the bra: i.e., $\sum_{mn} C_{mn} |m \rangle\langle n | \mapsto \sum_{mn} C_{mn} |m \otimes n \rangle$. MPOs can be time evolved using TEBD exactly like MPSs; the quality of the representation depends on the entanglement of the operator viewed as a state, i.e., the ``operator-space entanglement entropy'' (OSEE)~\cite{pizorn}. The evolution of the OSEE is very different from that of the entanglement of a \emph{state}, however: while a state entangles everywhere, an operator is close to the identity outside its lightcone, and its entanglement within the light-cone builds up continuously from zero. This observation was first exploited to study many-body dynamics in Refs.~\cite{enss2012, enss2014}, where the dynamics of local operators after a quantum quench, $\langle \Psi | O(t) | \Psi \rangle$, was computed by representing the operator as an MPO and evolving it backwards in time. To do this, suffices to represent the structure of the operator inside its light-cone; thus, for simple initial states, one can straightforwardly extract answers at short times that are in the thermodynamic limit.

One might wonder if even more drastic simplifications are feasible. To accurately \emph{represent} an operator near its lightcone, even at late times, one only needs modest bond dimensions. At late times, of course, the representation of the operator deep inside the lightcone becomes intractable; however, one might hope that truncation errors inside the lightcone do not affect motion at the lightcone, i.e., that the motion of the operator front is insensitive to dynamics deep inside the lightcone. 
This logic suggests the following method for establishing the rate at which operators spread: one evolves the operator using TEBD in operator space, with a relatively small bond dimension; then one uses data at the butterfly cone or beyond it -- which is assumed to be reliable -- to extrapolate the location of the butterfly cone, and thus the butterfly velocity. 

This program was carried out in Refs.~\cite{shenglong0, shenglong1, shenglong2}; they found clear signs of a ballistic-to-subballistic transition in the operator front, in both the random and quasiperiodic cases. This transition is accompanied by a divergence in the width of the operator front. Quite unexpectedly, in the quasiperiodic case, slow dynamics sets in even in the regime where the noninteracting system would be localized. While these results are suggestive, the question of exactly when one can quantitatively trust butterfly velocities extracted using this algorithm is unsettled at present~\cite{hemery2019tensor}. One danger is as follows: although one can \emph{represent} the exact time-evolved operator using a low-rank MPO near the butterfly cone, the nature of this operator might still depend on details of the dynamics deep inside the butterfly cone that are truncated out in the procedure we have described. Thus there might not be an accurate closed description of the evolution of the light-cone that does not rely on considerable information on the state deep inside the lightcone. Under what conditions this happens is still unclear~\cite{hemery2019tensor, lopes2019}. 

\emph{Other operator-evolution methods}.---We briefly remark on some other methods that have recently been applied to study autocorrelation functions in disordered spin chains. One of these is based on the continued-fraction expansion of autocorrelation functions. We will not discuss this method in detail here, as it is treated in various textbooks~\cite{forster2018hydrodynamic, viswanath2008recursion}. To summarize: we construct a vector space of operators, starting from the operator of interest $O$, by repeated applications of the Liouvillian $\mathcal{L}$. This gives a set of vectors $\{ O, \mathcal{L} O, \mathcal{L}^2 O \ldots \mathcal{L}^{n-1} O \}$, which we orthogonalize by the Gram-Schmidt procedure (using the Frobenius operator inner product introduced above). After the Gram-Schmidt procedure, we have a series of orthogonal operators $\{ | O_n ) \}$, which we will now treat as vectors in operator space. The Liouvillian---written as a matrix acting on this basis---is tridiagonal by construction, and one can show that the autocorrelation function at time $t$ is given by $( O_0 | e^{\mathcal{L} t} | O_0 )$. This formalism lends itself to various resummation and extrapolation schemes, some of which have been applied, e.g., to study dynamics in the subdiffusive phase~\cite{khait2016}. 

Other related methods include short-time series, which can be used to extract moments of the conductivity~\cite{svo}, as well as numerical linked-cluster expansions (NLCEs), which have recently been used to study dynamics~\cite{rigol2014, devakul2015}. All of these approaches build on the simplicity of the short-time series, but require uncontrolled extrapolations to reach the long-time limit.  
However, they are complementary to methods such as exact diagonalization, suffering from a different set of artifacts. These methods are also conceptually important beyond their numerical applications: for instance, the asymptotic structure of the continued fraction expansion---which is related to the high-frequency limit of the conductivity, and to the convergence properties of the short-time series---has been proposed as a diagnostic of quantum chaos~\cite{parker2018}. 

\emph{Boundary-driven Lindblad method}.---A powerful method to compute d.c. response is to consider a setup in which the system of interest is attached to leads at the boundary, at different chemical potentials~\cite{prosen_matrix_2009, znidaric_diffusive_2016, vznidarivc2018interaction}. The evolution of the density matrix then obeys a Lindblad master equation, which approaches (in a generic thermal phase) a unique steady state. The equation for the steady state is given by $\mathcal{L} \rho = 0$, where $\mathcal{L}$ is a ``superoperator'' that acts on the density matrix as follows:

\beq
\mathcal{L}\rho = - i [H, \rho] + \sum_\alpha \gamma_\alpha (2 O_\alpha \rho O^\dagger_\alpha - O^\dagger_\alpha O_\alpha \rho - \rho O^\dagger_\alpha O_\alpha).
\eeq
The operators $O_\alpha$ (sometimes called jump operators) describe how the system is coupled to the environment. In the case at hand, the system is only coupled to the environment at the boundaries, so (for a one-dimensional chain of length $L$) we need four jump operators:

\beq
O_{+, l} = c^\dagger_1, \,\, 
O_{-, l} = c_1 \,\,
O_{+, r}  =  c^\dagger_L \,\,
O_{-, r} = c_L 
\eeq
The operators $O_{+/-,l/r}$ describe the process where a fermion enters/leaves the system (indexed by $\pm$)  from the left/right edge (indexed by $l/r$). The coefficients are chosen so that at the left edge there is a net influx of particles into the system, and at the right edge there is a net outflow. This mimics a situation with a chemical potential difference across the system. Note that this Lindblad description is not necessarily a microscopically accurate description of the physics at the boundary: in general there are many subtleties in applying and interpreting Lindblad master equations for thermodynamically large many-body systems. However, since the bath only acts at the boundary, it does not affect the dynamics deep inside the system, and the master equation is expected to converge to a steady state that is reliable away from the boundaries. 

We have now reduced the problem to finding a good variational approximation to the density matrix for which $|\mathcal{L} \rho|$ is small in some appropriate norm. We work in the limit where the left and right rates are only slightly imbalanced, so one expects the system to come to a steady state that is locally near thermal equilibrium, i.e., $\rho \propto 1 + \epsilon R$ for some matrix $R$. We expect such a near-equilibrium density matrix to be described by a matrix-product operator of low bond dimension, which suggests a matrix-product operator ansatz for $\rho$. Starting from a generic matrix-product operator, one can then use the power method to find the steady state by repeated application of the superoperator $\mathcal{L}$: one time evolves the initial MPO by applying gates, then truncates the resulting (larger) MPO by dropping small Schmidt coefficients, and so on, until the algorithm converges to a steady state. This part of the procedure is standard and reminiscent of imaginary-time evolution to project to the ground state of a Hamiltonian; the main technical distinction is that under non-unitary evolution, initially orthogonal vectors do not remain orthogonal. This can be addressed by re-orthogonalizing the vectors every few steps. 

The Lindblad method is restricted in scope---it can address questions about the steady state, and conceivably about asymptotic convergence, but it does not directly offer a way to probe dynamics. Further, its convergence to the steady state depends on how long the system takes to reach its steady state; as one approaches the MBL transition this convergence slows down dramatically and the method ceases to be useful. Regardless, it is capable of addressing key questions about steady-state currents in large systems for which no other methods exist; many of the most definite existing results on diffusion and subdiffusion rely on this method. 

\subsection{Summary}

This section has surveyed a number of new ideas and algorithms that exploit the structure of either the MBL or thermal phase to describe late-time dynamics in an approximate way. These algorithms are a focus of much current research, and are undergoing rapid development. At present, none of them is able to access the critical properties of the MBL transition, but these approaches hold the promise of characterizing the thermal and localized phases in more detail than has been possible so far, and thus constraining possible theories of the MBL transition. 

Finally, we should remark that in addition to better numerical methods, a fruitful approach has been to ask sharp questions about simplified toy models.
One important instance of this is Ref.~\cite{ldrh}, which considered a toy model of a random matrix coupled to many single spins. The approach taken there was to test a very specific hypothesis---\emph{incorporating nearby spins helps a bath to incorporate more distant spins}---using exact small-system numerics. 
Such specific results constrain the space of possible theories of the MBL transition, ruling out certain scenarios entirely. 
RG approaches to the MBL transition, in particular, are built up piecemeal out of plausible hypotheses about single RG steps, and these can be tested even on relatively modest systems.

\section{Conclusions and Open Questions \label{sec:conclusions}}

In this review, we have described  how the techniques and conceptual framework of many-body localization may be fruitfully applied to systems in which, strictly speaking, MBL itself does not occur for a variety of reasons. The ``threshold of MBL" can be approached in a variety of ways. e.g., by tuning a system until it is proximate to an MBL transition; by working near a putative many-body mobility edge or a $d>1$ MBL phase (both of which are destroyed by rare-region effects); by considering multi-component systems some of whose components are constrained by symmetry to thermalize; or by driving systems into long-lived prethermal regimes. Various aspects of MBL --- such as the existence of local integrals of motion and the absence of transport of conserved quantities --- are clearly no longer strictly applicable in this setting, but nevertheless constrain the dynamical behaviour. In particular, features such as dynamical heterogeneity, slow relaxation to equilbrium, and  (in one dimension) subdiffusive transport remain quite robust even away from the rather rarefied settiings where strict MBL emerges, and indicate a degree of universality beyond what might na\"ively have been expected. We take the point of view that the most convenient manner in which to study such phenomena is to couple a picture of the  MBL phase `beyond the threshold'  with an understanding of the mechanism whereby the system in question avoids MBL --- and hence of the processes that eventually drive thermalization. We have discussed the common dynamical signatures that emerge across such problems, and surveyed an array of theoretical challenges in studying such systems and critically assessed existing methods devised to circumvent these.

To round out our discussion, we identify several  open questions raised by these considerations. Several of these have already  been alluded to in the preceding sections; however, here we flag those whose resolution we believe would be especially impactful --- particularly in light of existing  or near-term experiments and numerical simulations. 

First, in light of the relative ease with which cold atom experiments can probe quasiperiodic systems ---  including in $d>1$ --- and the extant data on apparent localization in such systems, it is clearly of great significance to establish whether quasiperiodic systems  indeed exhibit  MBL,  where  in even in $d=1$ there are no rigorous results at the level of Ref.~\cite{jzi}. The issue is, as we have noted, subtle.  On the one hand, quasiperiodic systems are naturally hyperuniform and hence free of the rare-region instabilities that plague MBL systems. On the other hand, the near-repetition of local motifs means that there can be other `coherent' routes to delocalization, absent in more standard disordered systems, whose role has not been carefully analyzed to date.

Second, by far the most striking prediction in the near-MBL regime of one-dimensional disordered systems is the existence of subdiffusion. The extent of the subdiffusive regime is as not fully settled. {Many early numerical studies, mostly relying on exact diagonalization, found subdiffusion for arbitrarily weak disorder, and for quasiperiodic as well as random systems~\cite{BarLev_Absence_2015, khait2016, lev2017transport}. Neither of these results is what one would expect theoretically if the Griffiths scenario were the true cause of subdffusion. However, the most recent large-scale simulations~\cite{znidaric_diffusive_2016, vznidarivc2018interaction, doggen1, doggen2} support the existence of a diffusive phase at small disorder. Nevertheless, even these simulations find a transition to subdiffusion at much weaker disorder than the MBL critical point. We are used to thinking of Griffiths effects as a property of the phase ``near'' the critical point, and these observations are in some tension with that picture (though it also appears that the critical regime of the MBL transition is very broad in parameter space)}. 
Also, in systems with multiple conserved charges, it is not clear whether the  time scales of subdiffusion of distinct charges all coincide, or whether there can be richer behaviour in this setting.

A third and related open issue is a more nuanced understanding of the ETH phase, and in particular, of the emergence of MBL viewed from the perspective of the thermal side of the transition. At present, the most fully developed theories of the transition begin with a typically (i.e. perturbatively) stable localized phase and examine how rare/non-perturbative thermal inclusions destabilize it --- this is the view that  leads us naturally to the KT picture  discussed in Sec.~\ref{sec:intro:MHRG}. Single-particle Anderson localization admits a distinct perspective from the delocalized side: namely, the weak localization correction to the conductivity identifies processes that lead to localization at strong disorder. It is far more challenging to develop a similar perspective on MBL, and a central challenge is the relatively coarse level of detail provided by ETH. For instance, as noted in Ref.~\cite{foini_kurchan, chalker_eth}, the phenomenon of scrambling and the butterfly effect and related ideas of operator spreading imply structure beyond that captured by ETH. Extending these considerations  to the Griffiths phase is a natural direction to explore: it is possible that such detailed considerations beyond `bare-bones' ETH might distinguish the Griffiths phase near the MBL transition it from a conventional diffusive system.  Also, a key tenet of ETH is the RMT ansatz for the energy levels; meanwhile, MBL systems  have  Poissonian level statistics. It is unclear if the behavior at the ETH-MBL transition, or indeed the entire  subdiffusive regime. should be characterized by some distinct universal random matrix ensemble intermediate between these regimes. If this is the case, then such new universal ensembles provide a new perspective on near-MBL dynamics.

Fourth, and finally, we note that there remain many open questions surrounding the MBL transition, the fractal dimension of thermal blocks inside the localized phase, the efficacy of thermal blocks in subsuming proximate localized spins etc., for which there  the answers are for the most  part inconclusive or poorly understood. Further investigation along these lines is clearly warranted, and as noted in Sec.~\ref{sec:conclusions}, a judicious design of numerical experiments and toy models may circumvent  some of the daunting technical challenges 
of addressing such questions.

\setcounter{secnumdepth}{0}

\section{Acknowledgements}
We thank K.~Agarwal, I.~Bloch, P.~Bordia, E. Demler, P.T. Dumitrescu, A.J.~Friedman, A.~Goremykina, D.A.~Huse, V.~Khemani, M.~Knap, M.D.~Lukin, M.~M\"uller, R.~Nandkishore, A.C.~Potter, U.~Schneider, M.~Serbyn, R.~Vasseur and others for 
helpful discussions and collaborations on related work. S.G. acknowledges support from NSF DMR-1653271. S.A.P. acknowledges support from NSF Grant DMR-1455366 for previous research on some of the topics discussed in this review, and from EPSRC grant EP/S020527/1 as this review was completed. Both authors acknowledge the hospitality of the Kavli Institute for Theoretical Physics, which is supported by NSF Grant PHY-1748958, where some of this review was completed.

\section{Glossary}

\begin{description}
\item[Area/volume law]
Succinct expression for the scaling of the (usually, von Neumann) entropy with the size of a subregion. Ground (highly excited) states of local Hamiltonians generically satisfy area (volume) laws for entanglement. MBL eigenstates satisfy area laws, while ETH dictates that eigenstates at finite energy density satisfy a volume law.
\item[Avalanche instability]
Nonperturbative instability of MBL systems in the presence of a thermal inclusion.
\item[ETH]
Eigenstate thermalization hypothesis. ETH posits that local expectation values of operators in any eigenstate of a many-body quantum chaotic system coincide with thermal expectation values at a temperature (and chemical potentials) set by the mean energy (and densities). 
\item[Frobenius inner product and Frobenius norm]
For two matrices $A$ and $B$, the Frobenius inner product $(A | B) = \mathrm{Tr}(A^\dagger B)$. The corresponding Frobenius norm is $\Vert A \Vert = \sqrt{(A | A)}$. 
\item[Griffiths effects/inclusions]
Rare regions in an inhomogeneous system in which the control parameter driving a phase transition is either unusually large or unusually small, so the system locally seems to be in the wrong phase.
\item[IPR]
Inverse participation ratio: defined for a normalized quantum state on a lattice as $\mathrm{IPR}_q = \sum_i |\psi_i|^{2q}$. When $q$ is not explicitly specified, it is assumed to be $2$. Generalizations can be defined for any normalized list.
\item[LIOM or l-bit]
A local integral of motion, i.e., a quasi-local operator $O$ that has a nonzero fraction of its support on a small, finite number of sites in the thermodynamic limit (i.e., has a finite inverse participation ratio in the Hilbert space of operators), and that \emph{exactly} commutes with the time-evolution operator.
\item[MBL]
Many-body localization (see main text). 
\item[MPO]
Matrix product operator. An MPO can be regarded as an operator written as a \emph{state} in the Hilbert space of operators, which is then compressed using MPS methods. One can equivalently regard an MPO as an operator that acts on the space of MPS's of a certain size. 
\item[MPS]
Matrix product state. A class of variational quantum states with low entanglement that are widely used in numerical simulations, because algorithms like TEBD rely on them.
\item[OTOC] 
Out-of-time-order correlator, defined as $\langle A(t) B(0) A(t) B(0) \rangle$ for arbitrary operators $A$ and $B$. Sometimes this acronym refers instead to the squared commutator, $\langle [A(t), B(0)]^2 \rangle$. While causal response functions measure the \emph{expectation value} of the commutator, the OTOC measures its variance. 
\item[perturbative resonance]
What happens in perturbation theory when the matrix element exceeds the energy denominator.
\item[R\'enyi entropy, Von Neumann entropy]
The $n$th R\'enyi entropy is defined as $S_n = (1 - n)^{-1} \mathrm{Tr}(\rho^n)$. The Von Neumann entropy is $S_1 \equiv \lim_{n \rightarrow 1} S_n = \mathrm{Tr} (\rho \ln \rho)$. Here, $\rho$ is a density matrix (typically the reduced density matrix when one is computing entanglement).
\item[RSRG, SDRG]
Real-space renormalization-group/strong-disorder renormalization group. A renormalization-group method that iteratively eliminates sites or bonds on a lattice and that is controlled for strong quenched randomness. Characterized by broad fixed-point distributions for various observables, as evinced by starkly congtrasting behavior of typical and disorder-averaged quantities.
\item[TEBD]
Time-evolving block decimation, a method for studying the dynamics of quantum states that are not too entangled, using a matrix-product representation. 

\end{description}

\section{References}

\setlength{\bibsep}{0.0pt}
\bibliography{library}
\bibliographystyle{iopart-num}

\end{document}